\documentclass[aps,preprint,nofootinbib,floatfix]{revtex4}
\usepackage{graphicx,color}
\usepackage[latin1]{inputenc}
\usepackage{amsmath,amssymb}
\usepackage{hyperref}
\usepackage{epstopdf}

\newcommand{\lsim}{\mathrel{\mathop{\kern 0pt \rlap
  {\raise.2ex\hbox{$<$}}}
  \lower.9ex\hbox{\kern-.190em $\sim$}}}
\newcommand{\gsim}{\mathrel{\mathop{\kern 0pt \rlap
  {\raise.2ex\hbox{$>$}}}
  \lower.9ex\hbox{\kern-.190em $\sim$}}}

\newcommand{\sigmav}{\ensuremath{\langle\sigma v\rangle}}
\newcommand{\sigsip}{\ensuremath{\sigma^{\rm{SI}}_p}}

\newcommand{\gev}{\ensuremath{\,\mathrm{GeV}}}

\def  \bcen   {\begin{center}}
\def  \ecen   {\end{center}}
\def  \beq    {\begin{equation}}
\def  \eeq    {\end{equation}}
\def  \beqa   {\begin{eqnarray}}
\def  \eeqa   {\end{eqnarray}}

\def\bea{\begin{eqnarray}}
\def\eea{\end{eqnarray}}

\begin{document}
\title{An Updated Analysis of Inert Higgs Doublet Model \\
in light of the Recent Results from\\ 
LUX, PLANCK, AMS-02 and LHC}
\author{
Abdesslam Arhrib$^{1,2}$, Yue-Lin Sming Tsai$^{2,3}$,
Qiang Yuan$^{4,5}$ and Tzu-Chiang Yuan$^2$ 
}

\affiliation{
$^1$AbdelMalek Essaadi University, Faculty of Science and Techniques, B.P 416 Tangier, Morocco\\
$^2$Institute of Physics, Academia Sinica, Nankang, Taipei 11529, Taiwan\\
$^3$Kavli IPMU (WPI), University of Tokyo, Kashiwa, Chiba 277-8583, Japan\\
$^4$Key Laboratory of Particle Astrophysics, Institute of High 
Energy Physics, Chinese Academy of Sciences, Beijing 100049, P. R. China\\
$^5$Key Laboratory of Dark Matter and Space Astronomy, 
Purple Mountain Observatory, Chinese Academy of Sciences, 
Nanjing 210008, P. R. China }

\date{\today}

\begin{abstract}
In light of the recent discovery by the ATLAS and CMS experiments at the 
Large Hadron Collider (LHC) of a Higgs-like particle
with a narrow mass range of 125-126 GeV, we perform an updated analysis on 
one of the popular scalar dark matter models, the Inert Higgs Doublet Model (IHDM).
We take into account in our likelihood analysis of various experimental
constraints, including recent relic density measurement, dark matter direct and 
indirect detection constraints as well as the latest collider constraints on the 
invisible decay width of the Higgs boson and monojet search at the LHC.
It is shown that if the invisible decay of the standard model Higgs boson is open, 
LHC as well as direct detection experiments like LUX and XENON100
could put stringent limits on the Higgs boson couplings to dark matter.  
We find that the most favoured parameter space for IHDM 
corresponds to dark matter with a mass less than 100 GeV or so.
In particular, the best-fit points are at the dark matter mass around 70 GeV
where the invisible Higgs decay to dark matter is closed.
Scalar dark matter in the higher mass range of 0.5-4 TeV is also explored in our study.
Projected sensitivities for the future experiments of monojet at LHC-14, XENON1T 
and AMS-02 one year antiproton flux are shown to put further constraints 
on the existing parameter space of IHDM.
 
\end{abstract}
\maketitle

\section{Introduction \label{section:1}}

The 7 TeV and 8 TeV run at the Large Hadron Collider (LHC) have revealed and confirmed the existence 
 of a Higgs-like particle $h$ in the standard model (SM) with mass in the narrow range 
 of 125-126 $\gev$  \cite{ATLAS,CMS}. This discovery is also verified by the recent Tevatron final results \cite{Tevatron}.
The observation of this new particle 
combines evidence in the decays $h\to \gamma\gamma$, $h\to ZZ^*$ and 
$h\to W^\pm W^{\mp *}$. Different signal strengths, defined as the product of Higgs boson 
production cross sections from different channels and the branching ratios 
for different decay modes normalized to the corresponding products in SM, 
have been measured with good precision 
by both experiments at ATLAS and CMS \cite{ATLASn,CMSn,CMS-PAS-HIG-12-045}. 
These measurements will be further improved in the future 13-14 TeV run at the LHC,
and perhaps at a future International Linear Collider (ILC) should this machine ever be built.
From these signal strengths measurement one can 
extract information on the couplings of this Higgs-like particle 
to the gauge bosons and SM fermions.  From the most recent measurements, 
extraction of this Higgs-like particle couplings to SM particles seem 
to be consistent to a great extent with those of the SM Higgs boson couplings
\cite{Carmi:2012yp}. Moreover, data collected both at ATLAS and CMS 
indicate that this Higgs-like particle has zero spin and is CP-even, {\it i.e.} $J^{P}=0^{+}$ is 
preferred \cite{atlaspin,cmspin}.

The discovery of the Higgs-like particle at the LHC indicates for 
the first time that fundamental scalar exists in Nature. 
Certainly, many phenomenological models that extend the SM scalar sector with just one scalar doublet existed already in the literature. Some of them are motivated by physics of the dark matter (DM) or
neutrinos masses. Among these extensions, we have 
models with multiple Higgs doublets, with one Higgs doublet and 
multiple singlets or triplets etc. All these extensions should have one light scalar with 
Higgs-like couplings to SM particles in the range tolerated by signal strength measurements. 
Indeed many studies (see for example the references in \cite{Carmi:2012yp}) have been done 
using these data to constrain various extensions of the scalar sector of the SM.

In this paper, we concentrate on the Inert Higgs Doublet Model (IHDM) 
which is a very simple extension of the SM. It was first proposed 
by Deshpande and Ma \cite{Deshpande:1977rw} in order to study the pattern of 
electroweak symmetry breaking.  
The IHDM is an attractive model due to its simplicity.
It is basically a Two Higgs Doublet Model (THDM) 
(see \cite{THDM} for a recent overview) with an imposed exact 
$\mathbb{Z}_2$ symmetry. Under the $\mathbb{Z}_2$ symmetry, 
all the SM particles are even representing the visible sector, while the new Higgs doublet
field is odd representing the inert dark sector. 
Imposing the $\mathbb{Z}_2$ symmetry forbids the second Higgs doublet developing a 
vacuum expectation value (VEV) and all the inert particles in this doublet
can only appear in pair in their interaction vertices. Indeed, 
recent studies \cite{globalCLT,globalBelangeretal1,globalBelangeretal2,globalOthers} 
of global fits of the LHC data suggest 
that the couplings between the $W$ and $Z$ gauge bosons with the new 125-126 GeV Higgs-like 
boson are very close to their SM values. The new 125-126 GeV boson may play 
the entire role of electroweak symmetry breaking (EWSB) and
leave no room for other Higgs fields to develop any VEVs. This favors the IHDM.
As a result, IHDM exhibits very interesting phenomenology. It predicts
the existence of a neutral scalar field, denoted generically by $\chi$ here, which 
is the Lightest Odd Particle (LOP) in this model and 
will play the role of DM candidate.  
The Higgs mechanism provides a portal for communication between the inert 
dark sector and the visible SM sector. Thus if kinematics allowed, the SM Higgs boson may
decay into a pair of DM $\chi$ and will contribute to the invisible SM Higgs boson width 
which is now constrained by the LHC data.
Moreover, annihilation of $\chi$ into SM particles will provide 
thermal relic density and the scattering of 
$\chi$ onto nucleons will lead to direct detection signatures.
Therefore, IHDM could be considered as a simple but competitive model in the market
with a weakly interacting massive particle (WIMP). As we will
see later, IHDM could predict correct DM relic density as well as 
a cross section for scattering of $\chi$ onto nucleons that is 
consistent with existing data from direct detection.
Almost three decades later, IHDM was extended further by Ma \cite{Ma:2006km}
to include three $\mathbb{Z}_2$ odd weak singlets of right-handed neutrinos with Majorana masses. 
In this extended model \cite{Ma:2006km}, a radiative seesaw mechanism for light neutrino masses 
was proposed and either $\chi$ or one of the right-handed neutrinos could be DM candidate.
We will not consider this extended version of IHDM in this work but
would like to return to this in the future \cite{wip}.

As mentioned earlier, there have been many attempts to introduce 
DM Higgs models by extending the SM scalar sector with 
more singlets or doublets \cite{Silveira-Zee,Cheung:2012xb,higgsportal-scalar1,higgsportal-scalar2,higgsportal-scalar4}.
In particular, the phenomenology of IHDM had been extensively discussed in 
the context of DM phenomenology 
\cite{Goudelis:2013uca,Gustafsson:2012aj,LopezHonorez:2010tb,idmdm,unitarity,maria} 
and also for collider phenomenology 
\cite{Lundstrom:2008ai,idm_lhc,Dolle:2009ft}. IHDM has been also advocated to 
explain the naturalness problem \cite{Barbieri:2006dq}. 
In the present study, we will reconsider the IHDM model in light
of the recent ATLAS and CMS discovery of a Higgs-like particle of 125-126 GeV.
We assume that the LOP must fulfill the recent relic density measurement
by PLANCK \cite{Ade:2013zuv}. As a good DM candidate test, we also
consider the constraints from DM direct and indirect detection.
For the constraint from DM direct detection search, we study the impact
from the most recent LUX upper limit \cite{Akerib:2013tjd}
which provides a robust constraint on the parameter space. 
As for indirect detection, we will take into
account the Fermi-LAT $\gamma$-ray observations of the dwarf spheroidal
galaxies (dSphs) \cite{Ackermann:2011wa} and the Galactic center (GC) \cite{Huang:2012yf}. 
In addition to $\gamma$-rays, we also include constraints from cosmic ray
electrons/positrons from AMS-02 \cite{Aguilar:2013qda}, PAMELA \cite{Adriani:2008zr},
and Fermi-LAT \cite{FermiLAT:2011ab,Ackermann:2010ij}, and cosmic ray
anti-protons from PAMELA \cite{Adriani:2010rc}. These constraints will be also supplemented by
the LHC constraints such as monojet and diphoton signal strength
measurement as well as constraint on the Higgs boson invisible decay width.

Some of the above aspects for IHDM have been discussed in recent studies 
\cite{Gustafsson:2012aj,Goudelis:2013uca}. 
The compatibility of a heavy SM Higgs boson with LHC results 
and XENON100 data \cite{Aprile:2012nq}
were discussed in Ref.~\cite{Gustafsson:2012aj}\footnote{ 
In a note added in \cite{Gustafsson:2012aj}, 
the consistency of IHDM with the 125-126 GeV Higgs-like particle observed at LHC and XENON100 were also discussed.}.
Similar issues for IHDM were discussed in Ref.~\cite{Goudelis:2013uca} with the inclusion of radiative
corrections to the scalar masses of the model. Ref.~\cite{Goudelis:2013uca} 
also included renormalization group effects for the quartic scalar couplings $\lambda_i$ in order to evaluate 
vacuum stability, perturbativity and unitarity constraints at a higher scale. 
In our analysis, we will go further by including also the following aspects:     
\begin{itemize}
\item[1.] Larger parameter space for 
DM mass: we will scan $m_\chi$ from 5 GeV to 4 TeV. 
\item[2.] LHC monojet constraint in the likelihood. 
\item[3.] Accurate DM indirect detection likelihood. 
\item[4.] Constraints from the first result of LUX in direct detection likelihood.
\item[5.] Future sensitivity to monojet search at LHC with 14 TeV 
at the planned luminosity of 100 fb$^{-1}$ and  300 fb$^{-1}$.
\item[6.] Sensitivity to AMS-02 anti-protons and XENON1T.
\end{itemize}

The layout of this paper is as follows. In section 2, 
we briefly review IHDM and its parameterization.
We then list the theoretical constraints such as perturbativity, 
perturbative unitarity and vacuum stability that must be satisfied by 
the scalar potential parameters. The constraints from collider searches   
that IHDM is subjected to are discussed in section 3. These include:
electroweak precision test constraints, $W$ and $Z$ width constraints, 
negative search for charginos and neutralinos from LEP-II that could restrict
the inert Higgs bosons masses, diphoton signal strength measurement as well as
monojet constraint from DM search at LHC.
In section 4, we will discuss the relic density measurement 
by PLANCK as well as DM direct detection and indirect detection constraints. 
In section 5, we present our methodology for 
likelihood analysis and explain how all the constraints are included.
We present our numerical results in section 6. Future experimental constraints from LHC-14,
XENON1T and AMS-02 are discussed in section 7. We conclude in section 8.

\section{Inert Higgs Doublet Model (IHDM) \label{section:2}}

In this section, we briefly review the salient features of IHDM and discuss some existing 
theoretical constraints.


\subsection{Parameterization of the IHDM scalar potential\label{section:2a}}

The IHDM \cite{Deshpande:1977rw} is a rather 
simple  extension of the SM Higgs sector. It contains the SM Higgs doublet 
$H_1$ and an additional Higgs doublet $H_2$. This model has a $\mathbb{Z}_2$
symmetry under which all the SM fields including $H_1$ are even while  $H_2$ 
is odd under $\mathbb{Z}_2$: $H_2 \to - H_2$. 
We further assume that $\mathbb{Z}_2$ symmetry is not spontaneously 
broken {\it i.e.} $H_2$ field does not develop VEV. 
These doublets can be parameterized as: 
\beq
H_1 =
   \left( \begin{array}{c}  G^+ \\ \frac{1}{\sqrt{2}} \left( v + 
h + i G^0 \right)  \end{array}
     \right)  ~~~~, ~~
H_2 =    \left( \begin{array}{c}  H^+ \\ 
\frac{1}{\sqrt 2} (S  + i A)  \end{array}  \right)
\eeq
where $G^\pm$ and $G^0$ are the charged and neutral 
Goldstone bosons respectively, which will be absorbed by the $W^\pm$ and $Z$ 
to acquire their masses.

The scalar potential with an exact $\mathbb{Z}_2$ symmetry 
forbids the mass term $-\mu_{12}^2 (H_1^\dagger H_2 + {\rm h.c.})$ which mixes $H_1$ and $H_2$.
Thus it has one fewer term than in THDM, {\it i.e.}
\begin{eqnarray}
\label{potential}
V &=& \mu_1^2 |H_1|^2 + \mu_2^2 |H_2|^2 + \lambda_1 |H_1|^4
+ \lambda_2 |H_2|^4 +  \lambda_3 |H_1|^2 |H_2|^2 + \lambda_4
|H_1^\dagger H_2|^2 \nonumber \\
&& \;\;\;\; + \; \frac{\lambda_5}{2} \left\{ (H_1^\dagger H_2)^2 + {\rm h.c.} \right\} \;\; .
\end{eqnarray}
The electroweak gauge symmetry is broken when $H_1$ doublet gets 
its VEV:
$\langle H_1^T \rangle = \left( 0 , \; v/\sqrt{2} \right)$ while $\langle H_2 \rangle = 0$.
This pattern of symmetry breaking ensures unbroken $\mathbb{Z}_2$ symmetry and
results in one more CP-even neutral scalar $S$, one CP-odd neutral
scalar $A$, a pair of charged scalars $H^+$ and $H^-$ in addition to the SM CP-even scalar Higgs $h$.
Note that since $h$ is the SM Higgs boson, it is $\mathbb{Z}_2$ even, 
while $S$, $A$ and  $H^\pm$ are $\mathbb{Z}_2$ odd.
Moreover, the exact $\mathbb{Z}_2$ symmetry naturally imposes the 
flavor conservation. Only SM Higgs boson couples to SM fermions while the inert Higgses 
$S$, $A$ and $H^\pm$ do not.
The $\mathbb{Z}_2$ symmetry also ensures the stability of the 
lightest scalar ($S$ or $A$) that can act 
as a DM candidate. DM phenomenology of IHDM had been studied extensively
in the literature 
\cite{Goudelis:2013uca,Gustafsson:2012aj,idmdm,unitarity,maria,idm_lhc,Dolle:2009ft,Gustafsson:2007pc}. 

The above scalar potential in Eq.~(\ref{potential}) has 8 real
parameters: 5 $\lambda_i$, 2 $\mu_i^2$ and the VEV $v$. 
Minimization condition 
for the scalar potential eliminates $\mu_1^2$ in favour of the Higgs mass 
and the VEV $v$ is fixed to be 246 $\gev$ by the weak gauge boson masses.
We are left with 6 independent real parameters.
The masses of all the four physical scalars can be written in terms of 
$\mu_2^2, \lambda_1, \lambda_2, \lambda_3,\lambda_4$ and $\lambda_5$ as the following
\begin{eqnarray}
\label{mh-lambda1}
&& m_h^2=-2 \mu_1^2=2 \lambda_1 v^2 \\
&& m_S^2=\mu_2^2+\frac{1}{2} (\lambda_3 +\lambda_4 + \lambda_5) v^2=
\mu_2^2+ \lambda_L v^2 \\
&& m_A^2=\mu_2^2+\frac{1}{2} (\lambda_3 +\lambda_4 - \lambda_5) v^2 = 
\mu_2^2+ \lambda_A v^2 \\
&& m_{H^\pm}^2= \mu_2^2+ \frac{1}{2} \lambda_3  v^2 
\end{eqnarray}
where 
\beq
\lambda_{L,A}=\frac{1}{2}(\lambda_3 +\lambda_4 \pm \lambda_5) \; .
\eeq
Four of the five quartic couplings can be written in terms of 
physical scalar masses and $\mu_2^2$ as the following expressions
\beqa
\lambda_1 &=& \frac{m_h^2}{2 v^2}\quad , \quad 
\lambda_3 = \frac{2}{v^2}\left(m_{H^\pm}^2 - \mu_2^2\right), \\
\lambda_4 &=& \frac{\left(m_S^2 + m_A^2 - 2 m_{H^\pm}^2\right)}{v^2} \quad ,
\quad 
\lambda_5 = \frac{\left(m_S^2 - m_A^2\right)}{v^2} \;\; .
\label{lambds}
\eeqa
We are then free to take $(\lambda_i)_{i=1,\ldots , 5}$ and $\mu_2^2$ as 6 independent parameters, or 
equivalently, the following set
\beq
\{m_h, m_S, m_A, m_{H^\pm}, \lambda_2 , \lambda_L \}
\label{eq:input_par_2}
\eeq
which is more convenient for our purposes to describe the full scalar sector. 
In our tree level parameterization,  $\lambda_A$ can be expressed as
\beq
\lambda_A = \lambda_L - \lambda_5 = \lambda_L + \frac{m_A^2-m_S^2}{v^2} \; .
\label{eq:lambda_A}
\eeq
It is clear from Eq.~(\ref{eq:lambda_A}) that $\lambda_A >\lambda_L$ for
$m_A>m_S$ and  $\lambda_A <\lambda_L$ for
$m_A<m_S$. 
In our systematic scan in the following numerical work, we will consider both cases 
where $\chi=S$ or $\chi=A$ being the LOP. Thus, the DM mass is defined as
\beq
m_{\chi}=\min\lbrace m_S,m_A\rbrace.
\eeq
In order to illustrate constraint on $\lambda_{L,A}$ with $S$ or $A$ 
being the LOP, we define the coupling $g_{h\chi\chi}$ as
\begin{eqnarray}
g_{h \chi \chi} = - 2 v \lambda_{\chi \chi} \; \; {\rm with} \;\; \lambda_{\chi\chi}=
\left\lbrace
\begin{array}{l}
\lambda_L\qquad {\rm if} \qquad \chi = S \; , \\
\lambda_A\qquad {\rm if} \qquad \chi = A  \; .
\end{array}
\right.
\label{eq:lamchi}
\end{eqnarray}
The coupling $g_{h \chi \chi}$ shows up directly in the relic density computation 
depends on whether $\chi = S$ or $\chi = A$.


\subsection{Theoretical constraints\label{section:2b}}

The parameters of the scalar potential of the IHDM are severely constrained
by theoretical constraints. First, to trust our perturbative calculations
we have to require all quartic couplings in the scalar potential of
Eq.~(\ref{potential}) to obey $|\lambda_i| \le 8 \pi $. Second,
in order to have a scalar potential bounded from below we must also demand 
the following constraints \cite{THDM}:
\begin{eqnarray}
\lambda_{1,2} > 0 \quad \rm{and} \quad \lambda_3 + \lambda_4 -|\lambda_5| +
2\sqrt{\lambda_1 \lambda_2} >0 \quad\rm{and} \quad\lambda_3+2\sqrt{\lambda_1
  \lambda_2} > 0 \; .
  \label{eq:lamCONS}
\end{eqnarray}
Third, to further constrain the scalar potential parameters of the
IHDM one can impose tree-level unitarity in a variety of scattering processes among the various scalars
and gauge bosons.
For the unitarity constraints, it is convenience to define the following twelve parameters $e_i$ \cite{unitarity}:
\begin{eqnarray}
&&e_{1,2}=\lambda_3 \pm \lambda_4 \quad , \quad
e_{3,4}= \lambda_3 \pm \lambda_5 \quad , \\
&&e_{5,6}= \lambda_3+ 2 \lambda_4 \pm 3\lambda_5\quad , \quad
e_{7,8}=-\lambda_1 - \lambda_2 \pm \sqrt{(\lambda_1 - \lambda_2)^2 + \lambda_4^2} \quad , 
\\
&& e_{9,10}= -3\lambda_1 - 3\lambda_2 \pm \sqrt{9(\lambda_1 - 
\lambda_2)^2 + (2\lambda_3 + \lambda_4)^2} \quad , 
\\ &&  e_{11,12}= -\lambda_1 - \lambda_2 \pm \sqrt{(\lambda_1 - \lambda_2)^2 + 
 \lambda_5^2} \quad .
\end{eqnarray}
The perturbative unitarity constraints are then imposed on all $e_i$ satisfying \cite{unitarity} 
\beq
|e_i| \le 8 \pi ~, \forall ~ i = 1,...,12.
\eeq 
We observe that $e_{9,10}$ give the strongest constraints on $\lambda_{1,2}$ when 
$2\lambda_3 + \lambda_4=0$, which translate into
\begin{equation}
\label{eq:lam2CONS}
\lambda_{1,2}\leq \frac{4 \pi}{3} \; .
\end{equation}
In fact, from Eq.~(\ref{mh-lambda1}) with $m_h = 126$ GeV and $v=246$ GeV 
we have $\lambda_{1} =\frac{m_h^2}{2 v^2}=0.13$ 
which is well below the unitarity bound given by Eq.~(\ref{eq:lam2CONS}).


\section{Collider Constraints} \label{section:3}


In this section, we discuss DM constraints from the collider search experiments.
We will focus on the constraints from the electroweak precision test (EWPT) experiments at LEP-II,
neutral and charged Higgs search at LEP-II, as well as the mass and the invisible width of the Higgs, 
diphoton signal strength and monojet search from the LHC.

\begin{itemize}

\item\underline{\bf Electroweak precision tests:} 

EWPT is a common approach to constrain 
physics beyond SM by using the global 
electroweak fit through the oblique $S$, $T$ and $U$ parameters
\cite{Ref:PeskinTakeuchi}. It is well known that in 
the SM the EWPT implies a close relation between the three masses $m_t$, $m_h$ and $m_W$. 
Similarly, in the IHDM, the EWPT implies constraints on the mass splitting among 
the Higgs boson masses \cite{Barbieri:2006dq}. 
In this study,  we will use the PDG values of
$S$ and $T$ with $U$ fixed to be zero \cite{Baak:2012kk}. 
We allow $S$ and $T$ parameters to
be within 95\% C.L. (Confidence Level). The central value of $S$ and
$T$, assuming a SM Higgs boson mass of $m_h = 126$
GeV, are given by \cite{Baak:2012kk} :
\beq
S = 0.05 \pm 0.09 \; , ~ ~ T = 0.08 \pm 0.07 \; .
\eeq
The correlation between $S$ and $T$ is 91\% in this fit. 
Analytic expressions for $S$ and $T$ in IHDM can be found in Ref.~\cite{Barbieri:2006dq}.

\item \underline{\bf LEP limits on neutral and charged Higgs bosons:}

Other LEP constraints come from the precise measurements of 
$W$ and $Z$ widths. In order not to affect  these decay widths 
we demand that the channels $W^\pm\to \{SH^\pm, AH^\pm\}$ 
and/or $Z\to \{SA,H^+H^-\}$ are kinematically not open. 
This leads to the following  constraints: 
$m_{S,A}+m_{H^\pm}>m_W$, $m_A+m_S>m_Z$ and $2 m_{H^\pm}>m_Z$ \cite{pdg}.

Additional constraints on the charged Higgs boson $H^\pm$, CP-even $S$ 
and CP-odd $A$ masses can be derived. Note that LEP, Tevatron and LHC
bounds on $H^\pm$ and $A$ can not apply because the standard 
search channels assumes that those scalars decays into a pair of fermions 
which are absent in the IHDM due to $\mathbb{Z}_2$ symmetry. 

In the IHDM, if $S$ is the LOP the CP-odd $A$ could decay like $A\to SZ$,
while the charged Higgs boson $H^\pm$ could decay into $W^\pm S$ 
and/or $W^\pm A$ followed by $A\to SZ$. Therefore the final states
of the two production processes $e^+e^-\to H^+H^-$ and 
$e^+e^-\to SA$ would be multi-leptons or 
multi-jets, depending on the decay products of $W^\pm$ and $Z$, 
plus missing energies. 
To certain extents, the signatures for the charged Higgs case 
would be similar to the supersymmetry searches for charginos and neutralinos at 
$e^+e^-$ or at hadron colliders \cite{idm_lhc,Dolle:2009ft}.
\footnote{The projection of the experimental limits from SUSY 
searches to IHDM has to be made with some care since the production cross 
sections for the fermionic chargino/neutralino pair in the SUSY case are different
from the scalar pairs of $H^\pm H^\mp$ and $SH^\pm$ in the IHDM case \cite{Pierce:2007ut}.
The cross sections for fermionic and scalar pair production are scaled by $\beta^{1/2}$ and
$\beta^{3/2}$ respectively, where $\beta$ is the velocity of the final state particle in the
center-of-mass frame. Hence, the scalar pair will be suppressed by an extra factor of $\beta$
as compared with the fermionic case.}
Taking into account these considerations, we will safely choose in our scan 
for the charged Higgs mass $m_{H^\pm}$ being always greater than $70\gev$.
For the neutral inert Higgses $S$ and $A$, neutralinos search at LEP-II via 
$e^+e^- \to \widetilde{\chi}_1^0\widetilde{\chi}_2^0$ followed by 
$\widetilde{\chi}_2^0 \to \widetilde{\chi}_1^0 f\bar{f}$ 
\cite{Acciarri:1999km} could apply here since the process
$e^+e^-\to SA$ followed by the cascade $A \to SZ \to S f\bar{f}$
would give similar signals. Such analysis had
been carefully done in Ref.~\cite{Lundstrom:2008ai}. 
Their limits on $m_S$ and $m_A$ can be summarized as
${\rm max}(m_A,m_S)\geq 100$ \gev. However, in the present study,
we will use the exact exclusion region as given in 
Ref.~\cite{Lundstrom:2008ai}.

\item \underline{\bf Higgs mass}:

In the IHDM, the SM Higgs boson $h$ have similar couplings to SM fermions and
gauge bosons. Therefore, as long as $h$ decays into SM final states, 
all the measurements from ATLAS and CMS experiments about SM Higgs boson
properties can be used. In particular, we will require the 
mass of the SM Higgs boson of the IHDM should lie within the 
measurement \cite{CMS-PAS-HIG-12-045}:
\begin{eqnarray}
m_h= 125.8 \pm 0.6 \ \ (\gev) \; .
\end{eqnarray}

\item \underline{\bf Invisible decay:}

The openings of one of the non-standard decays of the Higgs boson
such as $h\to SS$ or $h\to AA$ \footnote{
From our previous discussion of LEP-II constraints, 
we assume $m_{H^\pm} >$ 70 GeV in our numerical scan}, 
hence $h\to H^+H^-$ is not open.
can modify the total width of the Higgs boson and can have 
significant impact on LHC results.  Since either
$\chi=S$ or $A$ is the lightest $\mathbb{Z}_2$ odd particle, it will be stable 
and the decay $h\to \chi \chi$ will be invisible.
Both ATLAS and CMS had performed searches for invisible decay of the 
Higgs boson 
\cite{higgs-invisible-atlas,higgs-invisible-cms,higgs-invisible-cmsVBF}.  
Using the Higgs-strahlung SM cross section for $pp\to ZH$ with a 125 GeV 
SM Higgs boson, ATLAS \cite{higgs-invisible-atlas} has excluded 
an invisible branching ratio of the Higgs boson
larger than 65\% with 95\% C.L.. CMS also studied the invisible decay
of the Higgs boson produced via the vector boson fusion (VBF) mechanism and obtained 
an upper limit for the invisible branching ratio 
of 69\% with 95\% C.L.. When the two production mechanisms are
combined the upper limit becomes 54\% with 95\% C.L. \cite{higgs-invisible-cmsVBF}.
This constraint on the invisible decay is 
rather weak compared to the one derived from various works of global fits to 
ATLAS and CMS data \cite{globalCLT,globalBelangeretal1,globalBelangeretal2,globalOthers}. 
These global fits studies suggest that the 
branching ratio of the invisible decay of the Higgs boson should not exceed  
19\% at 95\% C.L. in the case where the Higgs boson has SM-like couplings to all
SM particles plus additional invisible decay mode
which is exactly the case as in IHDM. On the other hand, if one allows for
deviation in the $h\gamma\gamma$ 
(and $hgg$ as well on general grounds but not for IHDM)
coupling from its SM value the 95\% C.L. limit on the invisible Higgs decay 
branching ratio moves up to 29\% \cite{globalBelangeretal2}.


\item \underline{\bf{Diphoton signal strength $R_{\gamma\gamma}$ in the IHDM:}}

Assuming that the production cross section of the Higgs boson is dominated by the 
gluon gluon fusion process, the diphoton signal strength in the IHDM normalized to the
SM value can be simplified as
\begin{eqnarray}
R_{\gamma\gamma} \equiv  
\frac{\sigma_{h}^{\gamma\gamma}}{\sigma_{h_{\rm SM}}^{\gamma\gamma}} & = &
\frac{\sigma(gg\to h)\times {\rm BR}(h \to \gamma\gamma) }{\sigma(gg\to
  h)^{\rm SM}\times {\rm BR}(h \to \gamma\gamma)^{\rm SM} } \; \; , \nonumber \\
& = & \frac{{\rm BR}(h \to \gamma\gamma)^{\rm IHDM} }
{{\rm BR}(h \to \gamma\gamma)^{\rm SM} } \; \; ,
\label{ratio}
\end{eqnarray}
where in the first line we have used the narrow width approximation 
and in the second line we used the fact that $\sigma(gg\to h)$ is the same in both the 
SM and IHDM. Thus the signal strength $R_{\gamma\gamma}$ in IHDM is simply given  
by the ratio of the branching ratios, which is not necessarily one since the charged Higgs
boson in IHDM can provide extra contribution other than the SM particles 
to the triangle loop amplitude of $h \to \gamma \gamma$.

At ATLAS, the overall signal strength for diphoton is about 
$1.55_{-0.28}^{+0.33} $, which corresponds to about 2$\sigma$
deviation from the SM prediction \cite{ATLAS:2013oma}, while the other 
channels are consistent with SM. However, at CMS, 
the new analysis for diphoton mode based on multivariate analysis
\cite{Palmer:2013xza} gives a signal strength about $0.78\pm 0.28$, 
which is consistent with SM. 
Many proposals based on physics beyond SM, including IHDM, have been suggested to explain the diphoton
excess, but the actual disagreement between ATLAS and CMS does 
not allow to draw any definite conclusions yet, given the current level of statistics.
In the present analysis we will not try to explain the diphoton excess
but rather study the impact of the other constraints on the ratio $R_{\gamma\gamma}$.

\item \underline{\bf{LHC monojet search:}}

Besides using the invisible width of the Higgs decay, another strategy to look for 
DM at the LHC is to study high $p_T$ 
monojet balanced by a large missing transverse energy ${\not\!E}_T$ \cite{monojet1,monojet2}. 
Such kind of signature is possible in IHDM 
by producing the SM Higgs boson $h$ in association with an energetic jet followed by the invisible
decay of $h$. In our analysis we will consider the following parton processes:
\begin{itemize}
\item $gb\to hb \to \chi \chi + b$: $s$-channel and $t$-channel tree level diagrams with the
 Higgs boson radiated from $b$ quark legs, 
\item $qg \to h q \to \chi \chi +q$: $t$-channel diagram through tree level gluon-quark-anti-quark vertex
and one-loop $hgg$ effective vertex,
\item $gg \to h g \to \chi \chi +g$: $t$-channel diagram through tree level three gluon vertex and one-loop $hgg$
effective vertex, 
\item $q\bar{q} \to h g \to \chi \chi +g$: $s$-channel diagram through tree level gluon-quark-anti-quark vertex
and one-loop $hgg$ effective vertex.
\end{itemize}
In all these processes, the final state consists of a pair of invisible
DM particles plus a quark or gluon jet.  
For the experimental cuts, see the later discussion of the likelihood function 
for the monojet data in section \ref{section:5}.
\end{itemize}
%


\section{Relic density, direct detection and indirect detection constraints 
\label{section:4}}


It is well known that annihilation of $\chi$ into SM particles and other inert Higgs bosons
can contribute to thermal relic density as well as indirect DM signals of 
high energy gamma-rays, positrons, antiprotons or neutrinos, 
while the scattering of $\chi$ onto nuclei will lead to direct detection 
signals by measuring the recoil energy of the nuclei via scintillation light, 
heat or ionization or some combinations of these three different 
signals using different technologies.


\begin{itemize}

\item \underline{\bf Relic density constraint:}

Assuming a standard thermal evolution of our Universe, we compute the  
relic density from the following channels: 
$\chi \chi \to f\bar{f}$ ($f=t, b, c, \tau, \mu $),
$\chi \chi \to W^\pm W^\mp, ZZ, \gamma\gamma,\gamma Z$ and $\chi \chi \to H^\pm H^\mp$. 
Since $\chi$ can be either $S$ or $A$, we consider $SS$ or $AA$ annihilation.

Note that in the case where $m_\chi< m_{W,Z}$, 
we take into account the annihilation into 3-body final state from $VV^*$ or 
4-body final state from $V^*V^*$ ($V=W^\pm,Z$).
All the annihilation into SM particles channels proceed through 
$s$-channel Higgs boson exchange while the annihilation into inert Higgs particles such as 
$H^\pm H^\mp$, $hh$ and $AA$ will proceed through both $s$-channel and
$t$-channel Higgs boson exchange as well as the contact interactions with the quartic couplings for
the $\chi\chi H^\pm H^\mp$, $\chi\chi hh$ and $\chi\chi AA/SS$ vertices.
The calculation is done  using the public code \texttt{MicrOMEGAs} 
\cite{micro}. The outcome of our relic density calculation should be 
in agreement with the recent PLANCK measurement \cite{Larson:2010gs}:
\begin{equation}
\Omega_{\rm{CDM}} h^{2}=0.1199\pm0.0027 \; .
\label{Omega}
\end{equation}

As is well known if the mass splitting between the LOP and 
Next-Lightest Odd Particle (NLOP) is $\lesssim 10\gev$ or so, 
the number densities of these NLOPs have only 
slight Boltzmann suppression with respect to the LOP number density.
Therefore, the contributions to the relic density from the scattering of
LOP-NLOP and NLOP-NLOP have to be taken into account in order to have a more 
precise relic density prediction. These mechanisms are known as coannihilation~\cite{Griest:1990kh,Binetruy:1983jf}.
As these are implemented already in the package \texttt{MicrOMEGAs}~\cite{micro}, 
we can take the $S-A$, $S-H^\pm$ and $A-H^\pm$ coannihilation 
into account at ease.

\item \underline{\bf The LUX limit:}

At present the most stringent limit on the spin-independent component 
of elastic scattering cross section $\sigsip$ for $\chi p \to \chi p$ comes 
from LUX \cite{Akerib:2013tjd}. They improved the minimum $\sigsip$ upper limit 
obtained by XENON100 \cite{Aprile:2012nq} about an overall factor of 3. 
This result then sets the limit on the spin-independent cross section, 
$\sigsip<8\times 10^{-10}$ pb for DM mass $m_\chi \approx 33$ GeV.
In this study, we include the  90\% upper limit
obtained in \cite{Akerib:2013tjd} for $\sigsip$ versus the DM mass in our likelihood function. 
However, one should bear in mind that $\sigsip$ may be susceptible to
large theoretical uncertainties from the hadronic matrix elements.  
We will take into account the uncertainties in 
the hadronic matrix elements, as will be discussed later in section \ref{section:5}.

\item \underline{\bf Gamma-rays:} 

We consider the Fermi-LAT 
observations of $\gamma$-rays from 
dSphs \cite{Ackermann:2011wa,GeringerSameth:2011iw,Tsai:2012cs}
and GC \cite{Huang:2012yf}. 
The $10$ dSphs as adopted in \cite{Ackermann:2011wa} 
will be used in this work. Four years of the 
Fermi-LAT data \footnote{\url{http://fermi.gsfc.nasa.gov/ssc/data}}, recorded 
from 4 August 2008 to 2 August 2012 with the pass 7 photon selection, are 
employed in this analysis. The energy range of photons is chosen from $200$ 
MeV to $500$ GeV, and the region-of-interest (ROI) is adopted to be a 
$14^{\circ}\times14^{\circ}$ box centered on each dSphs. For the GC
analysis, a slightly smaller ROI region of $10^{\circ}\times10^{\circ}$
is chosen to avoid too many sources in the analysis. In the likelihood
analysis, the normalization of the 
diffuse background models \footnote{\url{http://fermi.gsfc.nasa.gov/ssc/data/access/lat/BackgroundModels.html}}
{\tt gal\_2yearp7v6\_v0.fits} and {\tt iso\_p7v6source.txt}, and the point 
sources located in the ROIs in the second LAT catalog \cite{Fermi:2011bm}
are left free to do the minimization. The Fermi-LAT data
are binned into $11$ energy bins logarithmically spaced between $0.2$ 
and $410$ GeV, and we calculate the likelihood map of Fermi-LAT dSphs 
and GC observations on the $E_{\rm bin}-$flux plane following the
method developed in \cite{Tsai:2012cs}. Such a method is very efficient
to derive the final likelihood of any specific $\gamma$-ray spectrum,
and is tested to be consistent with the standard analysis procedure using
Fermi Scientific Tool \cite{Tsai:2012cs}.

\item \underline{\bf Cosmic ray electrons and positrons:} 

The cosmic ray positron
fraction measured by PAMELA \cite{Adriani:2008zr} and most recently
by AMS-02 \cite{Aguilar:2013qda} show clear evidence of excess compared
with the secondary production as expected from the cosmic ray propagation
model. The fluxes of the total $e^+e^-$ measured by ATIC \cite{Chang:2008aa},
Fermi-LAT \cite{Ackermann:2010ij}, HESS \cite{Aharonian:2008aa} and 
MAGIC \cite{BorlaTridon:2011dk} also show the 
deviation from the extrapolation of the low energy PAMELA data 
\cite{Adriani:2011xv}, which further supports the existence of extra 
$e^+e^-$ sources. There are many models proposed to account for the
$e^+e^-$ excesses, including the astrophysical sources such as
pulsars and supernova remnants, and DM annihilation/decay
(see {\it e.g.} the review articles \cite{review}).

While the DM model would suffer from strong constraints from $\gamma$-ray 
observations \cite{Bertone:2008xr}, it has been shown that the pulsars with 
reasonable parameters can explain the positron fraction as well as the electron plus positron flux
data \cite{Grasso:2009ma}. Therefore in this work we first fit both data set with 
the background plus pulsar-like models, and then add the DM contributions from IHDM 
to calculate their likelihoods \cite{Bergstrom:2013jra}. The framework
for doing such calculation in this astrophysical setting
can be found in \cite{Yuan:2013eja}. Basically, a Markov Chain 
Monte Carlo (MCMC) based global fitting tool was used to determine
the model parameters. The observational data used in the fit include
the AMS-02 positron fraction \cite{Aguilar:2013qda}, PAMELA electron
spectrum \cite{Adriani:2011xv}, and the total $e^+e^-$ spectra of
Fermi-LAT \cite{Ackermann:2010ij} and HESS \cite{Aharonian:2008aa}. 
Note that for the background electron spectrum
we employ a three-piece broken power-law function in order to fit
simultaneously the above data \cite{Yuan:2013eja}.

The solar modulation affects the fluxes of the particles at low energy.
In this work we simply adopt the force-field approximation to account
for the solar modulation effect \cite{Gleeson:1968zza}. It was found
that the modulation potential $\Phi\approx970$ MV can fit both the
positron fraction and electron spectra. However, for the cosmic ray
protons, a smaller modulation potential $\Phi\approx500$ MV is favoured
by the PAMELA data \cite{Adriani:2011cu,Yuan:2013eja}. We leave this
as an open question because the solar modulation may indeed depend
on the mass-to-charge ratio of particles. The best-fitting positron
fraction and electron spectra compared with the observational data
are shown in Fig. \ref{fig:bkgep}. The model fits the data
well and the reduced $\chi^2$ is about $0.92$.

\begin{figure}[t!]
\centering
\includegraphics[width=3in]{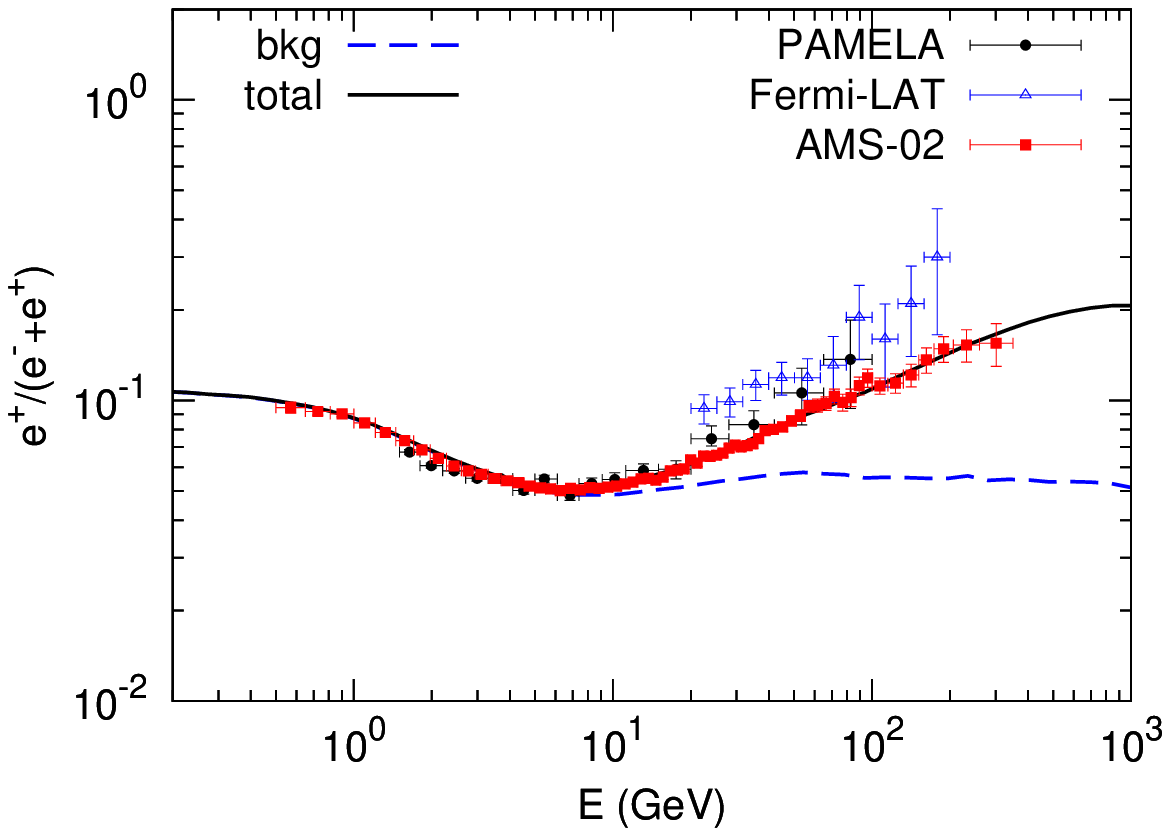}
\includegraphics[width=3in]{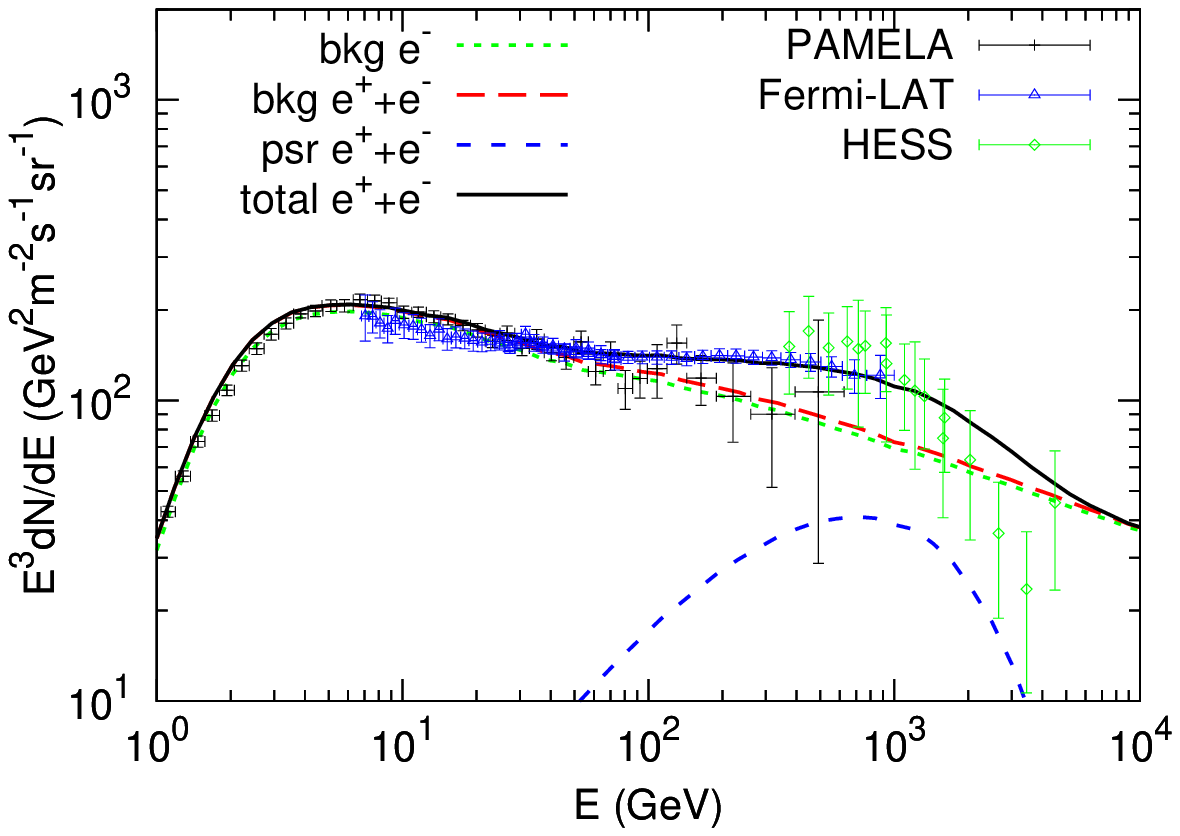}
\caption{The positron fraction ({\it left}) and electron spectra ({\it right})
for the background + pulsar model. Also shown are the positron fraction 
data from AMS-02 \cite{Aguilar:2013qda}, PAMELA \cite{Adriani:2008zr} and 
Fermi-LAT \cite{FermiLAT:2011ab}, 
and electron flux data from Fermi-LAT \cite{Ackermann:2010ij},
HESS \cite{Aharonian:2008aa} and PAMELA 
\cite{Adriani:2011xv}.
}\label{fig:bkgep}
\end{figure}

When calculating the likelihood after adding the DM contributions in IHDM, 
we further multiply a factor of $\alpha_iE^{\beta_i}$ to the fluxes of the
background components ($i=1,2,3$ for the $e^-$ background, $e^+$ background 
and pulsar $e^{\pm}$ respectively), in order to take the uncertainties of 
the modelling into account \cite{DeSimone:2013fia}. The parameters 
$\alpha_i$ and $\beta_i$ are left free and treated as nuisance parameters 
in our analysis. They are allowed to vary in the range of 
$0.1< \alpha_i <10$ and $-0.5 < \beta_i <0.5$
to calculate the maximum likelihood of a specific DM model point.

\item \underline{\bf Cosmic ray antiprotons:} 

The precise measurement of the 
antiproton-to-proton ratio and antiproton flux by PAMELA show relatively 
good agreement with the cosmic ray background model expectation 
\cite{Adriani:2008zq,Adriani:2010rc}, which leaves limited space for
the DM models \cite{Donato:2008jk}. We calculate the expected antiproton
flux in the same propagation model used to explain the $e^+e^-$ data, as
shown in Fig. \ref{fig:bkgap}. The solar modulation potential is adopted
to be $500$ MV as suggested by PAMELA \cite{Adriani:2011cu,Yuan:2013eja}. 
Same as the $\alpha_i$ and $\beta_i$ in $e^+e^-$ case, an adjustment factor 
$\alpha_{\bar{p}}E^{\beta_{\bar{p}}}$ with nuisance parameters 
$\alpha_{\bar{p}}$ and ${\beta_{\bar{p}}}$ varied in the same respective 
range is employed to account for the uncertainties of the background estimation.

\end{itemize}

\begin{figure}[t!]
\centering
\includegraphics[width=4in]{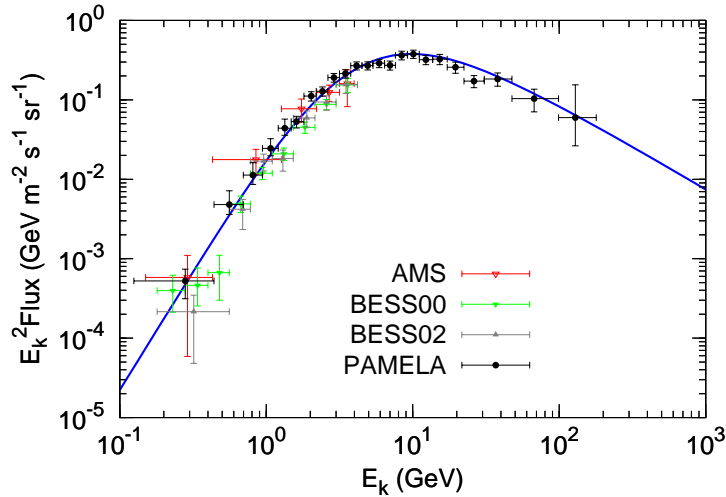}
\caption{Antiproton flux of the background model compared with the 
observational data of AMS \cite{Aguilar:2002ad}, BESS00 \cite{Asaoka:2001fv}, 
BESS02 \cite{BESS02} and PAMELA \cite{Adriani:2010rc}.
}\label{fig:bkgap}
\end{figure}

In Table \ref{tab:exp_constraints}, we summarize all the experimental constraints 
mentioned in this and previous section. To avoid words cluttering in later presentation,
we denote the first block of relic density and collider constraints together with the theoretical constraints
as \textbf{RC} (Relic density and Collider), the second block of LUX constraint
as \textbf{DD} (Direct Detection) and the third block of constraints as \textbf{ID} (Indirect Detection).
Additionally, we reject those points during our parameter scans which violate any one of the theoretical constraints on IHDM mentioned in section \ref{section:2b}. 
We note that the data in the \textbf{RC} block does not involve large theoretical 
uncertainties compared with the other
two blocks, so we will take special care of this block by including it only at the scan level.
\begin{table}[t]\footnotesize
\begin{center}
\begin{tabular}{|l|l|l|l|l|l|}
\hline
Measurement & Mean & Error:~Exp.,~Th. & Distribution & Refs.\\
\hline
$m_h$\ (by CMS) & $125.8\gev$ & $0.6\gev$, $0.0\gev$  & Gaussian &\cite{CMS-PAS-HIG-12-045} \\
$\Omega h^2$ & $0.1199$ & $0.0027$, $10\%$ & Gaussian &  \cite{Ade:2013zuv}\\
$S$ & $0.05$ & $0.09$, $0.0$ & Gaussian & \cite{Baak:2012kk}\\
$T$ & $0.08$ & $0.07$, $0.0$  & Gaussian &  \cite{Baak:2012kk}\\
${\rm BR}(h\rightarrow \rm{invisible})$ (by ATLAS) & $0.65$ & $5$\%, $10$\% & Error fn. 
& \cite{higgs-invisible-atlas}\\
$R_{\gamma\gamma}$& $0.78$ & $0.28$, $20$\% & Gaussian & \cite{Palmer:2013xza}\\
Monojet (by CMS $19.5$ fb$^{-1}$) & See text. & See text. & Poisson &\cite{monojet1}\\ 
\hline\hline
LUX (2013)  & See text. 	& See text. 	& Error fn. &\cite{Akerib:2013tjd}\\ 
\hline\hline
dSphs $\gamma$-ray  & See text. 	& See text. & Poisson &\cite{Ackermann:2011wa,Tsai:2012cs}\\ 
GC $\gamma$-ray flux & See text. & See text. & Half Poisson &\cite{Huang:2012yf}\\ 
$e^+$ fraction, $e^++e^-$ flux & See text. & See text. & Gaussian 
&\cite{Adriani:2008zr,Ackermann:2010ij,Aguilar:2013qda}\\
$\bar{p}$ flux  & See text. & See text. & Gaussian &\cite{Adriani:2010rc}\\
\hline
\end{tabular}
\caption{
The experimental constraints that we include in our likelihood 
functions to constrain the IHDM model. 
We denote the first block of relic density and collider constraints 
together with the theoretical constraints as 
\textbf{RC} (Relic and Collider) , 
LUX constraint as \textbf{DD} (Direct Detection) 
and the last block of constraints as \textbf{ID} (Indirect Detection).
} 
\label{tab:exp_constraints}
\end{center}
\end{table}


\section{Methodology \label{section:5}}


In this section, we will describe the statistical treatment 
of all the experimental constraints discussed in previous two sections 
and the numerical method used in our analysis.
At the fitting level, we use the following different likelihood distributions: 
Gaussian, Poisson, and error function, depending on which
experiments as shown in the fourth column in Table \ref{tab:exp_constraints}.   
For experiments that may lead to $5 \sigma$ discoveries, it is customary to use
Gaussian distribution if experimentalists can provide central values and errors. 
For counting experiments, Poisson distribution is a standard formula for the 
likelihood. However, for experimental data like LUX and invisible Higgs decay width where
only upper limits are provided, it is difficult to implement Poisson likelihood in the analysis.
Under these circumstances, we will follow the procedure described in \cite{de Austri:2006pe,Roszkowski:2012uf}, 
where the error function was used to smear the experimental bounds.


\begin{itemize}


\item 
\underline{\bf Gaussian likelihood distribution:}

The Gaussian likelihood distribution is related to the $\chi^2$ as
\begin{equation}
\label{eq:Gau}
\mathcal{L}_{\rm Gaussian}=e^{-\frac{\chi^2}{2}} \; ,
\end{equation} 
with the $\chi^2$ defined as usual
\begin{equation}
\chi^2=\frac{\left(\rm{prediction}-\rm{experimental\,central\,value}\right)^2}{\sigma^2+\tau^2},
\end{equation}
where $\sigma$ is an experimental error and $\tau$ is a theoretical 
uncertainty. We assume that the theoretical uncertainty $\tau$ owes to 
either the discrepancy between computations using different methods or unknown high order corrections
or non-perturbative uncertainties.
See Table \ref{tab:exp_constraints} for the list of experiments that we use the 
Gaussian likelihood distribution.


\item 
\underline{\bf Poisson likelihood distribution:}

Regarding the LHC monojet, we use a Poisson likelihood distribution 
augmented with an extra Gaussian distribution to account for the background 
uncertainties. The probability distribution for each $p_T^{\rm{miss}}$ threshold 
is then written as 
\begin{equation}
\mathcal{P}(s_i+b_i|o_i)= \max_{b'}{\Biggl\lbrace}\frac{e^{-(s_i + b')}\left(s_i + b' \right)^{o_i}}{o_i !}
\exp \left[ -\frac{(b'-b_i)^2}{2\delta b_i^2}\right]{\Biggr\rbrace}\,,\label{eq:monojet}
\end{equation}
where $i$ refers different missing 
transverse energy ${\not\!E}_T$ cuts described in \cite{monojet1}. 
We also use the values of background events $b$ with error $\delta b$ and observed 
events $o$ from Ref.~\cite{monojet1}. 
To simulate signal events $s$, we use \texttt{MadGraph 5} \cite{Alwall:2011uj} 
to compute the cross section at the parton level and apply the appropriate cuts. 
Following the CMS study \cite{monojet1}, we will use the following 
basic selection requirements for 
the transverse momentum ($p^j_T$) and pseudo-rapidity ($\eta^j$) of the monojet:
\begin{itemize}
\item at least one jet with $p_T^j>110$ GeV and $|\eta^j|<2.4$,
\item at most two jets with $p_T^j>30$ GeV,
\item and no isolated leptons in the final state.
\end{itemize}
CMS collaboration \cite{monojet1} also gave the events for seven different cuts on 
the missing transverse energy ${\not\!E}_T$ between 250 and 
550 GeV (in step of 50 GeV), which are largely dominated by the SM $(Z,W^\pm)$ background
where $Z$ decays to neutrinos and $W^\pm$ decays leptonically without 
reconstruction of the charged lepton. For a given ${\not\!E}_T$ threshold,
the signal event $s_i$ is the total number of monojet which is given by the 
monojet cross section after cuts times the CMS luminosity of 19.5 fb$^{-1}$.
Therefore, our likelihood function for CMS monojet search can be written as
\begin{equation}
\mathcal{L}_{\rm{LHC-monojet}}=  \prod_{i} \mathcal{P}(s_i+b_i|o_i) \; ,
\end{equation}
where $i$ runs over all the seven different cuts on
${\not\!E}_T ({\rm GeV}) >$ 250, 300, 350, 400, 450, 500 and 550 \cite{monojet1}.
We have tested that using the above likelihood function, we can reproduce the 
exclusion limits for the effective DM operators analyzed by the CMS \cite{monojet1}.
This justifies the use of this likelihood function for the IHDM.

While we also use the normal Poisson distribution to be the 
likelihood function for the $\gamma$-ray 
from dSphs, a half Poisson distribution is used for the $\gamma$-ray from GC.
In other words, if the signal events are
less than the number of events required for the maximum likelihood,
we set its likelihood to be the maximum likelihood. This is because
a normal Poisson likelihood of $\gamma$-ray from GC would give a very
significant signal at $m_\chi\sim$ a few GeV \cite{GCexcess}.
Since the GC is a very complicated astrophysical environment, it
is not clear whether such an excess is due to some kinds of astrophysical
background or genuine DM signals. In order to be less biased, we therefore adopt
a half Poisson distribution for our GC $\gamma$-ray likelihood.
Moreover, the halo profiles in GC can also lead to big uncertainties.
In this study, we use the isothermal halo profile in order to achieve 
a more conservative limit.


\item
\underline{\bf Error function:}

Instead of using a step function, we employ an error function (erfc)  likelihood 
with a theoretical error $\tau$ = 10\% 
to smear the upper bound on the branching ratio of the invisible Higgs decay width 
${\rm BR}(h\rightarrow \rm{invisible})$, 
\begin{equation}
\mathcal{L}_{{\rm BR}(h\rightarrow \rm{invisible})} = 
\frac{1}{2}\rm{erfc}\left( \frac{\rm{prediction}-\rm{experimental\,upper\,limit}}{\sqrt{2}\tau}\right) \; . 
\end{equation} 
%

For the upper limit of LUX for the $\chi$-nucleon cross section $\sigsip$ versus the DM mass,
we set $\tau=150\%$ for the hadronic uncertainties to account for the difference between the default 
value used in \texttt{MicrOMEGAs} and $1\sigma$ lower limit of the pion-nucleon sigma term
$\sigma_{\pi N}$ obtained from lattice calculation \cite{Stahov:2012ca}.

\end{itemize}

With the above set up of the likelihood distributions for each experiment, we are able to guide our
random scan of the parameter space to explore regions with high likelihood probability. 
The total likelihood is the product of all the individual likelihood from each experiment.
As noted earlier, since {\bf DD} and {\bf ID} experimental constraints could
suffer from large theoretical uncertainties ({\it e.g.} in the hadronic matrix elements in {\bf DD} and
DM halo profiles in {\bf ID}), we rather play safe and conservative by including only the first {\bf RC} block in Table \ref{tab:exp_constraints} in the 
likelihood at the scan level. 

Engaging with \texttt{MultiNest v2.18} \cite{Feroz:2008xx} of 20000 living points, 
a stop tolerance factor of $10^{-4}$, 
and an enlargement factor reduction parameter of 0.8, 
we perform 6 random scans in the six dimensional parameter space which will be 
restricted in the following ranges for the masses
\begin{eqnarray}
\label{domain}
122.0 &\leq \, m_h  / \textrm{ GeV} \, \leq &129.0 \; , \nonumber\\
5.0 &\leq \, m_S  / \textrm{ GeV} \, \leq &4\times 10^{3} \; , \nonumber\\
5.0 &\leq \, m_A  / \textrm{ GeV} \, \leq &4\times 10^{3} \; , \nonumber\\
70.0 &\leq \, m_{H^\pm}  / \textrm{ GeV} \, \leq & 4\times 10^{3} \; ,\nonumber
\end{eqnarray}
and the following ranges for the couplings
\begin{eqnarray}
-2.0 &\leq \lambda_L \leq& 2.0 \; , \nonumber\\
0.0 &\leq \lambda_2 \leq& 4.2 \; . \nonumber
\label{eq:scanrange}
\end{eqnarray}
Of the total 6 random scans, 3 of them we use flat priors for all the above six parameters,
while for the rest of the scans, we use flat priors for $m_h$ and $\lambda_L$ and log priors for
the other four parameters.
We note that coverage of the parameter space is the most important aspect for profile likelihood method.
We combine these 6 different scans to perform our analysis in order 
to achieve better coverage of the parameter space and obtain accurate best-fit points.
In order to scan the parameter space more efficiently, we set the range of 
$\lambda_2$ up to 4.2 allowed by the unitarity constraint of
Eq.~(\ref{eq:lam2CONS}). We finally collect $\sim 1.2\times 10^{6}$ points
in these scans.

In the next two sections, we will present our results mainly based on 
``Profile Likelihood'' method~\cite{Rolke:2004mj}. 
Under the assumption that all uncertainties 
follow the approximate Gaussian distributions, 
confidence intervals are calculated from the tabulated values of 
$\delta\chi^2\equiv -2\ln(\mathcal{L/L}_{\rm{max}})$. Thus,
for a two dimension plot, the 95\% confidence ($2\sigma$)
region is defined by $\delta\chi^2 \leq 5.99$.

We note that the best-fit points in either log or 
flat prior scan can have almost the same $\mathcal{L}_{\rm{max}}$ of 
individual scan but the locations of $m_S$ and $m_A$ are quite 
different from these 6 scans. 
This is due to the fact that we allow the LOP $m_\chi$ to 
be either $m_S$ or $m_A$. 
Same value of $m_\chi$ corresponds to two positions in $m_S$ 
for LOP is $S$ or not. Similar situation is found for $m_\chi = m_A$
depending on whether LOP is $A$ or not.
However, for the projection to $m_\chi$, the best-fits locate 
at the same region.


\section{Results and discussions \label{section:6}}


\subsection{Low dark matter mass scenario \label{section:6a}}

In this subsection, we will discuss the low DM mass scenario where the 
invisible Higgs boson decay is open either by 
$h\to SS$ or $h\to AA$ and study the implication from the LHC
constraint on such invisible Higgs boson decay as well as LUX and relic density constraints 
on the $h\chi\chi$ coupling.

Let us first give the analytical expression of the invisible Higgs boson decay branching ratio
\begin{equation}
{\rm BR}(h \to {\rm invisible})=\frac{\Gamma(h\to \chi \chi) }{\Gamma_{\rm tot}(h)} =
\frac{\Gamma(h\to \chi \chi) }{\Gamma_{\rm SM}(h)+\Gamma(h\to \chi \chi) } \; ,
\end{equation}
where $\Gamma_{\rm SM}(h)$ is the total width of the SM Higgs boson 
taken as $\Gamma_{\rm SM}(h)=4.02$ MeV in what follows, and
\begin{eqnarray}
\Gamma(h\to \chi \chi) &=& \frac{g_{h\chi\chi}^2 }{32 \pi m_h} 
\sqrt{1- \frac{4m_{\chi}^2}{m_h^2}} \; ,
\end{eqnarray}
with $g_{h\chi\chi}$ given by Eq.~(\ref{eq:lamchi}).

\begin{figure}[t!]
\centering
\includegraphics[width=3in]{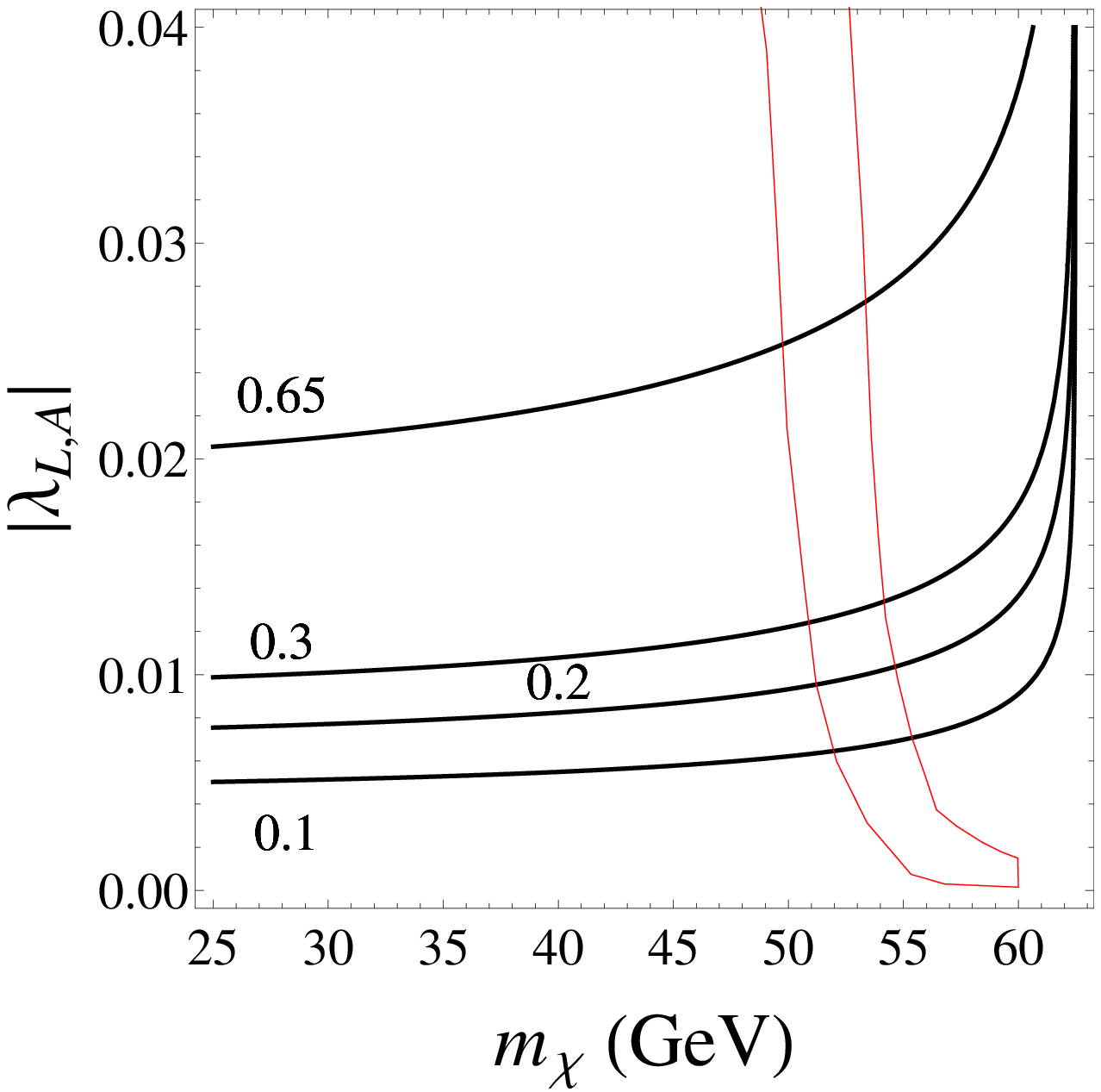}
\includegraphics[width=3in]{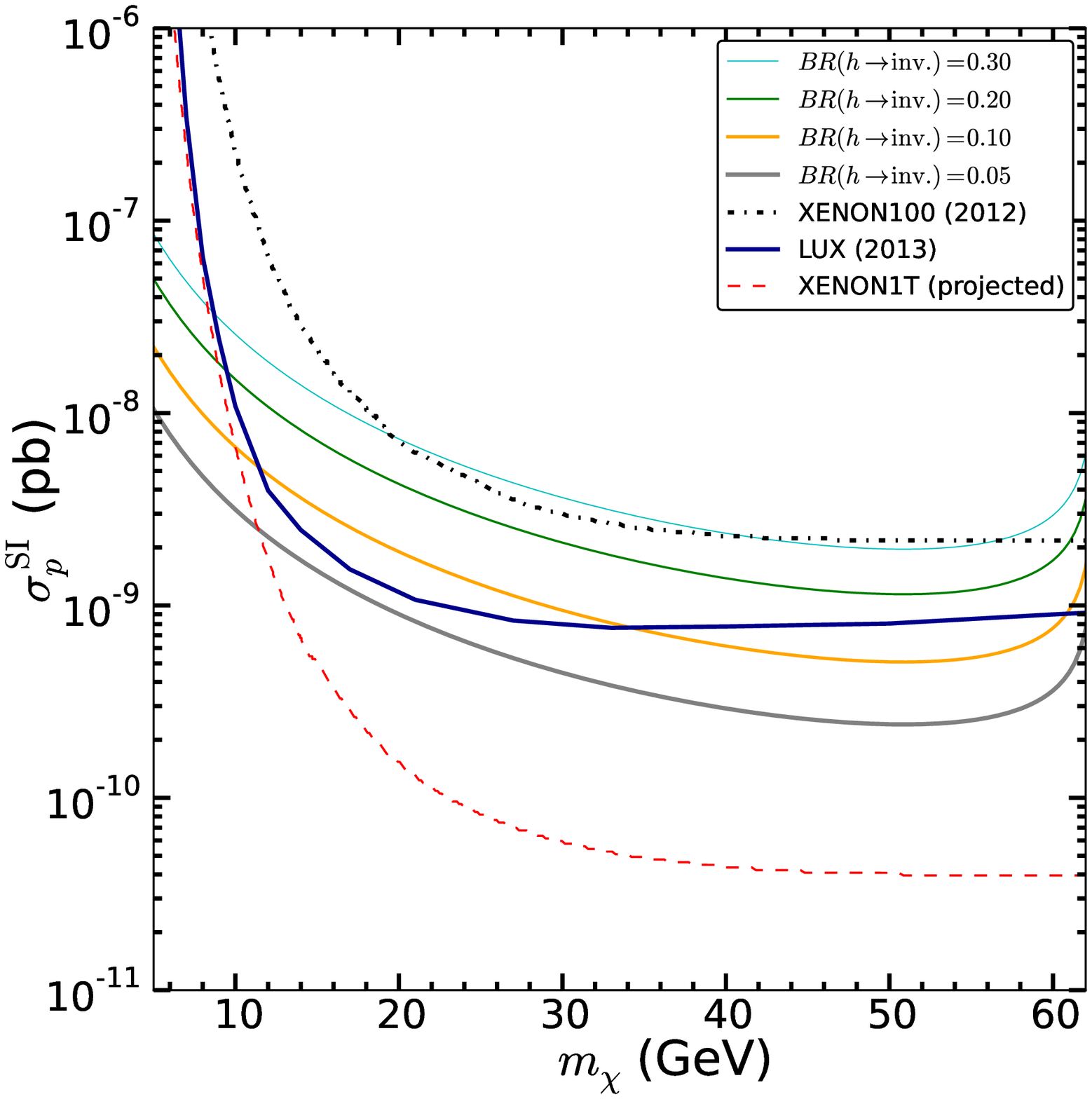}
\caption{({\it Left}) Contour plots for the invisible Higgs boson 
decay branching ratio
as a function of $\vert \lambda_{L,A} \vert$ and $m_\chi=m_{S,A}$.
The contours are, from top to bottom, 65\%, 30\%, 20\% and 10\%, while the red
curve is the 3$\sigma$ region allowed by PLANCK relic density value. ({\it Right}) 
Spin-independent cross section $\sigma^{\rm SI}_p$ as a function of the DM mass $m_\chi$ in IHDM
with fixed values of invisible Higgs decay branching ratio, 30\%, 20\%, 10\%
and 5\% from top to bottom. In both plots we take $m_h=125$ GeV.
}
\label{fig:hinv}
\end{figure}

In order to understand the correlation between the coupling
$g_{h\chi \chi} \varpropto \lambda_{L,A}$ and the invisible Higgs boson
decay branching ratio, we  illustrate in Fig.~\ref{fig:hinv} (left)
a contour plot for the ${\rm BR}(h\to {\rm invisible})$ 
in the $(m_\chi, \vert \lambda_{L,A} \vert)$ plane.
The contour lines are, from top to bottom, 65\%, 30\%, 20\% and 10\%.  
The domain bounded by the red curve is the 3$\sigma$ region allowed by PLANCK measurement of 
the relic density given by Eq.~(\ref{Omega}). It is clear that for 
$|\lambda_{L,A}|\approx  10^{-2}$, one can have an invisible decay 
branching ratio of the order of 20\% for $m_{\chi} \leq 55$ GeV. 
As one can see from the region inside the red curve, 10\% to 
20\% invisible Higgs boson decay is consistent with the relic density measurement 
only for  $m_\chi$ in the range of $50$-$56$ GeV. This is due to the fact that 
near the $s$-channel resonance of the Higgs boson where $m_h\approx 2 m_\chi$ 
the relic density can be significantly enhanced. It is clear that 
the smaller the invisible Higgs boson decay branching ratio is, the smaller the size of 
$\vert \lambda_{L,A} \vert$ unless the DM mass is close to the threshold 
region $m_h\approx 2 m_\chi$ where $\vert \lambda_{L,A} \vert$ could take larger values.

In fact, it is plausible to relate ${\rm BR}(h \to {\rm invisible})$ to the spin-independent 
cross section $\sigsip$ for direct detection. In the IHDM, the diagram contributed to $\sigsip$ is  
given by the $t$-channel Higgs boson exchange and so it is proportional 
to $g_{h\chi\chi}^2$ \cite{Kanemura:2010sh}. 
From the expression of ${\rm BR}(h \to {\rm invisible})$  one can then
eliminate the $g_{h\chi\chi}^2$ coupling in favour of 
$\sigsip$ and other parameters such as the
DM mass $m_\chi$, nucleon mass $m_N$, Higgs mass $m_h$, 
form factor $f_N$, Higgs total width $\Gamma_{\rm SM}(h)$, 
and the Higgs VEV $v$, {\it viz.},
\begin{equation}
{\rm BR}(h \to {\rm invisible})=\frac{\sigsip}{\sigsip + f(m_N,m_{\chi},m_h, f_N)} \; ,
\end{equation}
where
\begin{equation}
f(m_N,m_{\chi},m_h, f_N)=\frac{8\Gamma_{\rm SM}(h) m_N^2 f_N^2 }{
    m_h^3 v^2 (m_\chi+m_N)^2 \sqrt{1- \frac{4 m_\chi^2}{m_h^2}} } \; .
\end{equation}
For a given $f(m_N,m_{\chi},m_h, f_N)$ and ${\rm BR}(h \to {\rm invisible})$, one can then calculate 
$\sigsip$ (see \cite{Djouadi:2012zc} for a similar discussion in the framework of portal models).

We illustrate in Fig.~\ref{fig:hinv} (right) a contour plot for 
${\rm BR}(h \to {\rm invisible})$ in the plane $(m_\chi,\sigsip)$ where we have used 
$f_N= 260$ MeV which is roughly the default value used in \texttt{MicrOMEGAs}.
We show contour lines for ${\rm BR}(h \to {\rm invisible})$ = 30\%, 20\%, 10\% and 5\%.
Also shown is the actual limit from 
XENON100 and LUX as well as the projections from XENON1T experiments.
It is remarkable from this plot that 
${\rm BR}(h \to {\rm invisible}) > 30\%$ is excluded by 
XENON100 if $m_\chi$ is in the range of 20-60 GeV, 
while $m_\chi$ in the range of 12-32 GeV with ${\rm BR}(h \to {\rm invisible}) > 10\%$
is now excluded by LUX.
Combining these two plots of Fig.~\ref{fig:hinv}, we can conclude that 
$|\lambda_{L,A}|$ should be less than 
about $2\times10^{-2}$ for the three experimental 
constraints of Higgs invisible width from LHC, LUX limit on $\sigma^{\rm SI}_p$ and
relic density from PLANCK to be consistent with each other.
Future sensitivity of the XENON1T experiment would be able to 
exclude invisible Higgs decay branching ratio as low 
as 1\% or less, which can further constrain the couplings $\lambda_{L,A}$
that control the communication between the inert and visible sectors.


\subsection{Current experimental constraints and best-fit result \label{section:6b}}

\subsubsection{{\bf RC} \label{subsubsection:6b1}}

We will use the next three figures to discuss the two dimension profile likelihoods from the {\bf RC} block.

\begin{itemize}

\item[A:] Fig.~\ref{fig:mx}

First, in Fig.~\ref{fig:mx}, we present the two dimension profile 
likelihood on the ($m_S$, $m_A$) plane (left) 
and ($m_{\rm LOP}$, $m_{H^{\pm}}$) plane (right). The contours correspond 
to the 95\% C.L. of \textbf{RC} constraints. 
For the area above the red-dashed line, $S$ is the LOP; while below the 
red-dashed line,  $A$ is the LOP.  
Generally speaking, the relic density is a strong constraint, since
we treat it as a positive measurement with a very small experimental uncertainty 
rather than an upper limit. 
For all the parameter space, we found that the DM relic abundance $\Omega_\chi h^2$ 
is mostly too large, namely, its annihilation in the early Universe is too inefficient. 
Certain mechanisms, whether they are natural or not, 
have to play some peculiar roles to enhance the annihilation cross sections 
so as to reduce the relic abundance. 
These mechanisms can be clearly identified by the several different branches in the 
left panel of Fig.~\ref{fig:mx}:\\ 
(1) $m_A\approx m_S$ ($A-S$ coannihilation), \\
(2) $2 m_\chi\approx m_h$ (the SM Higgs boson resonance), and\\ 
(3) two small branches at $50\gev<m_A,m_S<70\gev$ 
(mixed $A-S$ coannihilation and the SM Higgs boson resonance).

\begin{figure}[t!]
\centering
\includegraphics[width=3in]{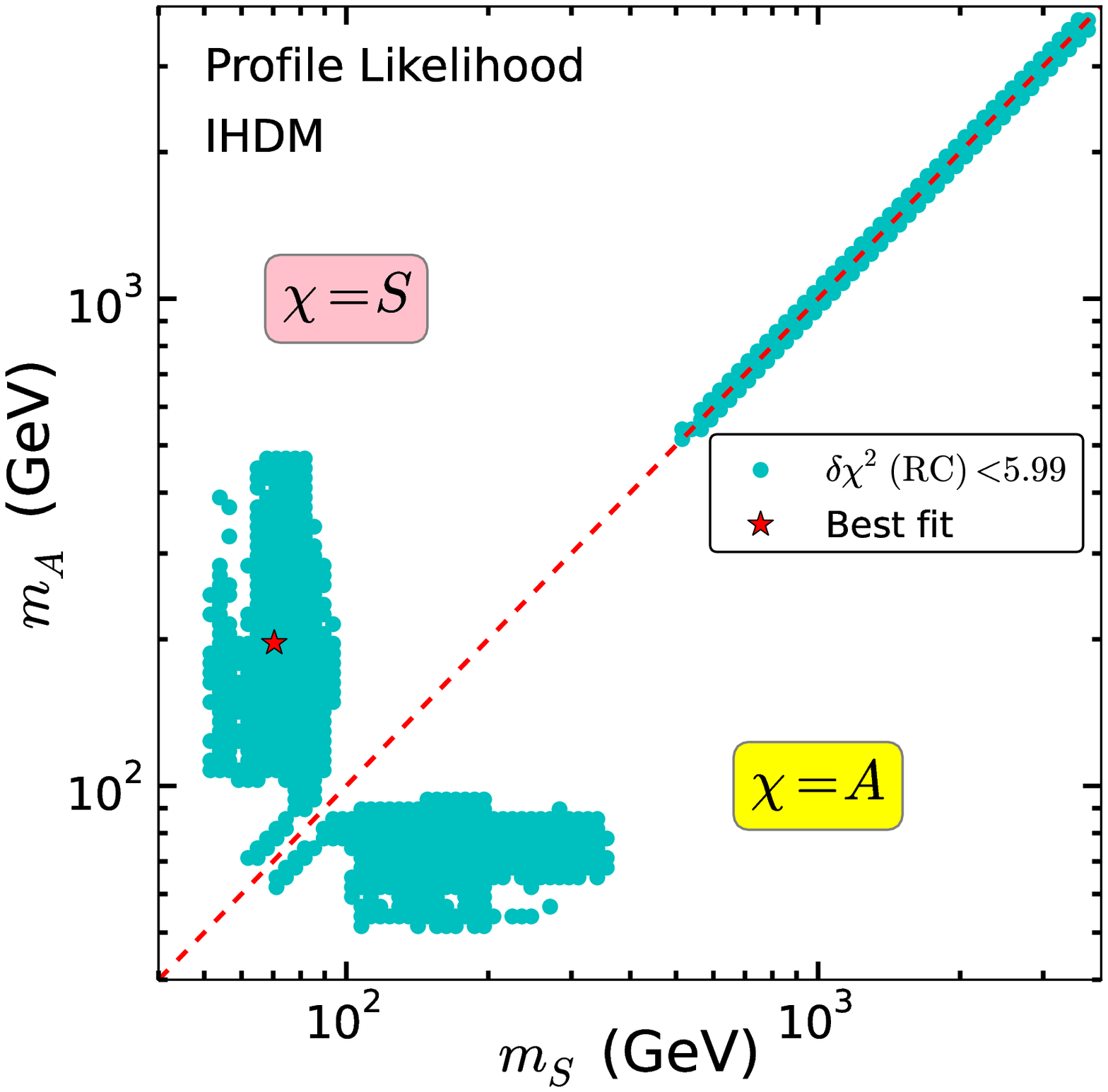}
\includegraphics[width=3in]{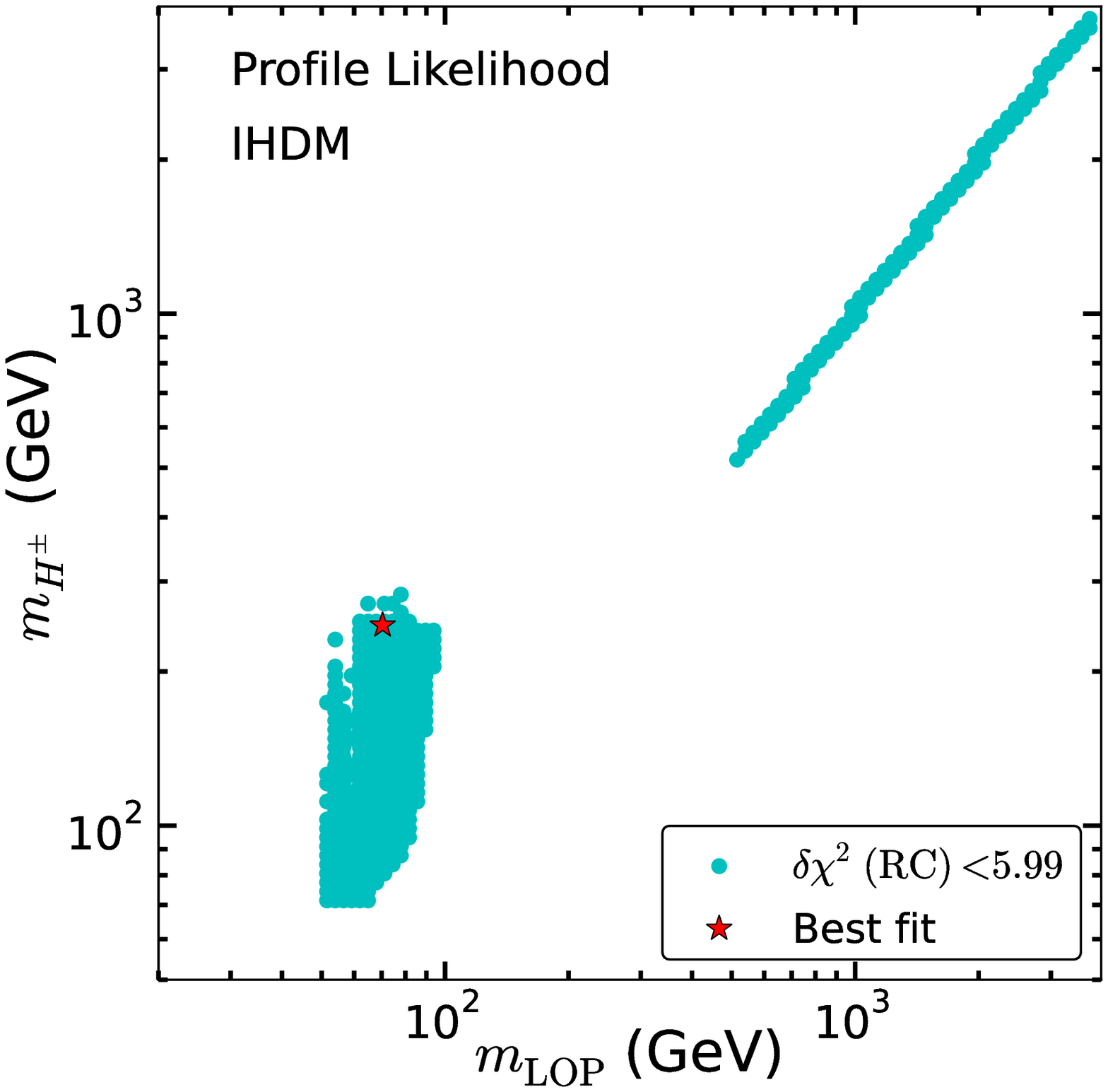}
\caption{\label{fig:mx}
The two dimension profile likelihood on the ($m_S$, $m_A$) 
plane ({\it left}) and ($m_{\rm{LOP}}$, $m_{H^{\pm}}$) plane ({\it right}).  
The cyan dots are 95\% C.L. (2$\sigma$).
The best-fit points are marked as the red stars in the plots. 
}
\end{figure}

However, inefficient annihilation is not the case at 
$100\gev<m_\chi<500 \gev$ which has actually too little relic density
(see also Fig. 6 of \cite{Goudelis:2013uca}). In fact, 
this is because the $W^+W^-$ final state is open so that 
the annihilation cross section can be dramatically enhanced. 
On the other hand, as the DM mass increases further, the
$s$-channel propagator will give rise to suppression in the cross section 
that can offset the enhancement from the opening of the $W^+W^-$ in the
final state. Thus, a correct DM relic density can be achieved again for $m_\chi>500\gev$
as indicated by the cyan dots along the diagonal lines in Fig.~\ref{fig:mx}.

In the right panel of Fig.~\ref{fig:mx}, we can see that the 
charged Higgs boson coannihilation at $m_\chi>600\gev$ and 
a nearly box-shaped region of smaller masses in the $(m_{\rm{LOP}},m_{H^\pm})$ plane. 
We found that in this nearly box-shaped region, the EWPT $T$ constraint will require 
the mass splitting $m_{H^\pm}-m_A\leq 250 \ \gev$ 
in the $3\sigma$  region.
However, there is another limit $m_A-m_{H^\pm}\leq 400\gev$ resulting 
from $\lambda_A<4$ 
(see later for further discussion of the $\lambda_A$ limit shown in the 
upper right panel of Fig. \ref{fig:mx_lam}). 
In addition, we impose the condition $m_A+m_S>m_Z$ in order to 
escape the precise measurement of the $Z^0$ decay width from LEP 
as well as the search for neutralinos at LEP adapted here to the IHDM process 
$e^+e^- \to S A$ as was done in \cite{Lundstrom:2008ai}.
             
\begin{figure}[t!]
\centering
\includegraphics[width=3in]{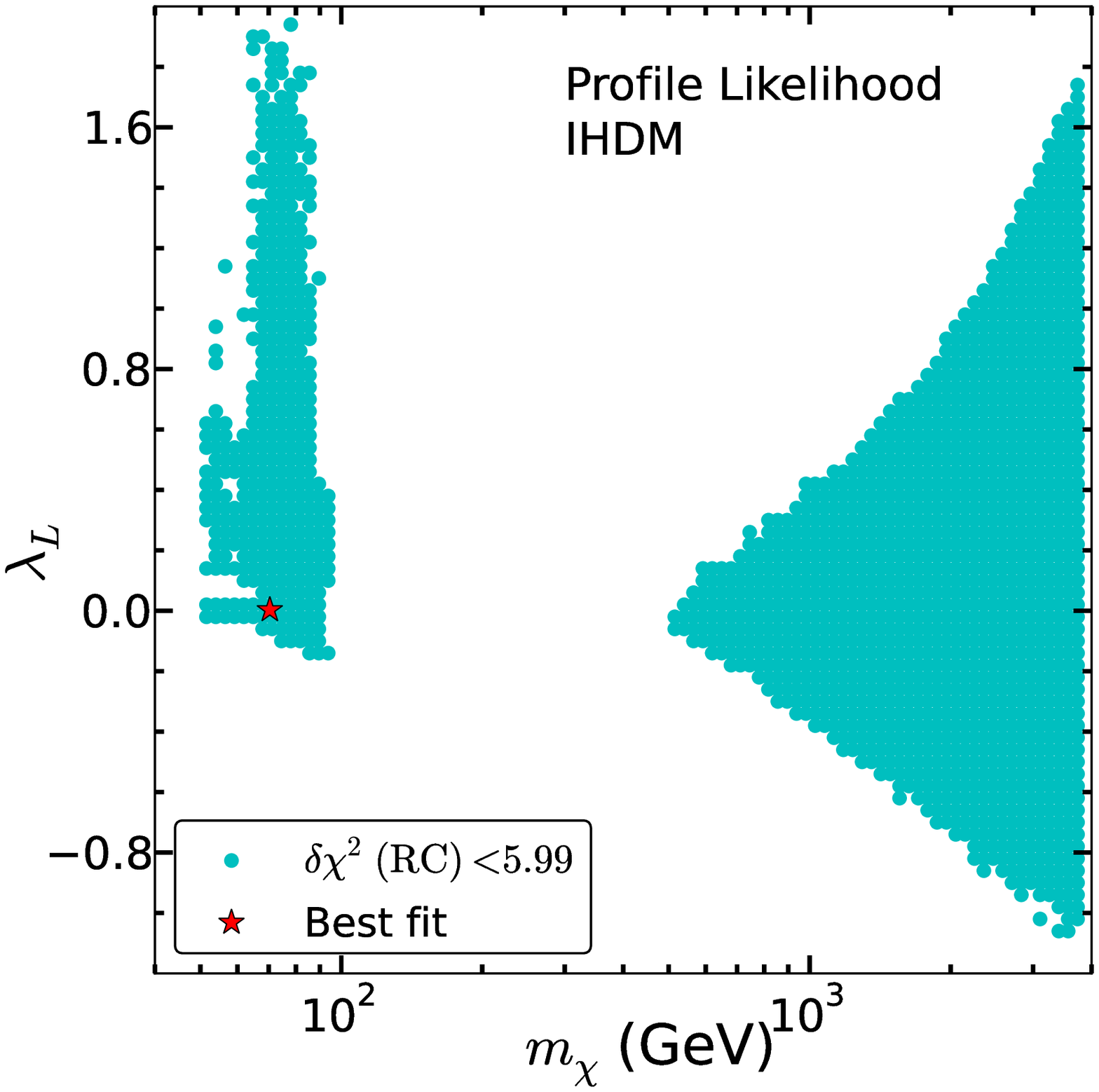}
\includegraphics[width=3in]{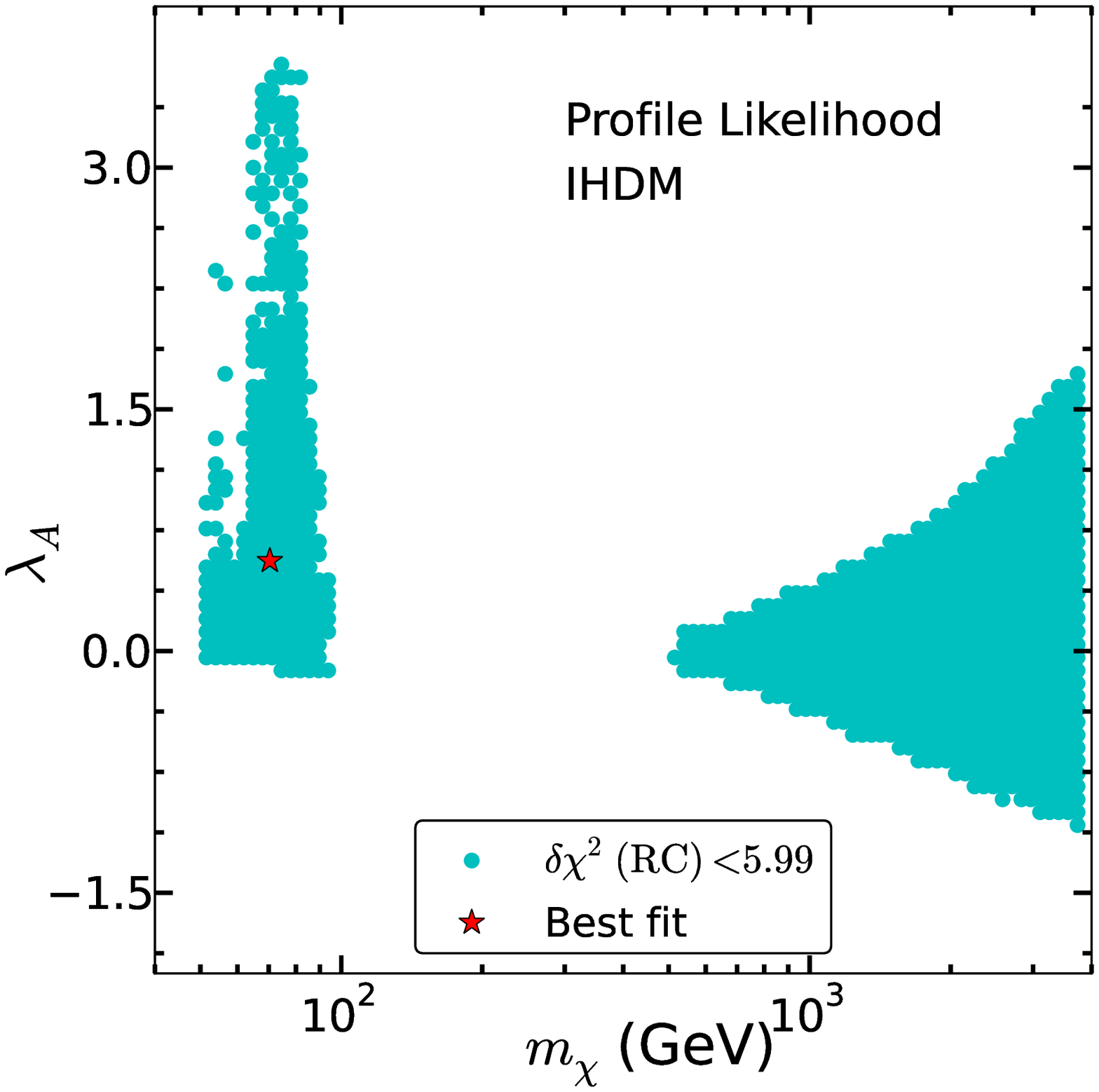}
\includegraphics[width=3in]{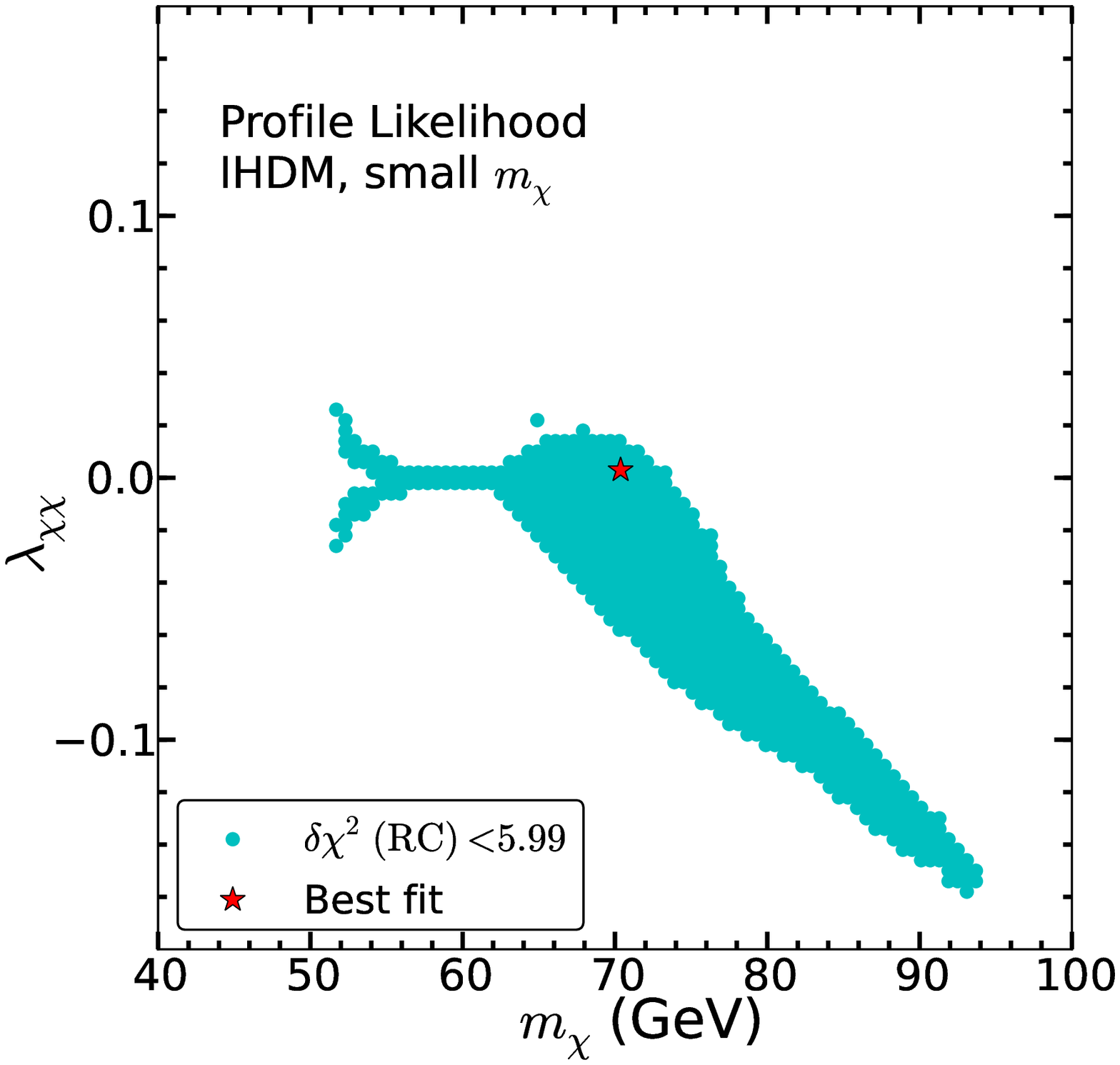}
\includegraphics[width=3in]{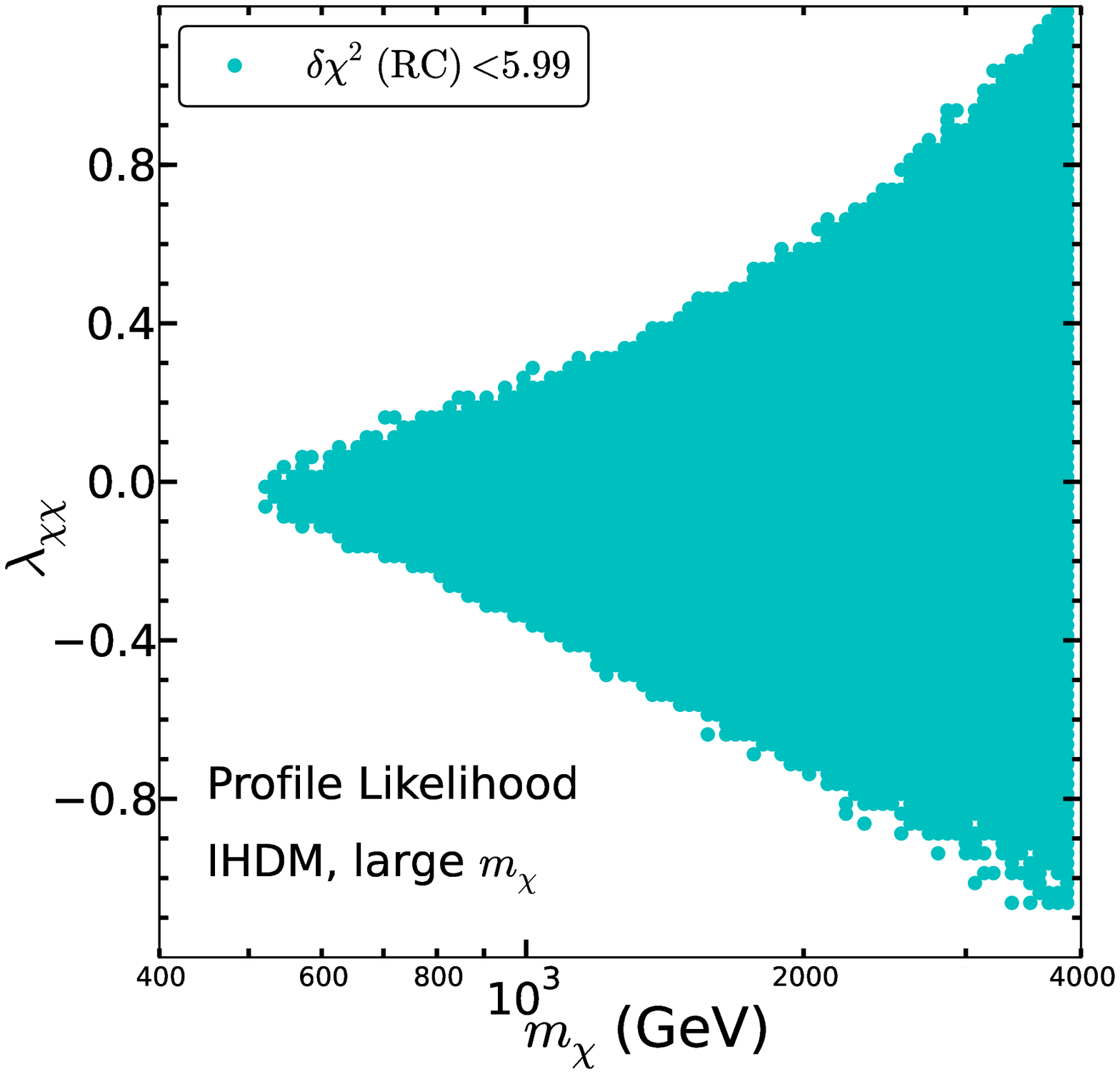}
\caption{\label{fig:mx_lam}
{\it Upper plots}: The two dimension profile likelihood on the ($m_\chi$, $\lambda_L$) 
plane ({\it left}) and  the ($m_\chi$, $\lambda_A$) plane ({\it right}).
{\it Lower plots}: The two dimension profile likelihood on the ($m_\chi$,
$\lambda_{\chi\chi}$) plane with low $m_\chi\leq110$ \gev  \, ({\it left}) and  
large $m_\chi\geq 500$ \gev  \, ({\it right}).
The cyan dots are 95\% C.L. (2$\sigma$).
The best-fit points are marked as the red stars in the plots.
}
\end{figure}

\item[B:] Fig.~\ref{fig:mx_lam}

Next, we discuss the limits on the two couplings $\lambda_L$ and $\lambda_A$ 
which play the role connecting the inert sector with the visible SM sector.
In the upper left and upper right panels of Fig.~\ref{fig:mx_lam}, 
we show the 95\% confidence level region (cyan dots) 
with the \textbf{RC} constraints projected on the ($m_\chi$, $\lambda_L$) 
and ($m_\chi$, $\lambda_A$) planes respectively.
In our analysis, we do not specify $S$ or $A$ must be a LOP before the scan.
We tolerate either $\chi=S$ or $\chi=A$ can be the LOP, fixed only by each model point
in the parameter space surviving the imposed constraints during the scan. 
By comparison of the upper left panel and the two lower panels of Fig.~\ref{fig:mx_lam},
one can identify the two regions of $|\lambda_{\chi\chi}|\leq 0.02$ with small $m_\chi\leq 110$ $\gev$
and of $|\lambda_{\chi\chi}|\leq 1.2$ with large $m_\chi\geq 500 \gev$ 
in which $\lambda_L > \lambda_{\chi\chi} = \lambda_A$ and hence $\chi=A$ being the LOP. 
Other regions in the upper left panel will have $\chi=S$ being the LOP.
Similar behaviors for $\lambda_A$ and $m_\chi$ can be found from the projected
($m_\chi$, $\lambda_A$) plane
in the upper right panel of Fig.~\ref{fig:mx_lam}, when comparing with the two lower panels.
The small differences seen from the two plots in the upper panel of Fig.~\ref{fig:mx_lam} 
near the Higgs resonance region can be traced back to the fact that we 
take $\lambda_L$ as input parameter while $\lambda_A$ 
as output parameter,  given by a combination of 
$\lambda_L$, $m_S$ and $m_A$ via Eq.~(\ref{eq:lambda_A}). 
From Eq.~(\ref{eq:lambda_A}), we can see that  
when $S$ is the LOP, $\lambda_5$ is negative so that $\lambda_A$ is
greater than $\lambda_L$. On the other hand, $\lambda_A<\lambda_L$ if $A$ is the LOP. 
Therefore, $\lambda_A$ will always have a wider range than $\lambda_L$.
From the two plots in the upper panel of Fig.~\ref{fig:mx_lam}, one can see that 
for \textbf{RC} constraints the allowed parameter space of $\lambda_L$ is also highly 
restricted compared to that of $\lambda_A$. 
The additional parameter space for $\lambda_A$ at Higgs resonance appears only
if $A$ is the NLOP. 
Because of $m_A-m_S>10\gev$, the 
$\lambda_A$ coupling being an output parameter 
according to Eq.~(\ref{eq:lambda_A}) implies a very low 
abundance of $A$ and
therefore is not very sensitive to the relic density likelihood function.

Generally speaking, the allowed ranges of $\lambda_L$ and $\lambda_A$ for a given mass
range of $m_\chi$ can be similar (but not identical) if one allows either $S$ or $A$ to be the LOP.
We illustrate this further using the two plots in the lower panel of Fig.~\ref{fig:mx_lam}
where the profile likelihood on the $(m_\chi , \lambda_{\chi\chi})$ is shown,
with left and right panels for $m_\chi \leq$ 100 GeV and $m_\chi \geq$ 500 GeV respectively. 
In these two plots, we can see that the 95\% confidence level regions 
for the \textbf{RC} constraints applied in the 
($m_\chi$, $\lambda_{\chi\chi}$) plane are mostly symmetric but with some 
small asymmetries, especially in the small mass region of $m_\chi \leq$ 100 GeV.
In the lower left plot of Fig.~\ref{fig:mx_lam} where $65 < m_\chi  < 100$ GeV, a 
negative $\lambda_{\chi\chi}$ is required to guarantee the cancellation
between the contributions from different diagrams in the $W^+W^-$ (or $ZZ$) channel 
such that a correct relic abundance can be achieved \cite{LopezHonorez:2010tb}.
From these two profile likelihood plots on the ($m_\chi$,$\lambda_{\chi\chi}$) plane,  
three different limits on $\lambda_{\chi\chi}$
can be summarized as follows:
\begin{itemize}

\item $m_\chi\leq 63 \gev$: 
The upper limit of $\lambda_{\chi\chi}$ in this region 
is due to the invisible Higgs boson decay width being too large. This limit is 
roughly $|\lambda_{\chi\chi}|\leq 0.026$. 
The impact from the current monojet
data is not strong. We have checked that it can constrain $|\lambda_{\chi\chi}|$ from 0.03 -- 0.026.

\item $63 \leq m_\chi\leq 95\ \gev$: Since the 
invisible Higgs boson decay is closed in this mass range, 
the $\lambda_{\chi\chi}$ can be in the range
$-0.16 \leq \lambda_{\chi\chi}\leq 0.02$
\footnote{It is worthy of mentioning that if we relax the relic density constraint 
to $3\sigma$ region, this strip will be extended to $m_\chi\sim 110\gev$ 
which is consistent with Ref.~\cite{Goudelis:2013uca,LopezHonorez:2010tb}.}.

\item $m_\chi\geq 500 \gev$:  
As seen in Fig.~\ref{fig:mx}, the relic density reduction at 
$m_\chi>500\gev$ region  is mainly resulting from the $S-A$, $S-H^\pm$ and $A-H^{\pm}$ 
coannihilation. 
In addition, the $W^+W^-$ final state is being suppressed by increasing $m_\chi$. 
Therefore, we can see $\lambda_{\chi\chi}$ is 
increasing with respect to $m_\chi$  in order to maintain correct relic density. 
The limit depends on $m_\chi$ and is in the 
range of $|\lambda_{\chi\chi}|\leq 1.1 $. The upper limit increases for higher $m_\chi$. 
One can work out these lower limits from 
the theoretical constraints, Eqs.~(\ref{eq:lamCONS}) and (\ref{eq:lam2CONS}). 

\end{itemize}  
%

\begin{figure}[t!]
\centering
\includegraphics[width=3in]{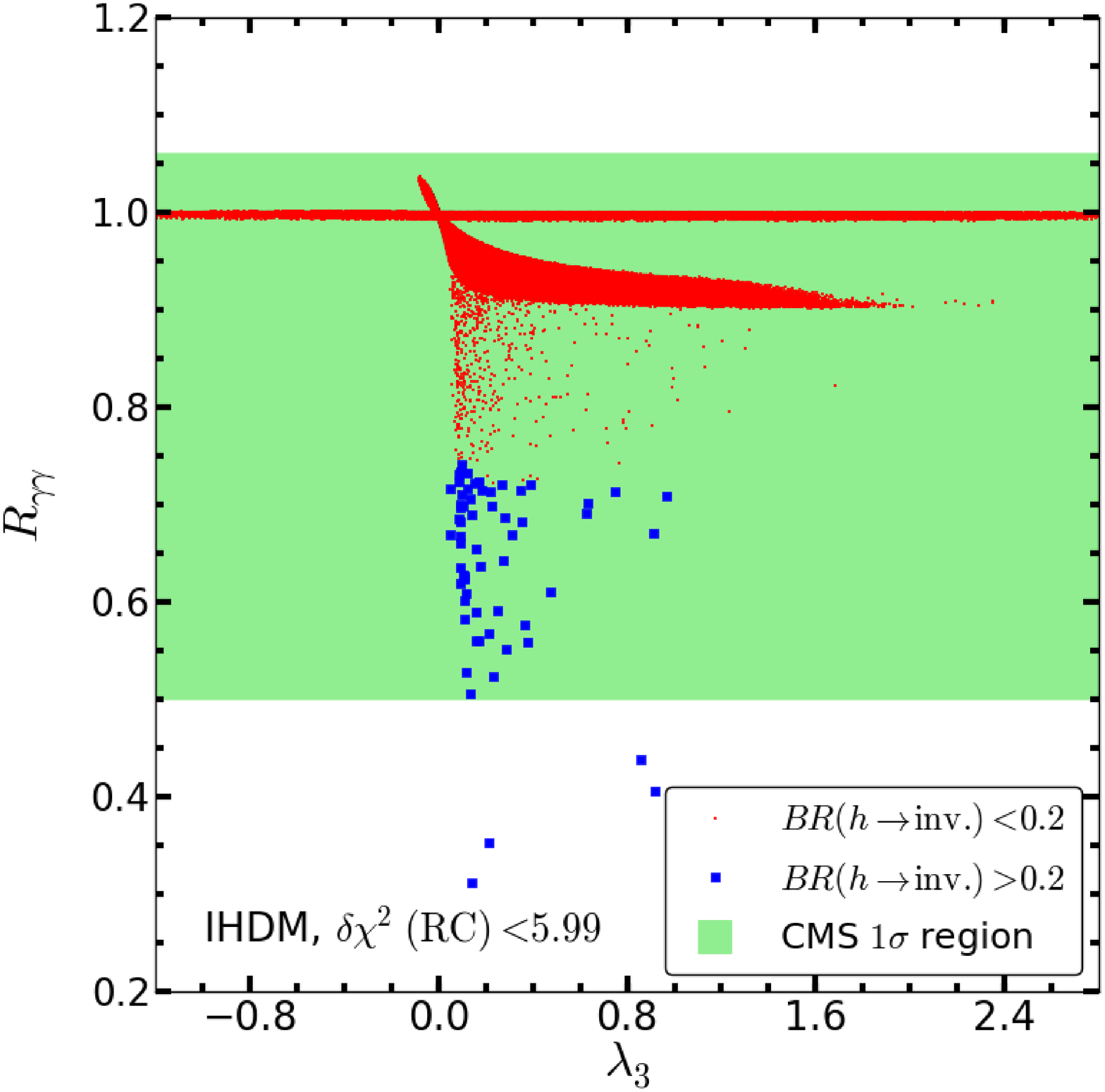}
\includegraphics[width=3in]{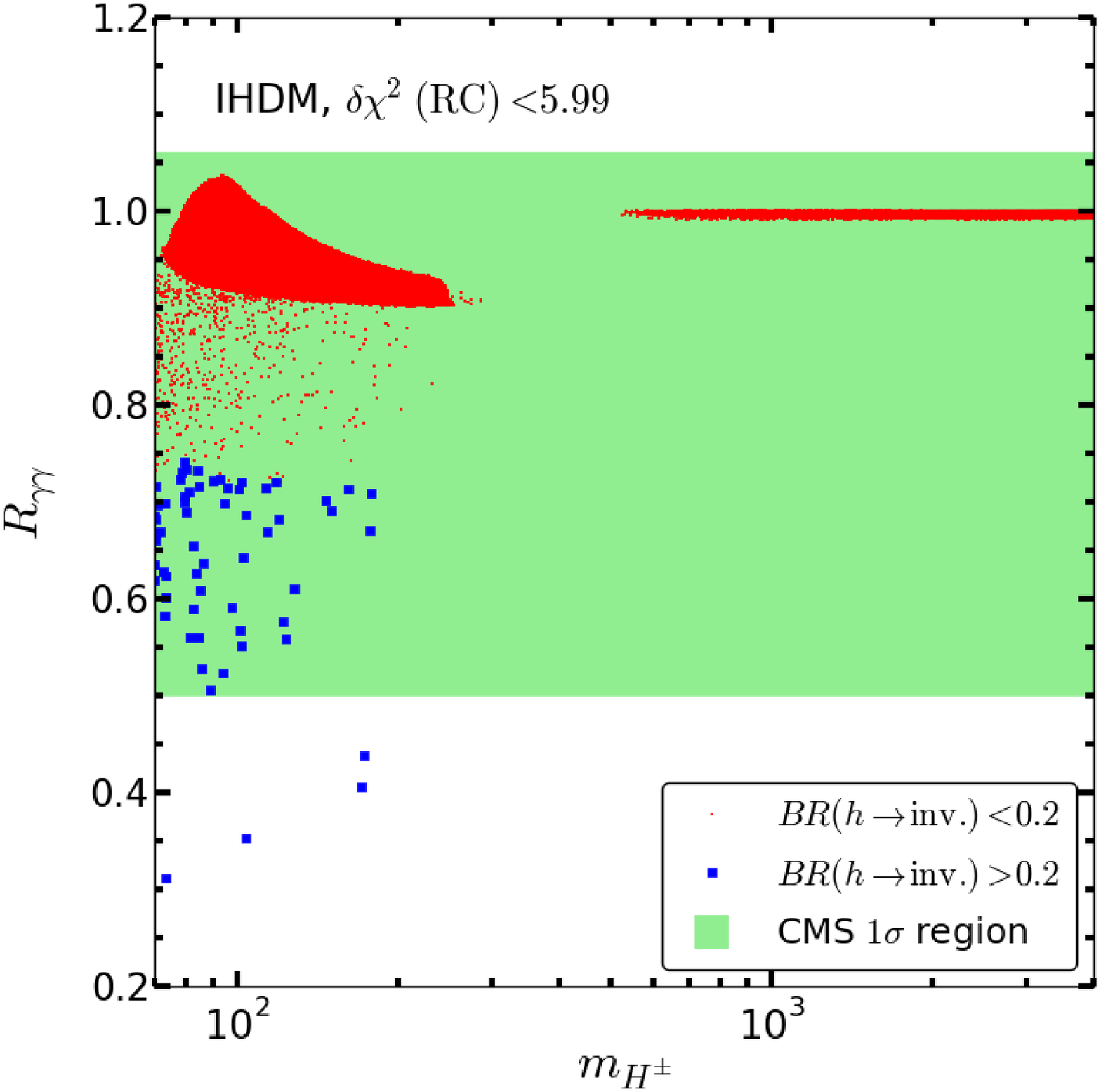}
\caption{\label{fig:mx_laL}
The two dimension scatter plots 
on the ($\lambda_3$, $R_{\gamma\gamma}$) plane ({\it left}) 
and the ($m_{H^\pm}$, $R_{\gamma\gamma}$) plane ({\it right}). 
The blue squares and the red dots correspond to ${\rm BR}(h\rightarrow \rm{invisible}) >$ 0.2 and  
$<$ 0.2 respectively. 
}
\end{figure}

\item[C:] Fig.~\ref{fig:mx_laL}

Third and last for this subsection of {\bf RC}, we discuss the diphoton signal strength 
constraint from the LHC. 
As reported by many studies, most of the ATLAS and CMS data are consistent with SM
predictions. However, there are some small discrepancies between 
ATLAS and CMS results as far as the diphoton channel is concerned. 
While the ATLAS result shows some small excesses with respect to SM 
value, the CMS result which is based on multivariate analysis is nevertheless consistent with SM. 
Here, we do not tempt to explain the ATLAS excess by the additional charged Higgs boson loops
in IHDM but instead we would like to show the 
points that satisfy $\delta\chi^2<5.99$.

It is well known that in the SM, $h\to \gamma\gamma$ is dominated by $W^\pm$
loops which interfere destructively with the subdominant top quark loop.
In IHDM, the charged Higgs boson loops can be constructive or 
destructive with the $W^\pm$ contributions depending on whether  $\lambda_3<0$ or $\lambda_3>0$
respectively \cite{unitarity,maria}. 
As we showed before $R_{\gamma\gamma}$ in the present case can be reduced to the ratio of the
IHDM and SM branching ratios (see Eq.~(\ref{ratio})). 
Thus once the invisible decay $h\to \chi\chi$ is open, 
as long as the partial width of $h\to \gamma\gamma$ has a comparable size with the SM one,
the ratio $R_{\gamma\gamma}$ 
will always be suppressed, {\it i.e.}  
$R_{\gamma\gamma}\approx \Gamma_{\rm SM}(h)/
(\Gamma_{\rm SM}(h)+\Gamma(h\to\chi\chi))<1$. 

In the left and right panels of Fig.~\ref{fig:mx_laL} we present the signal strength $R_{\gamma\gamma}$ 
as a function of the coupling between the SM Higgs boson 
and a pair of charged Higgs bosons $g_{hH^\pm H^\mp}=-v\lambda_3$ 
and of the charged Higgs boson mass respectively. 
The green band indicates the CMS result with 1$\sigma$ uncertainty.
In both panels of Fig.~\ref{fig:mx_laL}, 
we have the blue and red dots for the branching ratio of the 
invisible Higgs boson decay being larger and smaller than 20\% respectively.
On the other hand, if the invisible decay is close, one can see some small enhancements of 
$R_{\gamma\gamma}>1$ for negative $\lambda_3$ (left panel).
Taking the relic density within 2$\sigma$ range from Eq.~(\ref{Omega}) 
as well as the invisible decay branching ratio to be 
less than 65\% (with 11.18\% uncertainty obtained by adding  in quadrature the experimental and theoretical errors 
given in Table~\ref{tab:exp_constraints}), 
we find that $R_{\gamma\gamma}$ falls in the range 0.3 to 1.04. 
This upper limit of 1.04 for $R_{\gamma\gamma}$ from the relic density constraint
was already reported in \cite{unitarity,maria}.  
Most of the points are within CMS 1$\sigma$ band except for a few points with 
branching ratio of the Higgs invisible decay larger than 20\% which are already
excluded. Overall, our results agree with Ref.~\cite{Goudelis:2013uca}.

\end{itemize}

\begin{figure}[t!]
\centering
\includegraphics[width=4.0in]{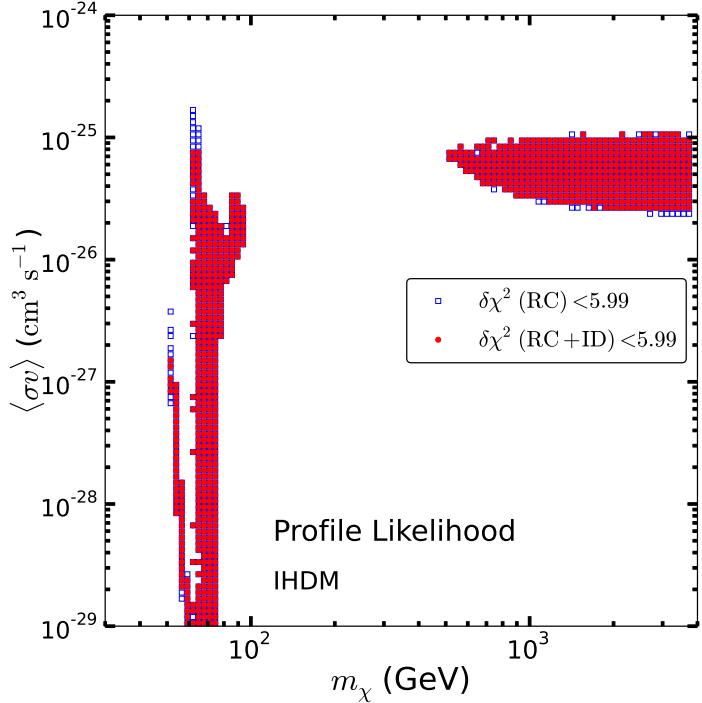}
\caption{\label{fig:mx_sigv}
The two dimension profile likelihood on the ($m_\chi$, $\sigmav$) plane.
The blue squares are 
$2\sigma$ allowed region by \textbf{RC} constraints and the red dots 
are $2\sigma$ allowed region by \textbf{RC+ID} constraints.
}
\end{figure}

\begin{figure}[t!]
\centering
\includegraphics[width=3in]{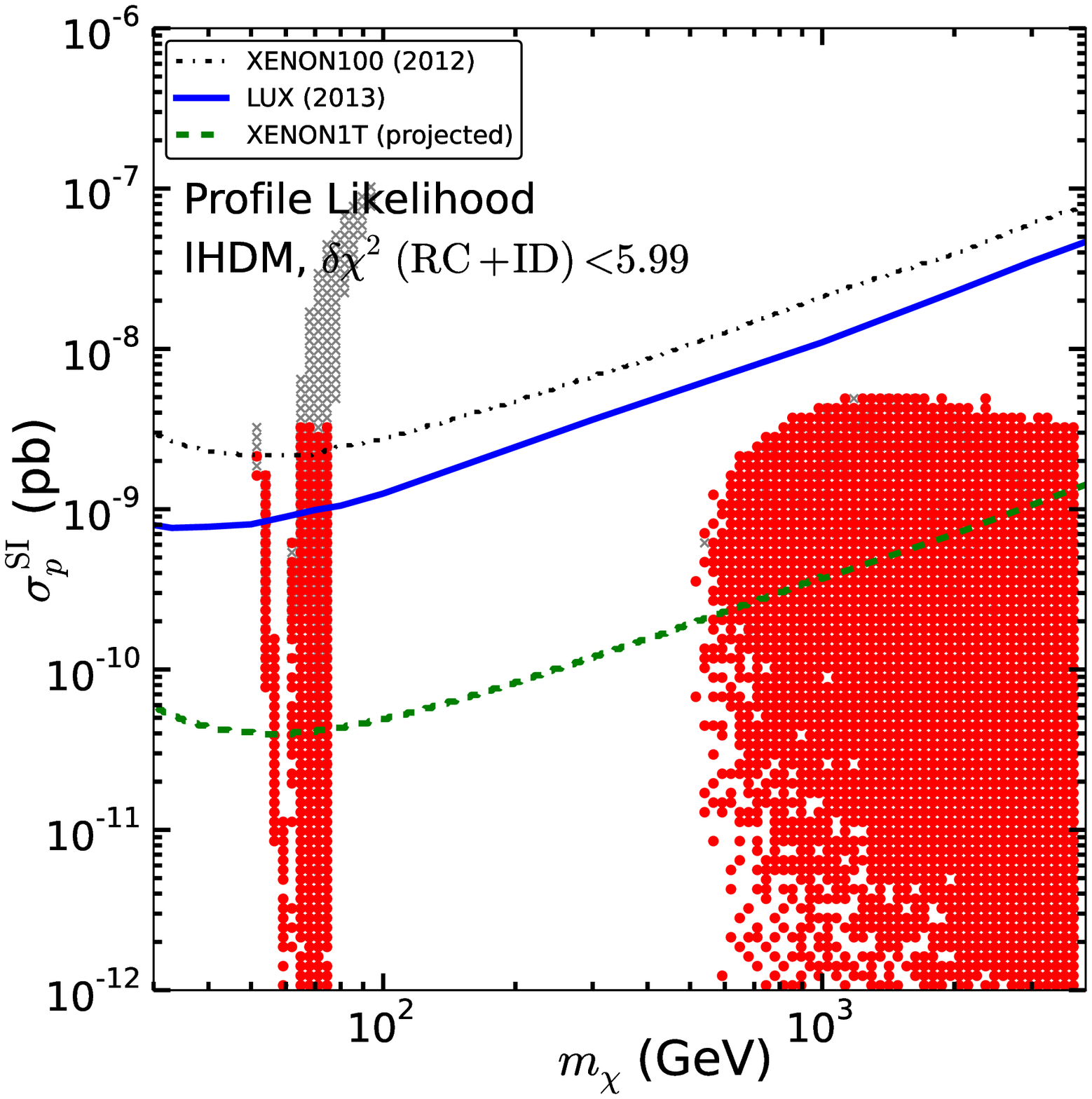}
\includegraphics[width=3in]{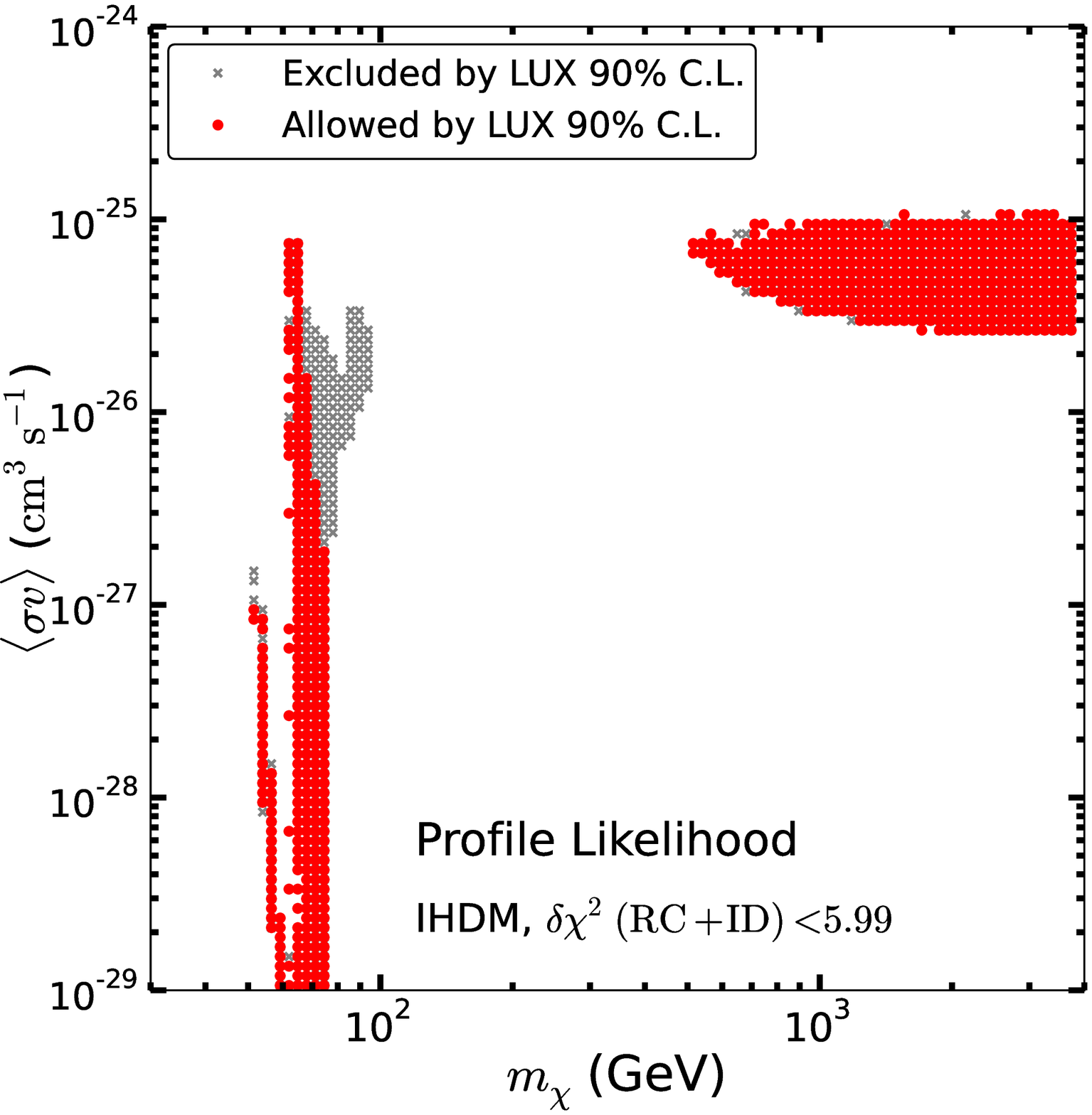}\\
\caption{\label{fig:mx_sigsip}
The two dimension profile likelihood on the ($m_\chi$, $\sigsip$) plane ({\it left})
and the ($m_\chi$, $\sigmav$) plane ({\it right}). 
All points (both red and gray colors) satisfy \textbf{RC+ID} 
constraints in $2\sigma$.
The regions of gray crosses in both panels are excluded in 2$\sigma$ level by 
LUX where the uncertainty on the hadronic matrix elements is taken
into account, while the regions of red dots are allowed. 
The theoretical uncertainty on LUX can weaken the experimental limit 
so that the red dotted region in the left panel can overshoot the LUX (or even XENON100) limit 
in the Higgs resonance region.
}
\end{figure}

\subsubsection{{\bf RC+ID} \label{subsubsection:6b2}}

We now move on to study the impact of DM indirect detection on the  
parameters $m_\chi$ and velocity averaged annihilation cross section $\sigmav$, 
where $v$ is the relative velocity of the annihilating DM. Nowadays,
the DM relative velocity is non-relativistic, 
one can simply use the approximation $\sigmav=\sigma v|_{v\rightarrow 0}$. 
In Fig.~\ref{fig:mx_sigv}, we show the two dimensional profile likelihood 
on the ($m_\chi$, $\sigmav$) plane. The blue squares are 
$2\sigma$ allowed region by \textbf{RC} constraints and the red dots 
are $2\sigma$ allowed region by \textbf{RC+ID} constraints. 

First, we can see three main branches, two vertical branches at 
$2 m_\chi\sim m_h$ region 
and one horizontal at $m_\chi \gtrsim 500\gev$ region. 
They are corresponding to two different mechanisms to produce the correct relic density
as discussed before. Comparing with Fig.~\ref{fig:mx}, 
the first vertical branch at $m_\chi<60\gev$ is $S-A$ coannihilation 
but the second vertical branch at $60\gev<m_\chi<100\gev$ is 
Higgs resonance region plus the openings of the $W^+W^-$ and $ZZ$ channels.
The horizontal branch is again the coannihilation region. 
The thermal averaged $\sigmav_T$ of the vertical branches are more $p$-wave (velocity dependent)
so that most of the points can have wider spread values of  $\sigmav$. 
On the other hand, the horizontal branch is more $s$-wave (velocity independent)
so that $\sigmav\sim {\rm a \; few} \, \times 10^{-26}$ $\rm{cm}^3 \cdot s^{-1}$. 

Clearly, we can see that 
at the low $m_\chi$ region (the two vertical branches) where the IHDM DM has larger $\sigmav$
the constraints from \textbf{ID} can further reduce the parameter space 
from the \textbf{RC} block only.
In the first branch, \textbf{ID} constraints can even make $\sigmav$ having a value  
as low as $10^{-27}$ $\rm{cm}^3 \cdot s^{-1}$.

\subsubsection{{\bf RC+DD+ID} \label{subsubsection:6b3}}

On top of the \textbf{RC+ID} constraints, we can further include the \textbf{DD} constraint 
from LUX.
In the left panel of Fig. \ref{fig:mx_sigsip}, the $2\sigma$ allowed region 
on the $(m_\chi , \sigma^{\rm SI}_p)$ plane by \textbf{RC+DD+ID} 
constraints is shown in red circles. 
The region of gray crosses was excluded by the LUX result with the hadronic uncertainties included.
In IHDM, only $t$-channel with the $h$ exchange can contribute to DM-quark elastic scattering. 
Therefore, one can expect the gray crosses region is due to the coupling $\lambda_{L,A}$ being too large.
Interestingly, in the right panel where we map to the $(m_\chi , \langle \sigma v \rangle)$
plane, we can see that the LUX limit 
can only remove some regions (gray crosses) with low $m_\chi$ which have large $\sigsip$ (left panel) but 
the region (red dots) with $\sigmav$ as high as $8 \times10^{-26} \, \rm{cm}^3 \cdot s^{-1}$ is still allowed!
In addition, in the left pane we also plot the projected sensitivity of XENON1T which 
has the potential to probe the higher mass coannihilation region. 
We will discuss this impact further in section \ref{section:7}.
      
\begin{figure}[t!]
\centering
\includegraphics[width=3in]{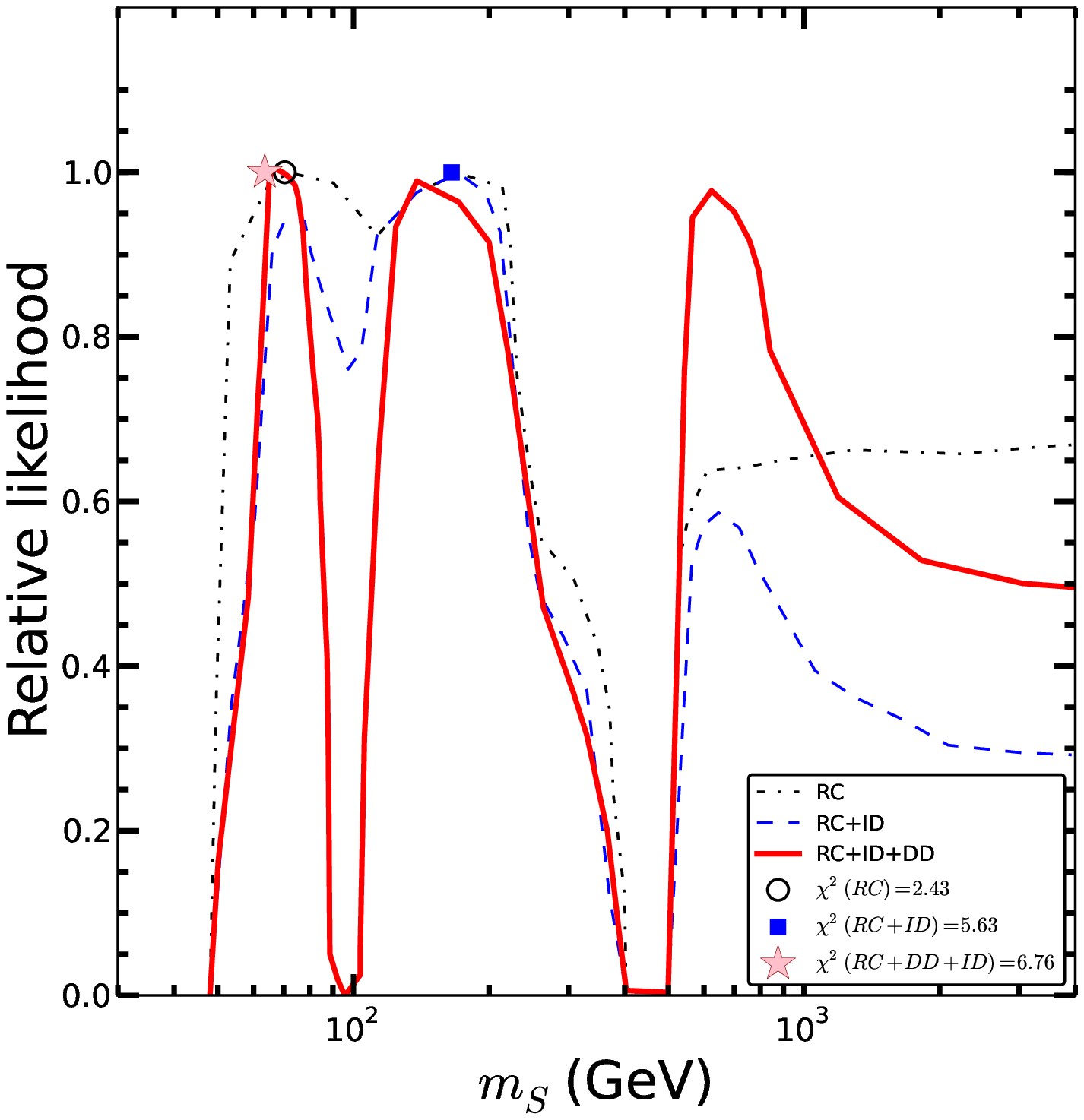}
\includegraphics[width=3in]{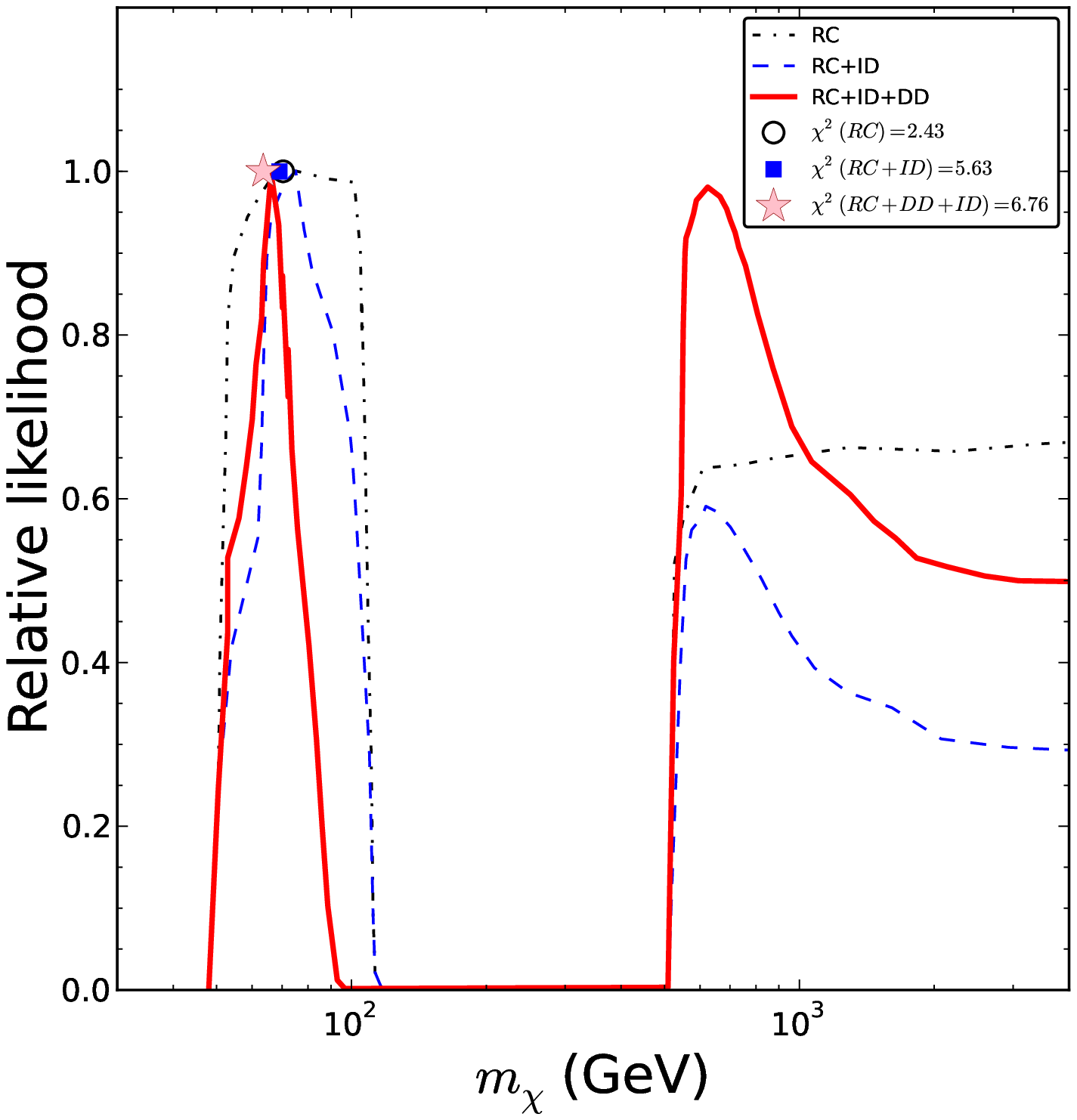}
\includegraphics[width=3in]{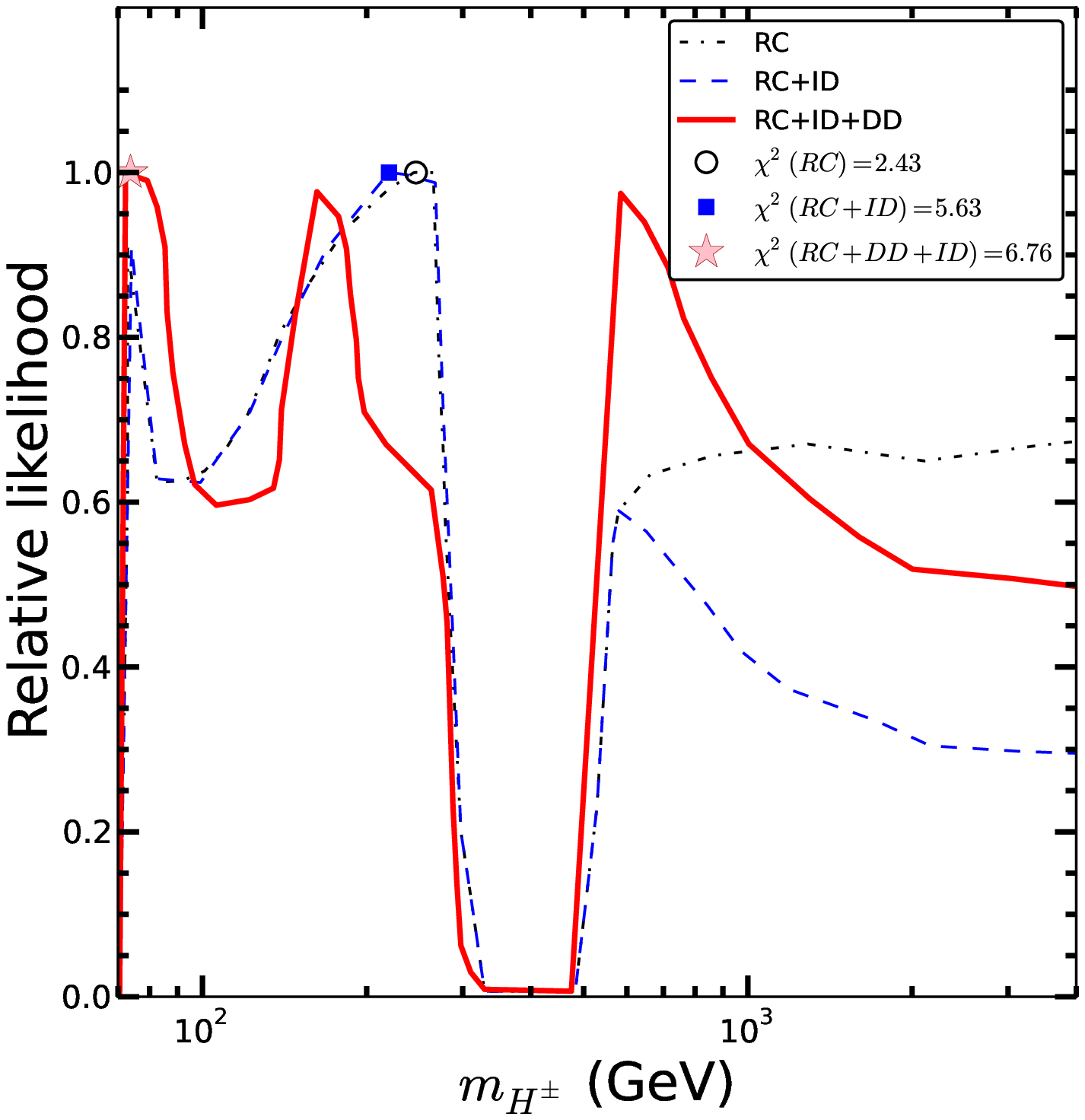}
\includegraphics[width=3in]{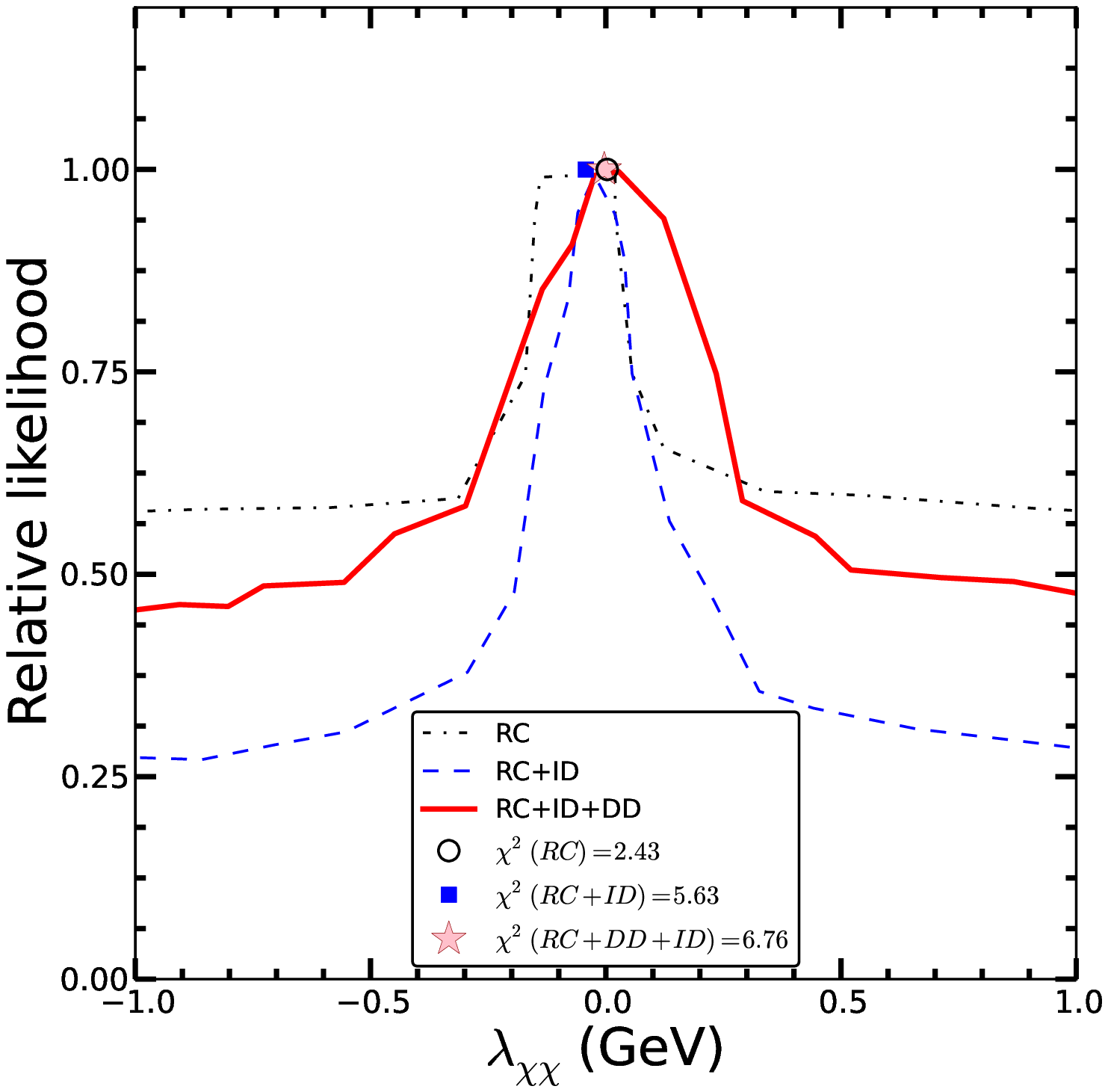}
\caption{\label{fig:1D}
One-dimensional profile likelihood distribution for $m_S$, $m_\chi$, $m_{H^\pm}$, 
and DM effective coupling $\lambda_{\chi\chi}$. 
The distribution of $m_A$ is almost identical to $m_S$. 
The best-fits are presented by black cycles, blue squares, and 
red stars for the three likelihood combination blocks, 
\textbf{RC}, \textbf{RC+ID}, and \textbf{RC+DD+ID}, respectively. 
}
\end{figure}

In Fig.~\ref{fig:1D}, we show the 1D relative likelihood distributions 
for $m_{S}$ (upper left), $m_{\chi}$ (upper right),  
$m_{H^\pm}$ (lower left) and $\lambda_{\chi\chi}$ (lower right) 
in the three blocks of \textbf{RC}, \textbf{RC+ID}, and \textbf{RC+ID+DD}, 
marked by black dash-dot, blue dash, and red solid lines, respectively. 
The relative likelihood in each case is defined as $\mathcal{L/L}_{\rm{max}}$ where  
$\mathcal{L}_{\rm{max}}$ is the likelihood at the best-fits. 
We do not show the distribution of $m_{A}$ since it is almost identical to $m_{S}$.
As aforementioned that $m_\chi$ can be either 
$m_S$ or $m_A$, the peaks at $m_\chi<100\gev$ in the upper right panel 
 correspond actually to two separated peaks with almost 
the same height at $m_S<300\gev$ in the upper left panel. 
The first peak owes to $m_\chi=m_S$ and the second $m_\chi=m_A$. 
We can see clearly that there is no preference of $m_{\chi}=m_A$ or $m_\chi=m_S$.

Since the $\gamma/e^+/\bar{p}$ fluxes are inversely proportional to $m_\chi^2$, 
the impact of \textbf{ID} constraint is mainly on the lower $m_\chi$ region. 
On the other hand, if $m_\chi$ turns out to be too large suppressing the DM signal, 
the total \textbf{ID} $\chi^2$ will be the same as consideration of background only.      
From Fig.~\ref{fig:1D}, we found that with the additional DM signal, 
the $\chi^2$ can be improved to at most $1\sigma$ significance. For example, 
in the upper right plot, we can see that at the $m_\chi>500\gev$ region
there is a flat \textbf{RC} likelihood distribution while the \textbf{RC+ID} one is decreasing.  
Because of $\chi-H^\pm$ coannihilation, we can see similar 
decrease of the \textbf{RC+ID} likelihood distribution 
for the $m_{H^\pm}>500\gev$ region in the lower left plot. 
Note that this large $m_\chi$ region can not be constrained by the current LUX data. 
There is a third peak at $m_\chi\sim 500\gev$ because 
the $m_\chi<100\gev$ region is less favored by LUX. 
Even though the best-fit point still locates at this lower mass region, 
the minimum $\chi^2$ is roughly increased by one unit. 
As a result of increasing the minimum $\chi^2$,
the relative likelihood of the $m_\chi>500\gev$ region becomes statistically more significant.         

\begin{table} 
\scriptsize
\centering
\begin{tabular}{| l | l |  l | l | l |}
\hline
& $S-A$ Co-ann. & Higgs Resonance & $S-A-H^{\pm}$ Co-ann. & $S-H^{\pm}$ Co-ann.\\
\hline
\multicolumn{5}{|c|}{Basic Parameters}\\
\hline
$m_h$ (\gev) &  125.6  & 125.92 &126.51 &	125.44   \\
$m_S$ (\gev) &  70.17    & 64.46 &522.13 &	971.88  \\
$m_A$ (\gev) &  78.55  & 130.83 &521.98 &	985.56   \\
$m_{H^\pm}$ (\gev) &  97.47 & 76.11 &523.27 &	979.49   \\
$\lambda_L$ &  -0.032  & -0.009 & -0.052&	 -0.248   \\
$\lambda_2$ &  0.60  & $9.3\times10^{-4}$ & 1.61 &	2.81   \\
\hline
\multicolumn{5}{|c|}{DM Observables}\\
\hline
$\Omega_\chi h^2$ & 0.107 & 0.123 &   0.105 & 0.097\\
$\sigmav$ (cm$^3 \cdot$ s$^{-1}$) & $9.81\times10^{-27}$ & $2.14\times10^{-26}$ & $7.51\times10^{-26}$ & $4.58\times10^{-26}$\\
$\sigsip$ (pb) & $7.11\times10^{-9}$ & $6.96\times10^{-10}$ &   $3.59\times10^{-10}$ & $2.62\times10^{-9}$\\
\hline
\multicolumn{5}{|c|}{Channels Contributed to $1/\Omega h^2$}\\
\hline
Dominant
&  $S A \to q \bar{q}$   (41\%)    
&  $S S \to b \bar{b}$   (49\%)    
&  $S S/AA/H^+H^- \to W^+ W^-$   (39\%)    
&  $H^\pm H^\pm \to W^\pm W^\pm$   (16\%)\\
Subdominant
&  $S S \to b \bar{b}$   (31\%)    
&  $S H^\pm \to \gamma W^\pm$   (17\%)    
&  $SS/AA/H^+H^- \to Z Z$   (21\%)    
&  $S S \to W^+W^-,ZZ, hh$   (9\%,15\%,14\%)\\
& & & $SH^\pm/AH^\pm \to W^\pm \gamma/W^\pm Z$ (21\%) & $SH^\pm \to W^\pm Z,W^\pm h$ (6\%,9\%) \\
\hline
\multicolumn{5}{|c|}{ID Annihilation Cross Section $\sigmav$}\\
\hline
Dominant 
&  $S S\to b \bar{b}$   (75\%)    
&  $S S\to b \bar{b}$   (77\%)
&  $A A\to W^+ W^-$   (51\%)
&  $S S\to Z Z$   (39\%) \\
Subdominant
&  $S S\to g g$   (12\%)    
&  $S S\to g g$   (10\%)
&  $A A\to Z Z$   (44\%)
&  $S S\to h h$   (36\%) \\
\hline
\multicolumn{5}{|c|}{Pulls for \textbf{ID} Observables}\\
\hline
$\delta\chi^2_{\gamma,\rm{dSphs}}$ & 0.83  & 0.22 &  1.05 & 1.99\\
$\delta\chi^2_{\gamma,\rm{GC}}$    & 0.87 & 0.61 &  0.88 & 0.93\\
$\delta\chi^2_{e^+}$  & 0.38 & 0.25 &  0.19 & 0.49\\
$\delta\chi^2_{\bar{p}}$ & 0.62 & 1.17 &  0.04 & 0.15\\
\hline
\end{tabular}
\caption{\label{table:BMpts} 
Values of some {\bf RC+ID+DD} parameters and observables as well as the 
$\delta \chi^2$ for the {\bf ID} observables at several benchmark points.
}
\end{table}

In Table~\ref{table:BMpts}, we show the results of some of the basic parameters and observables in the
{\bf RC+ID+DD} block at a few benchmark points. 
These benchmark points correspond to 
the main mechanisms that reduce the relic density, namely
the $S-A$ coannihilation, Higgs resonance, $S-A-H^{\pm}$ ($S-A$ and $\chi-H^{\pm}$) coannihilation, 
and $S-H^{\pm}$ coannihilation. Moreover, these four mechanisms 
can also represent four different relevant $m_\chi$ regions. 
For the two columns of the $S-A$ coannihilation and Higgs resonance in Table~\ref{table:BMpts},
the $b\bar{b}$ final state plays an important role in the annihilation channels. 
However, as long as the $W^+W^-$ and $ZZ$ channels are open, 
they will contribute significantly to the annihilation cross section,  as clearly seen 
in the last two columns of this table.
One may notice that in the last column of $S - H^\pm$ coannihilation case, 
the distribution of different channels contributed to
the relic density is quite spread out in this case.
Thus, beside the Higgs resonance point, the other coannihilation channels 
can lead to effective relic density reduction as well.
Finally, we also show in Table~\ref{table:BMpts} 
the \textbf{ID} $\delta\chi^2$ for these four benchmark points. 
We note that poorer values of $\delta\chi^2_{\bar p}$ are obtained
at the first two benchmark points ($S-A$ coannihilation and Higgs resonance) 
where the $b \bar b$ final state dominates in the antiproton flux. 
While the $b\bar{b}$ mode can be significant for the antiproton flux from
the fragmentation of $b$ and $\bar b$ into antiproton, 
the smallness of the $b$-quark parton distribution inside the nucleon makes the $b$-quark
has very small impact on the $\sigsip$. On the other hand, 
values of $\sigsip$ at these two benchmark points
are quite acceptable.

\begin{table} [t]
\begin{center}
\begin{tabular}{|c|c|c|c|}
\hline\hline
Parameter& $\mathcal{L}$(\textbf{RC}) & $\mathcal{L}$(\textbf{RC+ID})& $\mathcal{L}$(\textbf{RC+DD+ID})  \\
\hline\hline
$m_h$ (GeV)& 125.76 & 125.91 & 126.016 \\
\hline
$m_{S}$ (GeV)& 70.37 & 165.12 & 63.54 \\
\hline
$m_{A}$ (GeV)& 196.58 & 69.04 & 166.16 \\
\hline
$m_{H^{\pm}}$ (GeV)& 246.28 & 219.85 & 73.78 \\
\hline
$\lambda_L$ & $2.96\times 10^{-3}$ & 0.33 & $-3.29\times 10^{-3}$ \\
\hline
$\lambda_2$ & 3.50 & 3.58 & $5.67\times 10^{-4}$ \\
\hline\hline
$\sigmav$ ($\rm{cm}^3 \cdot s^{-1}$)& $7.97\times 10^{-28}$ & $2.44\times 10^{-26}$ & $2.18\times 10^{-26}$ \\
\hline
$\sigsip$ (pb)& $5.92\times 10^{-11}$ & $1.22\times 10^{-8}$ & $8.89\times 10^{-11}$ \\
\hline\hline
$\chi^2$ & 2.43 & 5.63 & 6.76 \\
\hline\hline
\end{tabular}
\end{center} 
\caption{
Table displaying the properties of our best-fit points for the three different blocks. Note that the invisible Higgs decay is closed at these three best-fit points. 
}
\label{tab:bestpts}
\end{table}
In Table~\ref{tab:bestpts}, we summarize the best-fit points for the three different blocks from our scan. 
The second column is only with \textbf{RC} constraints in the likelihood 
while the third and fourth columns are with \textbf{RC+ID} and \textbf{RC+DD+ID} 
in the likelihood, respectively. 
We would like to stress that 
there is no preference of $\chi=S$ or $\chi=A$ 
due to the symmetry between $S$ and $A$ in the model. 
We find that the maximum likelihood of $\mathcal{L}(\chi=S)$ and 
$\mathcal{L}(\chi=A)$ are roughly the same.  
Therefore, the fact that the best-fit points are located at $\chi=S$ or 
$\chi=A$ region is just due to the fact that
we collected the maximum likelihood 
before hitting the sampling stop criteria. 
In other words, we cannot tell the dark matter in IHDM must be a scalar or 
pseudoscalar from this analysis.
However, we can see that the best-fit points of three 
sets of constraint are
all located at the lower $m_\chi$ region. 
The reasons for this are mainly due to the EWPT and 
indirect detection constraints. 
First, the nearly degeneracy between $m_{H^\pm}$ and $m_\chi$ required by 
relic density constraints at the $m_\chi>500\gev$ region 
implies $S$ and $T$ are always negligible. 
On the other hand, $S$ and $T$ can be enhanced in the $m_\chi\lesssim 100\gev$ region
due to larger mass splitting between $H^\pm$ and $\chi$. 
Hence, we can obtain the better likelihood in small $m_\chi$ region.
Second, the lower $m_\chi$ region may amplify the indirect detection signal too, 
because the signal fluxes of indirect detection experiments are inversely proportional to $m_{\chi}^{2}$.
Hence, lower $m_\chi$ region can have stronger signals to fit current 
\textbf{ID} constraints than astrophysical background only. 
On the other hand, the fluxes from DM annihilation in the $m_\chi>500\gev$ region 
can be suppressed and may be lower than the astrophysical background. 
As seen from Figs.~\ref{fig:bkgep} and \ref{fig:bkgap},
there are still some rooms for DM in indirect detection 
experiments. Perhaps not a discovery with large statistical significance, but a weak signal 
usually fits the likelihood better than using the background only hypothesis. 
Certainly, our understandings of the astrophysical backgrounds could be too naive.


\section{Future Experimental Constraints from LHC-14, XENON1T and AMS-02 \label{section:7}}


\begin{figure}[t!]
\centering
\includegraphics[width=4in]{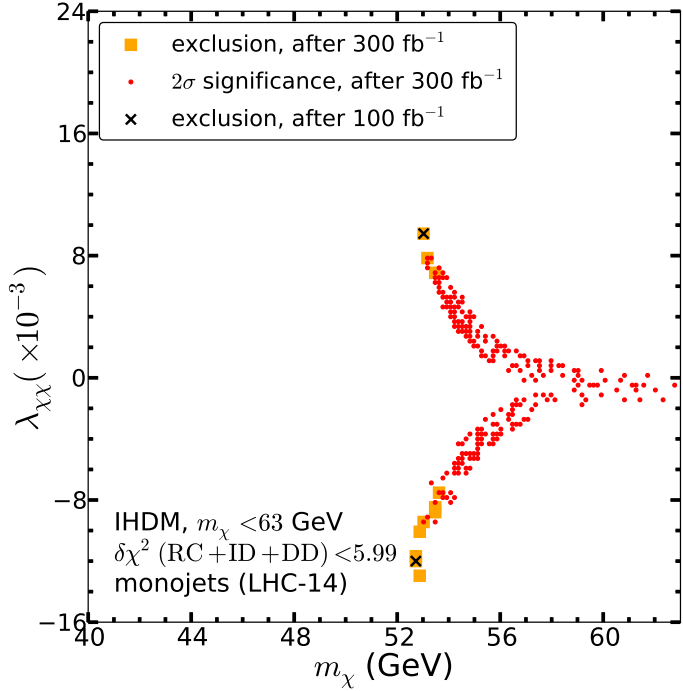}
\caption{\label{fig:mx_lamL_LHC14}
The $2\sigma$ profile likelihood of the \textbf{RC+ID+DD} constraints 
projected on the ($m_\chi$, $\lambda_{\chi\chi}$) plane
for the monojet result at LHC-14. 
The black crosses will be excluded by the future 100 $\rm{fb}^{-1}$ data. 
The orange square will still be allowed by 100 $\rm{fb}^{-1}$ data  
but disfavoured by 300 $\rm{fb}^{-1}$ data. 
The red dots will be allowed by  300 $\rm{fb}^{-1}$ data.
They are all in $2\sigma$ significance. 
}
\end{figure}
 
In this section, we will consider the sensitivities of three future experiments: 
(1) LHC-14 monojet with luminosities 100 fb$^{-1}$ and 300 fb$^{-1}$, 
(2) XENON1T,
and 
(3) AMS-02 one year antiproton data, given the 
parameter space obtained in previous section that satisfies the {\bf RC+DD+ID} constraints.

\subsection{LHC-14 monojet \label{section:7a}}

At the 14 TeV run of LHC (LHC-14), we will discuss 
the impact of the monojet search with luminosities of 100 fb$^{-1}$ and 300 fb$^{-1}$. 
We assume a null measurement of DM with background events same as observed, 
{\it i.e.} $b=o$ where $b$ and $o$ are the background and observed events at 100 fb$^{-1}$ or 300 fb$^{-1}$
obtained by scaling the current CMS data at luminosity of $19.5 \, \rm{fb}^{-1}$. 
Using the standard sensitivity formula, $s/\sqrt{b}$, we compute the significance of monojet for LHC-14.
      
In Fig.~\ref{fig:mx_lamL_LHC14}, we present the potential power of 
LHC monojet search with 100 fb$^{-1}$ and 300 fb$^{-1}$ on the ($m_\chi$,$\lambda_{\chi\chi}$) plane.
In linear scale, we zoom into the region $m_\chi<63\gev$
where the invisible Higgs decay is open. All the points shown satisfy the  
\textbf{RC+DD+ID} constraints in $2\sigma$.
With 100 fb$^{-1}$ of data, only the orange boxes will be allowed while 
the few black crosses located near the boundary where $|\lambda_{\chi\chi}|\sim10^{-2}$ 
will be disfavoured in the $2\sigma$ subset.
However, with 300 fb$^{-1}$ of data, the range of $\lambda_{\chi\chi}$ will be extended to
the region of red dots in Fig.~\ref{fig:mx_lamL_LHC14} where
$|\lambda_{\chi\chi}|\lesssim 6\times 10^{-3}$.

\subsection{XENON1T \label{section:7b}}

\begin{figure}[t!]
\centering
\includegraphics[width=3in]{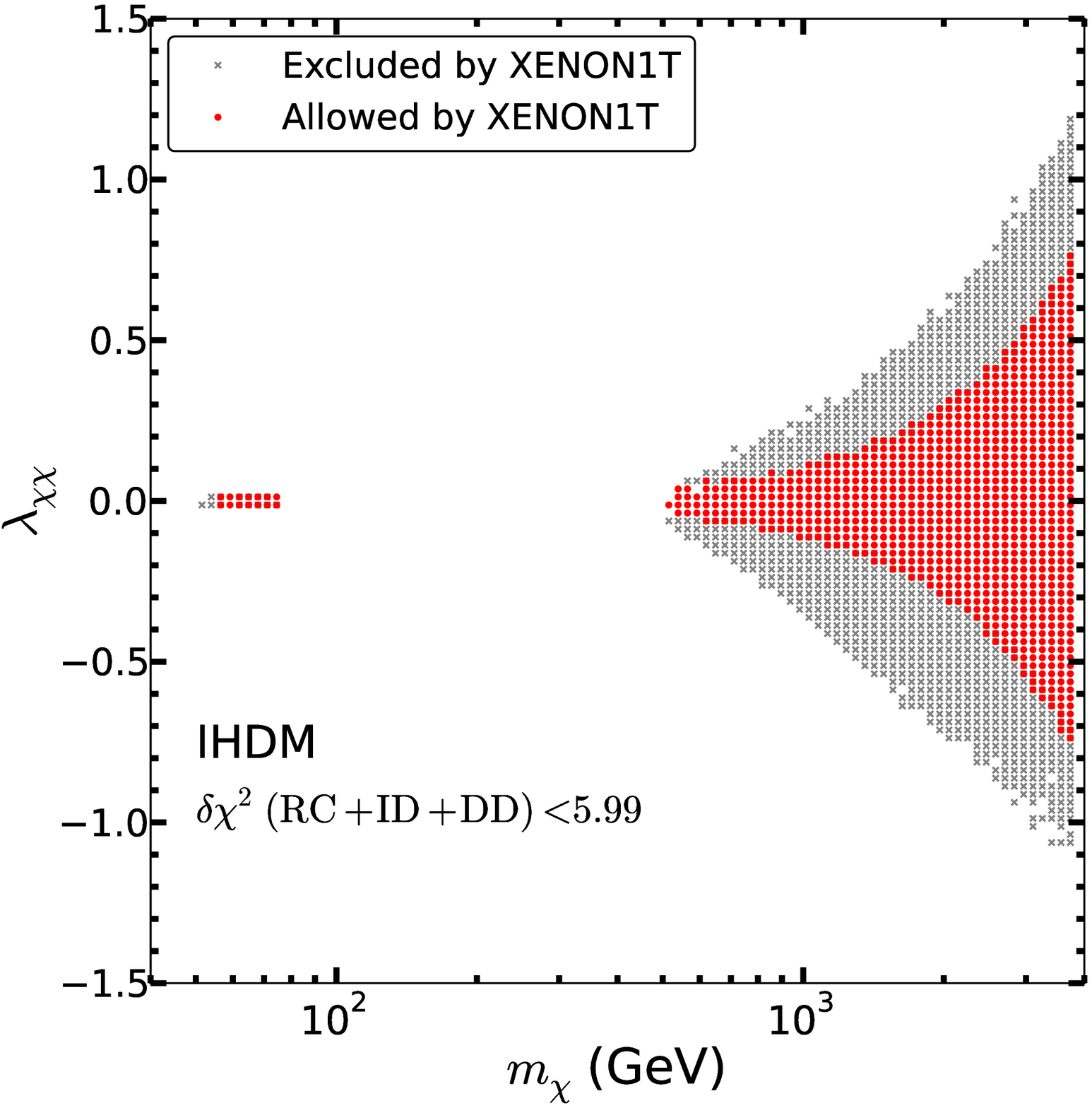}
\includegraphics[width=3in]{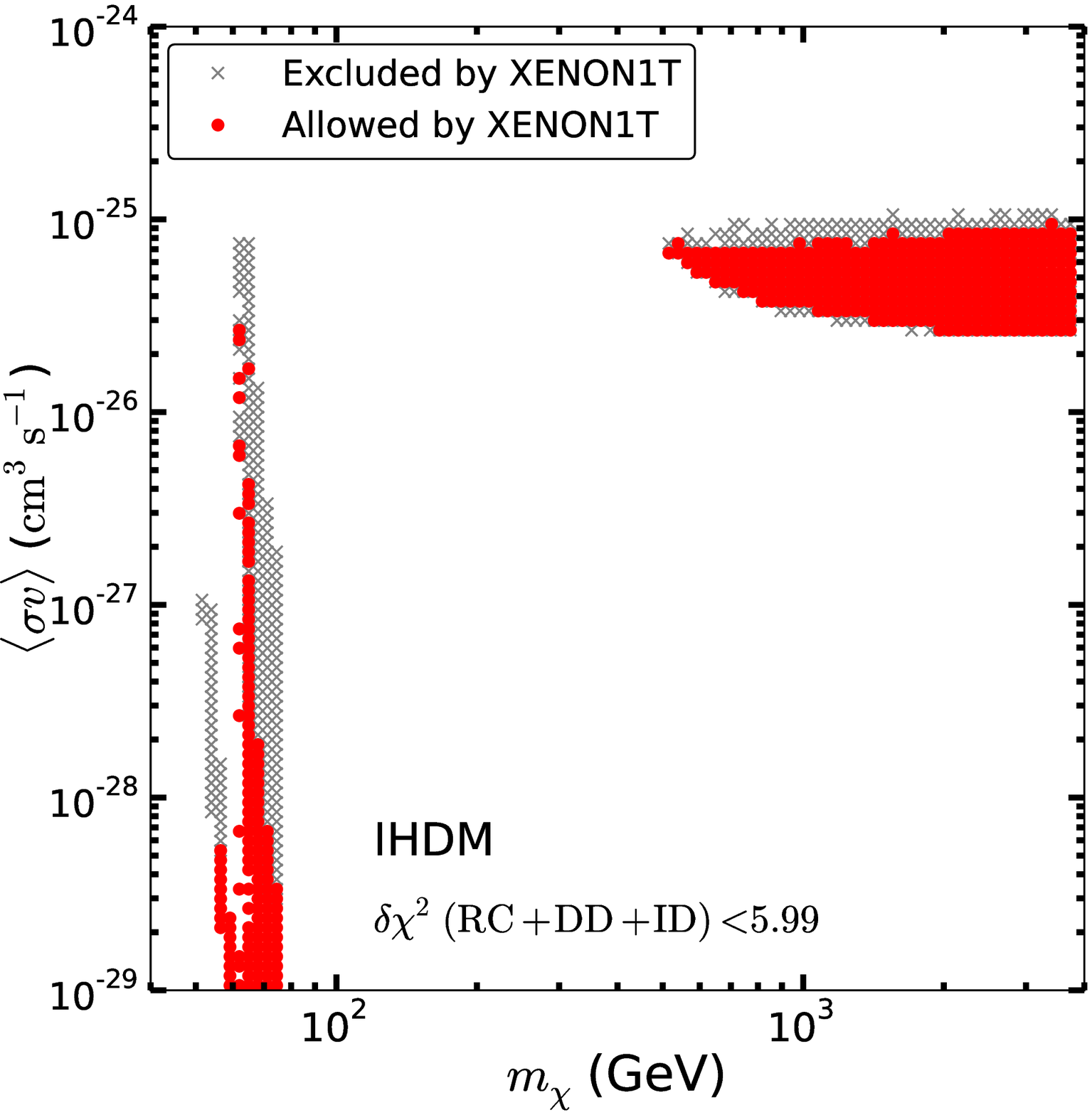}
\caption{\label{fig:mx_sigv_XENON1T}
The left and right panels show the sensitivity of XENON1T on the ($m_\chi$, $\lambda_{\chi\chi}$) and 
($m_\chi$, $\sigmav$) planes respectively subjected to the \textbf{RC+ID+DD} constraints
 in $2\sigma$ significance.
The red (gray) area will be allowed (excluded) by XENON1T sensitivity.
}
\end{figure}

In Fig.~\ref{fig:mx_sigv_XENON1T}, we show the disfavoured region by future XENON1T sensitivities
subjected to the \textbf{RC+ID+DD} constraints in $2\sigma$ significance.
The left panel is for $(m_\chi , \lambda_{\chi\chi})$ and the right panel is for ($m_\chi$, $\sigmav$).  
The red dots are favoured but gray dots/crosses are disfavoured by XENON1T limit.
Although the region of $m_\chi>500\gev$ can not be entirely 
ruled out by XENON1T from our global analysis based on tree level calculation, 
a recent paper~\cite{Klasen:2013btp} pointed out that electroweak corrections can significantly 
alter the theoretical prediction of $\sigsip$, especially for large $m_\chi$ region.
As shown in their computation, $\sigsip$ is not expected to be 
lower than $10^{-11}$ pb even when one loop corrections are included~\cite{Klasen:2013btp}.
We thus expect next generation of ton-sized detectors for DM direct detection 
can probe most of the parameter space of IHDM.

\subsection{AMS-02 antiproton \label{section:7c}}

\begin{figure}[t!]
\centering
\includegraphics[width=3in]{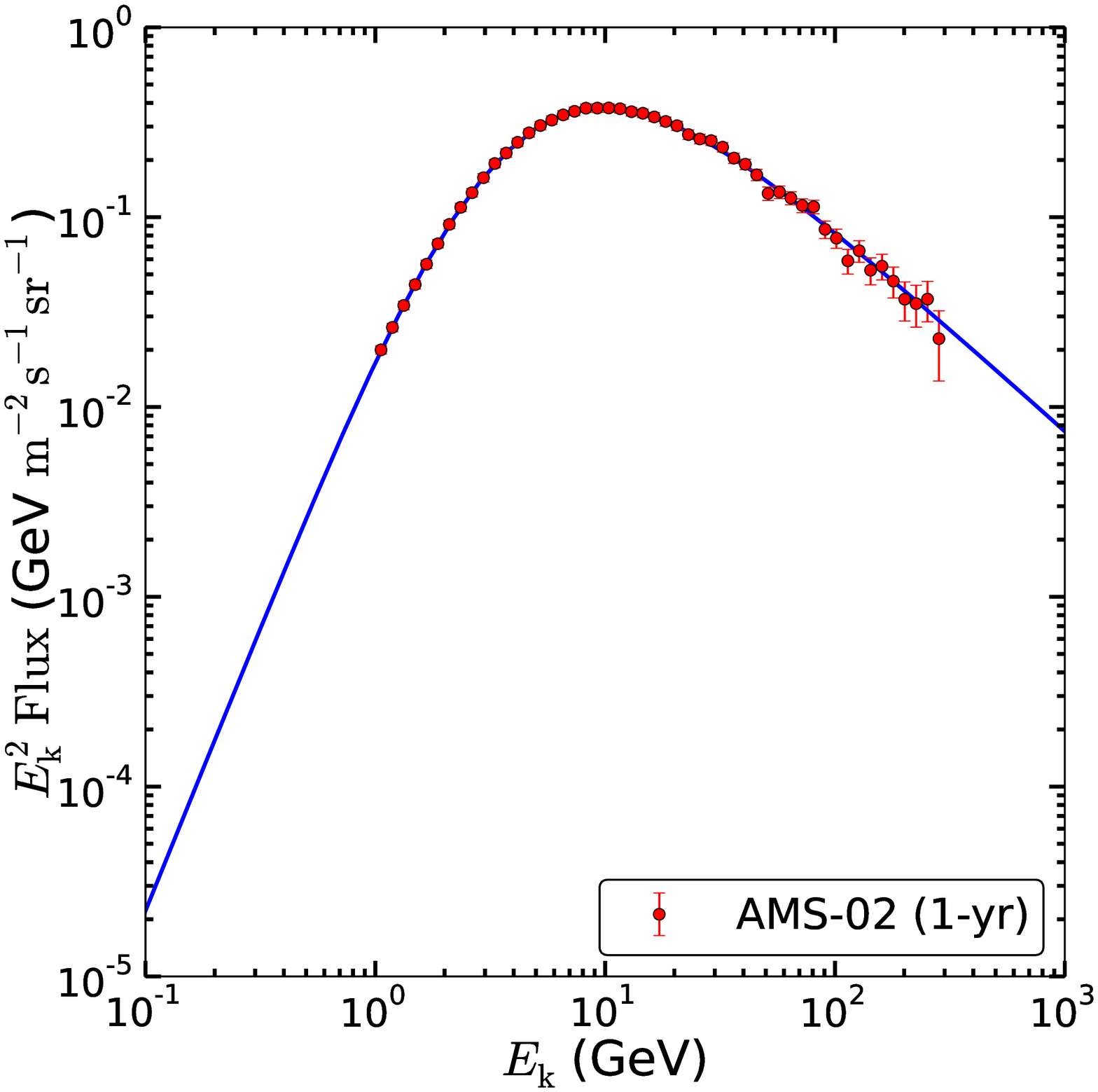}
\includegraphics[width=3in]{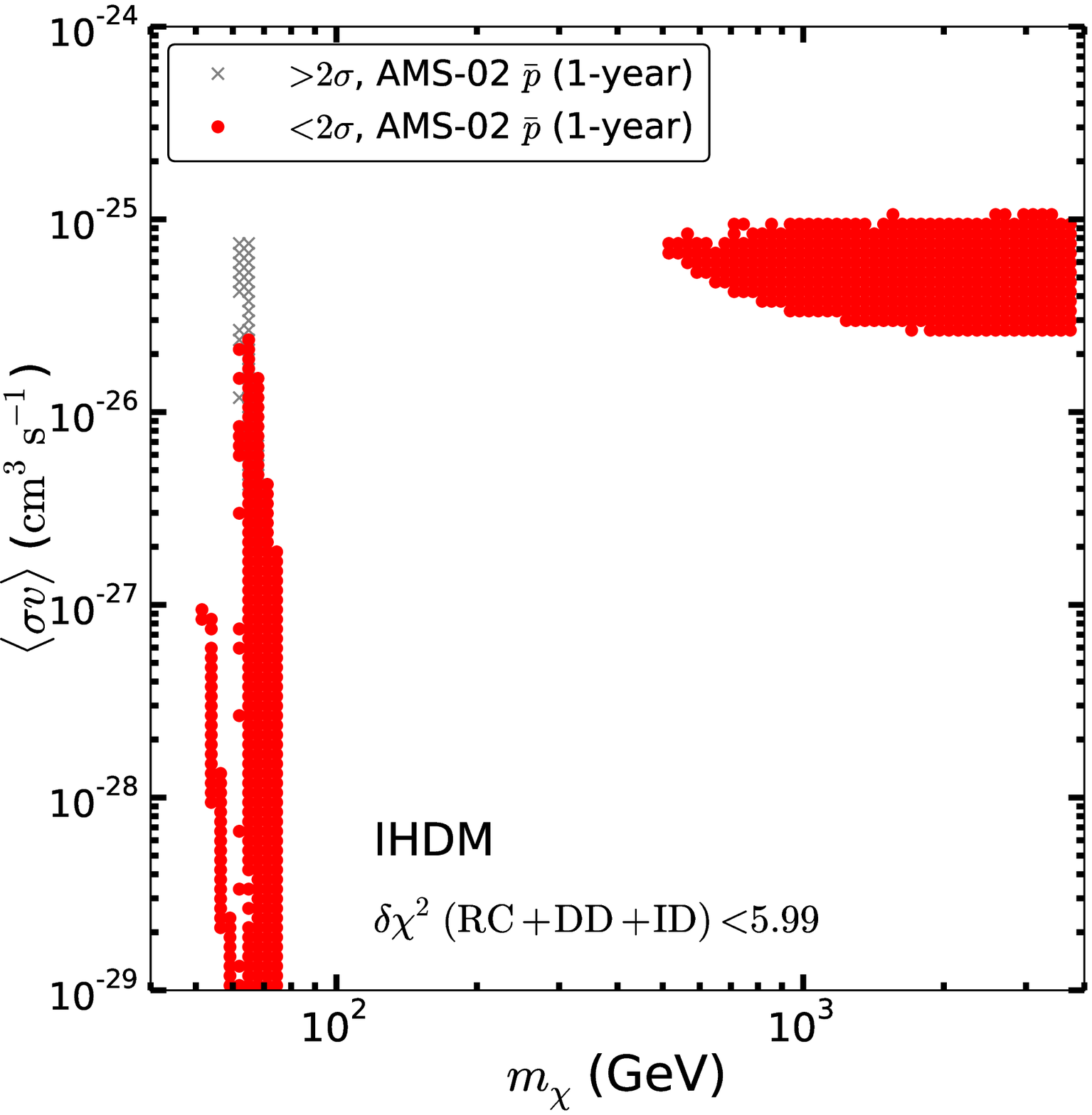}
\caption{Simulated antiproton flux for one year observation of AMS-02 ({\it left}).
The AMS-02 anti-proton one year sensitivity on the ($\sigmav$,$m_\chi$) plane ({\it right}).
The red dots (gray crosses) will be allowed (excluded) by AMS-02 anti-proton one year sensitivity.
}\label{fig:ap_ams02}
\end{figure}

We generate the simulated  one year AMS-02 data of the antiprotons 
following Ref. \cite{Pato:2010ih}. The expected antiproton flux $\phi$ is 
adopted to be the one described in section \ref{section:2} (see Fig. \ref{fig:bkgap}). 
The number of antiproton events in a given energy bin is approximately
\begin{equation}
N_{\bar{p}}(E_k)=\Delta t\int_{\Delta E}{\rm d}E_k\,\phi(E_k)\times A(E_k) \; ,
\end{equation}
where $A(E_k)$ is the simulated geometry factor given in \cite{geofac},
$\Delta E$ is the width of the energy bin and $\Delta t$ is the exposure
time. We generate the data from 1 to 300 GeV, with 50 bins logarithmically 
evenly distributed according to the binning of the positron fraction
measurement by AMS-02 \cite{Aguilar:2013qda}. For the ``observed'' number
of events we apply a Poisson fluctuation on $N_{\bar{p}}$, with statistical
error $\approx 1/\sqrt{N_{\bar{p}}}$. The systematic error is simply
adopted to be $\sim 5\%$ \cite{Pato:2010ih}, which is added quadratically 
to the statistical error. 
The simulated antiproton flux for one year
observation of AMS-02 is shown in the left panel of Fig. \ref{fig:ap_ams02}. Using the
simulated antiproton data, we calculate the $\chi^2$ of each DM model
point with the same method described in sections \ref{section:4} and \ref{section:5}.
In the right panel of Fig.~\ref{fig:ap_ams02}, 
the exclusion power of the AMS-02 one year antiproton data is shown. 
The red dots are allowed but gray crosses are disfavoured by future sensitivity. 
Comparing with the exclusion power of XENON1T sensitivity 
(right panel of Fig.~\ref{fig:mx_sigv_XENON1T}),
we can see most of parameter space excluded by AMS-02 antiproton data are 
also excluded by XENON1T. 
However, for the $\sigmav \sim 2\times 10^{-26} \rm{cm}^3 \cdot s^{-1}$ at 
the Higgs resonance region (near the top of the second vertical branch from the left), 
one can find some small fractions of red dots that
can be excluded by AMS-02 antiproton data but not yet ruled out by XENON1T.  
This is because the DM annihilation channels are 
dominant by $\tau^+\tau^-$ or $b\bar{b}$ final state at these points. However 
the $h\tau^+\tau^-$ coupling is irrelevant to direct detection and the $h b\bar{b}$ 
coupling can contribute to direct detection only by integrating this heavy $b$ quark to
obtain the $h gg$ coupling. Hence its contributions to 
$\sigsip$ is also small as compared with light quarks.

 \subsection{LHC-14 Monojet + XENON1T + AMS-02 antiproton flux \label{section:7d}}

\begin{figure}[t!]
\centering
\includegraphics[width=3in]{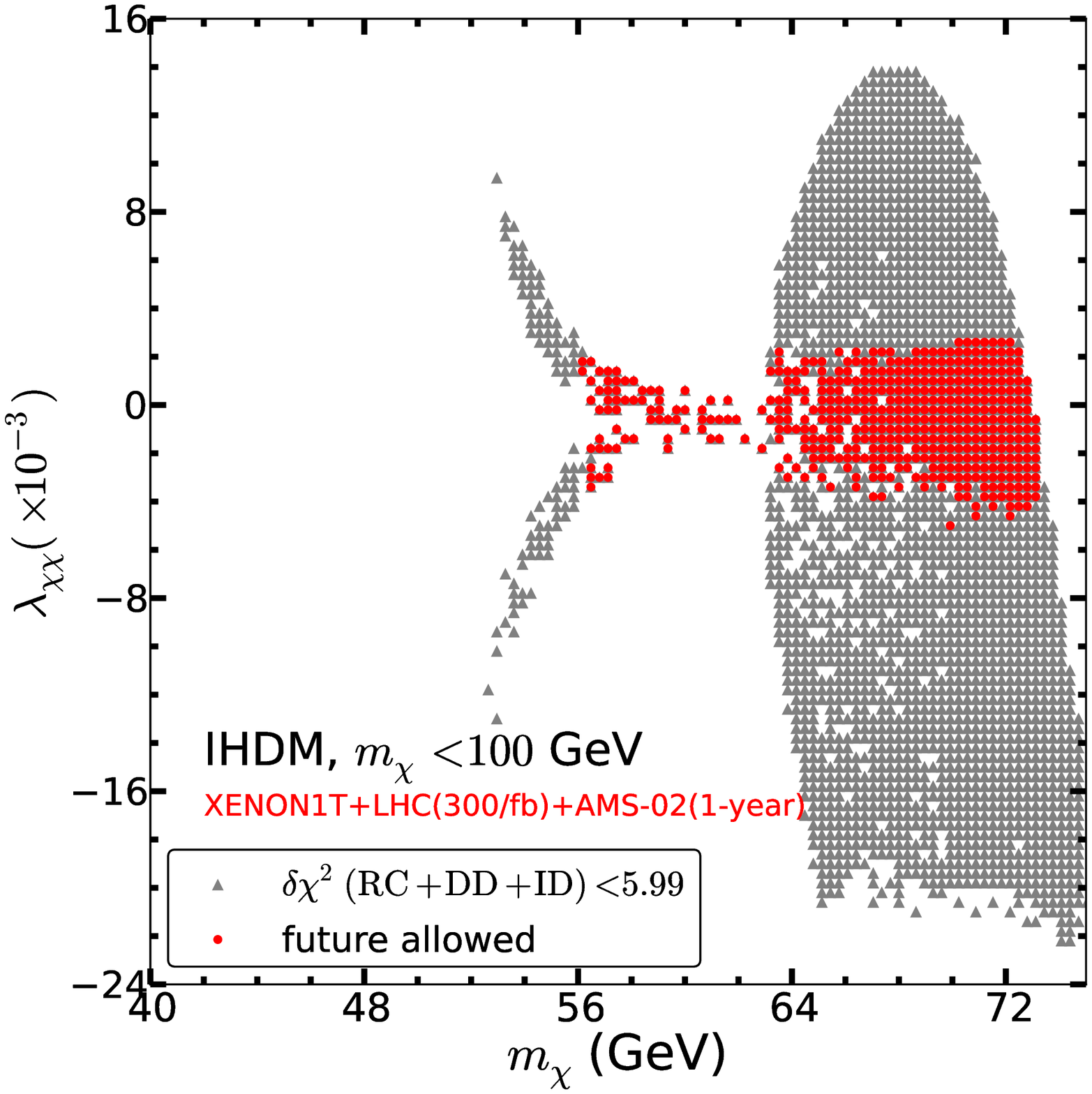}
\includegraphics[width=3in]{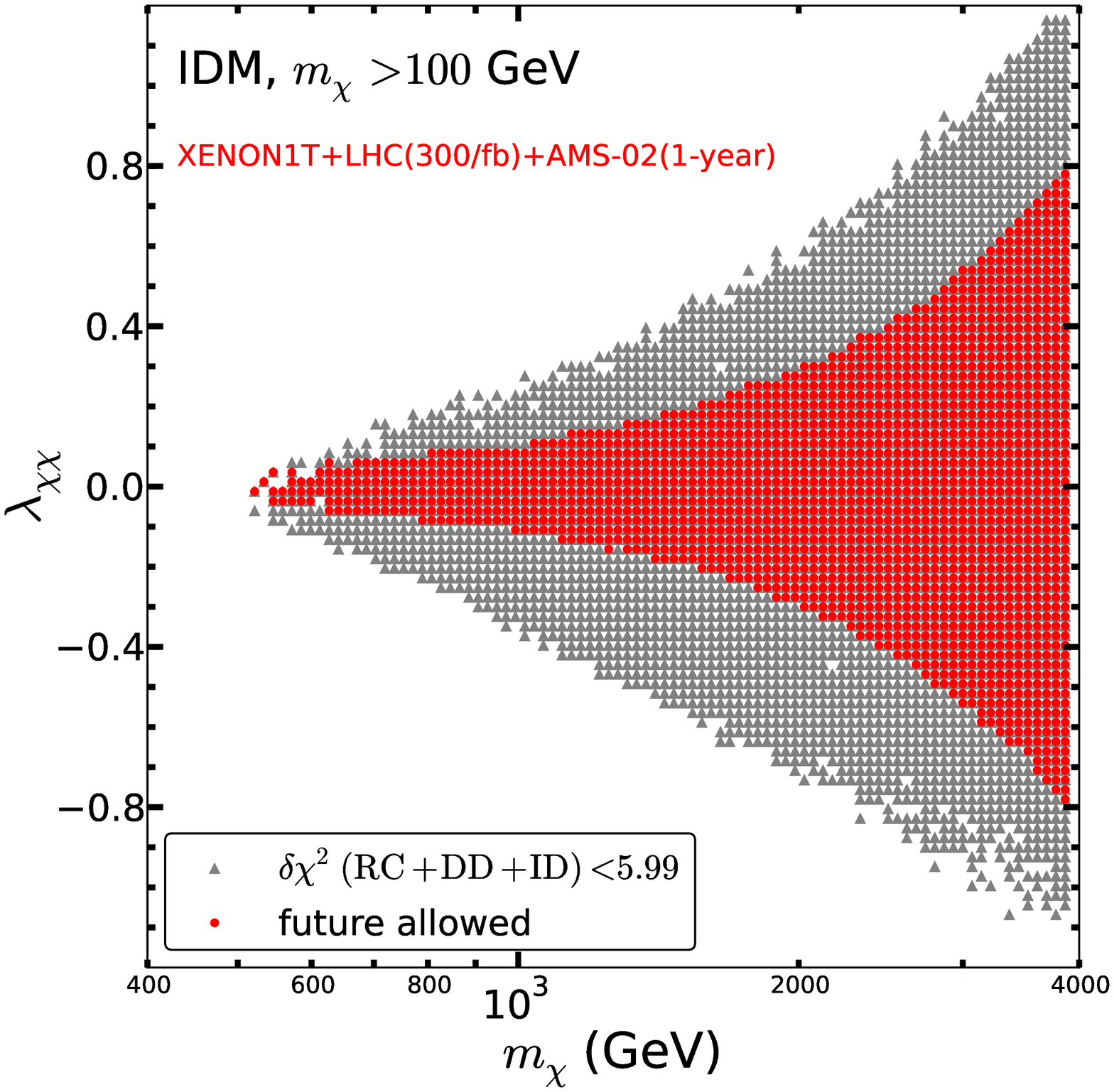}\\
\caption{\label{fig:mx_lamL_allfuture}
The scatter plots to present the future combined sensitivity from 
XENON1T, LHC-14 monojet (300 fb$^{-1}$), and AMS-02 antiproton flux measurements 
on ($m_\chi$,$\lambda_L$) plane 
subjected to the \textbf{RC+ID+DD} constraints in $2\sigma$ significance.
The left and right panels are for $m_\chi<100$ GeV and $m_\chi>500$ GeV respectively. 
}
\end{figure}

We finally show in Fig.~\ref{fig:mx_lamL_allfuture} 
the total impact from the combined sensitivities from the above three future experiments on the 
($m_\chi$,$\lambda_{\chi\chi}$) plane.
For the LHC-14 monojet, we will assume 300 fb$^{-1}$ of data in making these two plots.
Left and right panels correspond to
$m_\chi$ less than 100 GeV and greater than 500 GeV respectively.  
With all three future experiments sensitivities, 
from the left panel of a zoomed-in view of low $m_\chi$ region, we see that 
the lower limit of $m_\chi$ is lifted slightly from $52\gev$ to $55\gev$. 
In this low mass region where the invisible decay $h \to \chi \chi $ is open, 
while we do not found any upper limit on $m_\chi$, 
$\lambda_{\chi\chi}$ is found to lie between $-5\times 10^{-3}$ 
and $3\times 10^{-3}$. On the other hand, if the invisible mode is closed, the upper and lower limit 
of $\lambda_{\chi\chi}$ is varied with respect to $m_\chi$ (right panel). 
Comparing with current experimental data, these three future experiments sensitivities 
are robust but neither the lower $m_\chi$ region nor the larger $m_\chi$ region can be 
entirely ruled out.    

\section{Conclusions \label{section:8}}


Despite IHDM was proposed more than three decades ago, it is still 
one of the most simplest models and yet viable for scalar dark matter.
We have performed a global fit analysis on this model.
This analysis has been performed in light of
the recent ATLAS and CMS discovery of a  125-126 GeV Higgs-like particle,
taking into account the recent relic density measurement
by PLANCK, DM direct detection from LUX and indirect detection
from PAMELA, Fermi-LAT  and AMS-02.

We have shown that the constraint from DM direct detection search,
such as the latest LUX upper limit of year 2013,  provides a robust constraint on
the parameter space. 
In particular, if the invisible decay of the SM Higgs boson is open,
the upper limit of the Higgs invisible width from LHC together with the LUX constraint 
could put some interesting limits on the SM Higgs boson couplings to the DM in IHDM. Indeed,
an invisible decay of the SM Higgs boson with a branching ratio larger than 30\% 
and a scalar dark matter mass within the range of
20-60 GeV are excluded by current data.

We emphasize that there is no preference of $\chi=S$ or $A$ in our study. 
However, we found that $m_\chi\lesssim 100\gev$ region is slightly favoured 
by EWPT and \textbf{ID} constraints than $m_\chi\gtrsim 500\gev$ region. 
In addition, in the 95\% C.L. of \textbf{RC+DD+ID} constraints, 
$m_\chi$ has the lower limit around $52\gev$.

The exclusion power of the AMS-02 one year antiproton data on the model 
parameters can be inferred from Figs.~\ref{fig:ap_ams02} and \ref{fig:mx_lamL_allfuture}.
The results show that AMS-02 does have the potential to constrain
certain parameter space of IHDM, especially in the low mass region.
However, compared with the direct detection limit expected from XENON1T, the
constraint from AMS-02 antiproton data is weaker as comparing the two 
Figs.~\ref{fig:mx_sigv_XENON1T} and \ref{fig:ap_ams02}.

Nevertheless, future data from LHC-14 monojet, XENON1T direct detection and AMS-02 antiproton flux can 
further reduce the IHDM parameter space constrained by the existing {\bf RC+DD+ID} data, as is shown inevitably in 
the red regions as compared with the gray regions in the two plots at the left and right panels of Fig.~\ref{fig:mx_lamL_allfuture} for the low and high DM mass
regions respectively.

We also note that the likelihoods obtained in this work were obtained 
using the tree level relation for $\lambda_L$ and $\lambda_A$ 
(Eq.~(\ref{eq:lambda_A})) and the tree level formula for the 
coupling $g_{h \chi \chi}$ (Eq.~(\ref{eq:lamchi})). 
Higher order corrections will necessarily modify these relations and 
hence the relic density prediction will be affected. The profile likelihoods for the IHDM will be 
modified as well. 
Although loop corrections
have been shown to affect significantly the DM scattering cross-section
on nucleons in the IHDM~\cite{Klasen:2013btp} and thus modify the impact of 
direct detection searches on the viable parameter space of the model,
it is beyond the scope of the present analysis to take them into
account. One should bear in mind that such corrections will also
probably affect the positions of the best-fit points.

In summary, we note that the overall shapes of the 95\% C.L. contours presented in this work 
are mainly determined by the PLANCK relic density measurement.  
LHC monojet is only relevant when the invisible decay of the Higgs 
is open, in which case we obtain the limit for the coupling $\lambda_{L,A} \approx 10^{-1}$ 
which is not very stringent.
On the other hand, our best-fit points are
located at $m_\chi \approx$ 70 GeV and so do not allow the opening for 
invisible decay of the SM Higgs into DM.
The current \textbf{ID} and \textbf{DD} data are only sensitive to 
Higgs resonance region. Except for relic density constraint, currently
no other experimental data sets are sensitive to the IHDM parameter space at 
$m_\chi\gtrsim 500\gev$.
However, one expects future XENON1T can probe this region. 
Moreover, future instruments such as DAMPE, GAMMA-400 and CTA 
will test this higher $m_\chi$ region as well \cite{Garcia-Cely:2013zga}.

As mentioned previously, 
experimental search for the inert Higgs (neutral and charged) behaves like SUSY 
search for charginos and neutralinos. Therefore, one of the most popular 
signatures for inert Higgs searches would be also trilepton and/or dilepton 
plus missing $E_T$
\cite{Ball:2007zza,Dolle:2009ft}. 
A dedicated analysis for the IHDM has been
performed in \cite{Dolle:2009ft} where it has been demonstrated that 
the experimental reach for the inert Higgses is only about 300 GeV
which is somewhat smaller than the LHC reach for the charginos and neutralinos. 
The main reason is  that the cross section for the scalar pair production 
$pp\to H^\pm A^0$ is smaller than the gaugino pair production 
$pp\to \chi_1^\pm \chi_2^0$.
As we have seen in our analysis,
inert Higgses with masses $\geq 0.5-4$ TeV are consistent with all 
experimental and
theoretical constraints. One concludes that the IHDM
is here to stay for another decade.

The IHDM can be further extended by including inert right-handed neutrinos 
with Majorana masses \cite{Ma:2006km}. Masses for the 
SM light neutrinos can be generated through radiative processes 
with only inert particles running inside the loop. 
This extension of IHDM would exhibit intricate interplay 
between dark matter and neutrino physics.
Detailed global analysis of this extended model is also quite interesting and 
will be presented in a future publication.

\newpage

\section*{Acknowledgments}
AA would like to thank the National Science Council (NSC) of Taiwan 
for financial support during his stay at the Institute of Physics (IOP), Academia Sinica (AS). 
ST would like to thank the hospitality and support from the 
High Energy Theory Group at the IOP, AS.
This work is supported by National Natural Science Foundation of China under 
grant number 11105155 (QY) and by the NSC of Taiwan under project number 
102-2811-M-001-032 (AA) and grant number 101-2112-M-001-005-MY3 (TCY).



\begin{thebibliography}{99}
\bibitem {ATLAS}
G.~Aad {\it et al.}  [ATLAS Collaboration],
  Phys.\ Lett.\ B {\bf 716}, 1 (2012)
  [arXiv:1207.7214 [hep-ex]].

\bibitem{CMS}
S.~Chatrchyan {\it et al.}  [CMS Collaboration],
  Phys.\ Lett.\ B {\bf 716}, 30 (2012)
  [arXiv:1207.7235 [hep-ex]].

\bibitem{Tevatron}
T.~Aaltonen {\it et al.}  [CDF and D0 Collaborations],
  Phys.\ Rev.\ Lett.\  {\bf 109}, 071804 (2012)
  [arXiv:1207.6436 [hep-ex]];
  Phys.\ Rev.\ D {\bf 88}, 052014 (2013)
  [arXiv:1303.6346 [hep-ex]].
  
\bibitem {ATLASn}
G.~Aad {\it et al.}  [ATLAS Collaboration],
  Phys.\ Lett.\ B {\bf 726}, 88 (2013)
  [arXiv:1307.1427 [hep-ex]];
  ATLAS-CONF-2013-079, 19-July-2013.

\bibitem {CMSn}
S.~Chatrchyan {\it et al.}  [CMS Collaboration],
  JHEP {\bf 06}, 081 (2013)
  [arXiv:1303.4571 [hep-ex]];
 CMS-PAS-HIG-13-005, 17-April-2013.

\bibitem{CMS-PAS-HIG-12-045}
[CMS Collaboration],
  CMS-PAS-HIG-12-045;
[The ATLAS \& CMS Collaborations],
ATLAS-CONF-2011-157/CMS-PAS-HIG-11-023.

\bibitem{Carmi:2012yp} 
  D.~Carmi, A.~Falkowski, E.~Kuflik and T.~Volansky,
  JHEP {\bf 1207}, 136 (2012)
  [arXiv:1202.3144 [hep-ph]];
A.~Azatov, R.~Contino and J.~Galloway,
  JHEP {\bf 1204}, 127 (2012)
  [Erratum-ibid.\  {\bf 1304}, 140 (2013)]
  [arXiv:1202.3415 [hep-ph]];
M.~Klute, R.~Lafaye, T.~Plehn, M.~Rauch and D.~Zerwas,
  Phys.\ Rev.\ Lett.\  {\bf 109}, 101801 (2012)
  [arXiv:1205.2699 [hep-ph]];
I.~Low, J.~Lykken and G.~Shaughnessy,
  Phys.\ Rev.\ D {\bf 86}, 093012 (2012)
  [arXiv:1207.1093 [hep-ph]];
A.~Azatov, S.~Chang, N.~Craig and J.~Galloway,
  Phys.\ Rev.\ D {\bf 86}, 075033 (2012)
  [arXiv:1206.1058 [hep-ph]];
J.~R.~Espinosa, C.~Grojean, M.~Muhlleitner and M.~Trott,
  JHEP {\bf 1205}, 097 (2012)
  [arXiv:1202.3697 [hep-ph]];
J.~Ellis and T.~You,
  JHEP {\bf 1209}, 123 (2012)
  [arXiv:1207.1693 [hep-ph]];
P.~P.~Giardino, K.~Kannike, M.~Raidal and A.~Strumia,
  Phys.\ Lett.\ B {\bf 718}, 469 (2012)
  [arXiv:1207.1347 [hep-ph]].

\bibitem {atlaspin}
G.~Aad {\it et al.}  [ATLAS Collaboration],
  arXiv:1307.1432 [hep-ex];
G.~Aad \textit{et al.} [ATLAS Collaboration]: 
  ATLAS-CONF-2013-040.

\bibitem {cmspin}
S.~Chatrchyan {\it et al.}  [CMS Collaboration],
  Phys.\ Rev.\ Lett.\  {\bf 110}, 081803 (2013)
  [arXiv:1212.6639 [hep-ex]];
  CMS-PAS-HIG-13-002.

\bibitem{Deshpande:1977rw} 
N.~G.~Deshpande and E.~Ma,
Phys.\ Rev.\ D {\bf 18}, 2574 (1978).

\bibitem{THDM} 
  For a review of the general two Higgs doublet models, see
  G.~C.~Branco, P.~M.~Ferreira, L.~Lavoura, M.~N.~Rebelo, M.~Sher and J.~P.~Silva,
  Phys.\ Rept.\  {\bf 516}, 1 (2012)
  [arXiv:1106.0034 [hep-ph]].
 
\bibitem{globalCLT}
K.~Cheung, J.~S.~Lee and P.~-Y.~Tseng,
  JHEP {\bf 1305}, 134 (2013)
  [arXiv:1302.3794 [hep-ph]].
  
\bibitem{globalBelangeretal1}  
G.~Belanger, B.~Dumont, U.~Ellwanger, J.~F.~Gunion and S.~Kraml,
  Phys.\ Lett.\ B {\bf 723}, 340 (2013)
  [arXiv:1302.5694 [hep-ph]].

\bibitem{globalBelangeretal2}
  G.~Belanger, B.~Dumont, U.~Ellwanger, J.~F.~Gunion and S.~Kraml,
  Phys.\ Rev.\ D {\bf 88}, 075008 (2013)
  [arXiv:1306.2941 [hep-ph]].
  
\bibitem{globalOthers}  
J.~R.~Espinosa, M.~Muhlleitner, C.~Grojean and M.~Trott,
  JHEP {\bf 1209}, 126 (2012)
  [arXiv:1205.6790 [hep-ph]];
O.~Lebedev, H.~M.~Lee and Y.~Mambrini,
  Phys.\ Lett.\ B {\bf 707}, 570 (2012)
  [arXiv:1111.4482 [hep-ph]];
C.~Englert, M.~Spannowsky and C.~Wymant,
  Phys.\ Lett.\ B {\bf 718}, 538 (2012)
  [arXiv:1209.0494 [hep-ph]].

\bibitem{Ma:2006km} 
  E.~Ma,
  Phys.\ Rev.\ D {\bf 73}, 077301 (2006)
  [hep-ph/0601225].

\bibitem{wip}
  A. Arhrib, Y.-L. Sming Tsai, Q. Yuan and T.~C. Yuan, work in progress.
  
\bibitem{Silveira-Zee} 
  V.~Silveira and A.~Zee,
  Phys.\ Lett.\ B {\bf 161}, 136 (1985).

\bibitem{Cheung:2012xb} 
  K.~Cheung, Y.~-L.~S.~Tsai, P.~-Y.~Tseng, T.~-C.~Yuan and A.~Zee,
  JCAP {\bf 1210}, 042 (2012)
  [arXiv:1207.4930 [hep-ph]].

\bibitem{higgsportal-scalar1}
J.~McDonald,
  Phys.\ Rev.\ D {\bf 50}, 3637 (1994);
C.~P.~Burgess, M.~Pospelov and T.~ter Veldhuis,
  Nucl.\ Phys.\ B {\bf 619}, 709 (2001);
M.~Cirelli, N.~Fornengo and A.~Strumia,
  Nucl.\ Phys.\ B {\bf 753}, 178 (2006).

\bibitem{higgsportal-scalar2}
 M.~C.~Bento, O.~Bertolami, R.~Rosenfeld and L.~Teodoro,
 Phys.\ Rev.\  D {\bf 62}, 041302 (2000); 
 V.~Barger, P.~Langacker, M.~McCaskey, M.~J.~Ramsey-Musolf and G.~Shaughnessy,
 Phys.\ Rev.\  D {\bf 77}, 035005 (2008); 
  M.~Aoki, S.~Kanemura and O.~Seto,
  Phys.\ Rev.\ Lett.\  {\bf 102}, 051805 (2009)
  [arXiv:0807.0361 [hep-ph]];
  Phys.\ Rev.\ D {\bf 80}, 033007 (2009)
  [arXiv:0904.3829 [hep-ph]].

\bibitem{higgsportal-scalar4}
  K.~Cheung and T.~C.~Yuan,
  Phys.\ Lett.\  B {\bf 685}, 182 (2010);
   X.~G.~He, T.~Li, X.~Q.~Li, J.~Tandean and H.~C.~Tsai,
   Phys.\ Lett.\  B {\bf 688}, 332 (2010);
  M.~Farina, D.~Pappadopulo and A.~Strumia,
  Phys.\ Lett.\  B {\bf 688}, 329 (2010);
  M.~Kadastik, K.~Kannike, A.~Racioppi and M.~Raidal,
  Phys.\ Lett.\  B {\bf 685}, 182 (2010)
  arXiv:0912.3797 [hep-ph];
 M.~Aoki, S.~Kanemura and O.~Seto,
  Phys.\ Lett.\  B {\bf 685}, 313 (2010);
 M.~Asano and R.~Kitano,
 Phys.\ Rev.\  D {\bf 81}, 054506 (2010)
 [arXiv:1001.0486 [hep-ph]];
  A.~Bandyopadhyay, S.~Chakraborty, A.~Ghosal and D.~Majumdar,
  JHEP {\bf 1011}, 065 (2010)
  [arXiv:1003.0809 [hep-ph]];
S.~Andreas, C.~Arina, T.~Hambye, F.~-S.~Ling and M.~H.~G.~Tytgat,
  Phys.\ Rev.\ D {\bf 82}, 043522 (2010)
  [arXiv:1003.2595 [hep-ph]].

\bibitem{Gustafsson:2012aj} 
  M.~Gustafsson, S.~Rydbeck, L.~Lopez-Honorez and E.~Lundstrom,
  Phys.\ Rev.\ D {\bf 86}, 075019 (2012)
  [arXiv:1206.6316 [hep-ph]].

\bibitem{Goudelis:2013uca} 
  A.~Goudelis, B.~Herrmann and O.~St{\aa}l,
  JHEP {\bf 1309}, 106 (2013)
  [arXiv:1303.3010 [hep-ph]].
  
\bibitem{LopezHonorez:2010tb}
  L.~Lopez Honorez and C.~E.~Yaguna,
  JCAP {\bf 1101} (2011) 002
  [arXiv:1011.1411 [hep-ph]].

\bibitem{idmdm}
  L.~Lopez Honorez, E.~Nezri, J.~F.~Oliver and M.~H.~G.~Tytgat,
  JCAP {\bf 0702}, 028 (2007)
  [hep-ph/0612275];
  T.~Hambye and M.~H.~G.~Tytgat,
  Phys.\ Lett.\ B {\bf 659}, 651 (2008)
  [arXiv:0707.0633 [hep-ph]];
%
  P.~Agrawal, E.~M.~Dolle and C.~A.~Krenke,
  Phys.\ Rev.\ D {\bf 79}, 015015 (2009)
  [arXiv:0811.1798 [hep-ph]];
  T.~Hambye, F.~-S.~Ling, L.~Lopez Honorez and J.~Rocher,
  JHEP {\bf 0907}, 090 (2009)
  [Erratum-ibid.\  {\bf 1005}, 066 (2010)]
  [arXiv:0903.4010 [hep-ph]];
  E.~Nezri, M.~H.~G.~Tytgat and G.~Vertongen,
  JCAP {\bf 0904}, 014 (2009)
  [arXiv:0901.2556 [hep-ph]];
%
  S.~Andreas, M.~H.~G.~Tytgat and Q.~Swillens,
  JCAP {\bf 0904}, 004 (2009)
  [arXiv:0901.1750 [hep-ph]]; 
%
  C.~Arina, F.~-S.~Ling and M.~H.~G.~Tytgat,
  JCAP {\bf 0910}, 018 (2009)
  [arXiv:0907.0430 [hep-ph]]; 
%
  L.~Lopez Honorez and C.~E.~Yaguna,
  JHEP {\bf 1009}, 046 (2010)
  [arXiv:1003.3125 [hep-ph]]; 
%
  A.~Melfo, M.~Nemevsek, F.~Nesti, G.~Senjanovic and Y.~Zhang,
  Phys.\ Rev.\ D {\bf 84}, 034009 (2011)
  [arXiv:1105.4611 [hep-ph]];
  M.~Krawczyk, D.~Sokolowska, P.~Swaczyna and B.~Swiezewska,
  JHEP {\bf 1309}, 055 (2013)
  [arXiv:1305.6266 [hep-ph]];
  M.~Gustafsson, S.~Rydbeck, L.~Lopez-Honorez and E.~Lundstrom,
  Phys.\ Rev.\ D {\bf 86}, 075019 (2012)
  [arXiv:1206.6316 [hep-ph]];
%
  E.~M.~Dolle and S.~Su,
  Phys.\ Rev.\ D {\bf 80}, 055012 (2009)
  [arXiv:0906.1609 [hep-ph]];
S.~Kanemura, Y.~Okada, H.~Taniguchi and K.~Tsumura,
  Phys.\ Lett.\ B {\bf 704}, 303 (2011)
  [arXiv:1108.3297 [hep-ph]].

\bibitem{unitarity}
A.~Arhrib, R.~Benbrik and N.~Gaur,
  Phys.\ Rev.\ D {\bf 85}, 095021 (2012)
  [arXiv:1201.2644 [hep-ph]].

\bibitem{maria}
B.~Swiezewska and M.~Krawczyk,
  Phys.\ Rev.\ D {\bf 88}, 035019 (2013)
  [arXiv:1212.4100 [hep-ph]];
  arXiv:1305.7356 [hep-ph].

\bibitem{idm_lhc}
  Q.~-H.~Cao, E.~Ma and G.~Rajasekaran,
  Phys.\ Rev.\ D {\bf 76}, 095011 (2007)
  [arXiv:0708.2939 [hep-ph]].

%
\bibitem{Dolle:2009ft} 
  E.~Dolle, X.~Miao, S.~Su and B.~Thomas,
  Phys.\ Rev.\ D {\bf 81}, 035003 (2010)
  [arXiv:0909.3094 [hep-ph]];
%
  X.~Miao, S.~Su and B.~Thomas,
  Phys.\ Rev.\ D {\bf 82}, 035009 (2010)
  [arXiv:1005.0090 [hep-ph]].

\bibitem{Lundstrom:2008ai} 
  E.~Lundstrom, M.~Gustafsson and J.~Edsjo,
  Phys.\ Rev.\ D {\bf 79}, 035013 (2009)
  [arXiv:0810.3924 [hep-ph]].

\bibitem{Barbieri:2006dq} 
  R.~Barbieri, L.~J.~Hall and V.~S.~Rychkov,
  Phys.\ Rev.\ D {\bf 74}, 015007 (2006)
  [hep-ph/0603188].

 \bibitem{Ade:2013zuv} 
  P.~A.~R.~Ade {\it et al.}  [Planck Collaboration],
  arXiv:1303.5076 [astro-ph.CO].

\bibitem{Akerib:2013tjd} 
  D.~S.~Akerib {\it et al.}  [LUX Collaboration],
  arXiv:1310.8214 [astro-ph.CO].

\bibitem{Ackermann:2011wa} 
  M.~Ackermann {\it et al.}  [Fermi-LAT Collaboration],
  Phys.\ Rev.\ Lett.\  {\bf 107}, 241302 (2011)
  [arXiv:1108.3546 [astro-ph.HE]].

\bibitem{Huang:2012yf} 
  X.~Huang, Q.~Yuan, P.~-F.~Yin, X.~-J.~Bi and X.~Chen,
  JCAP {\bf 1211}, 048 (2012)
  [Erratum-ibid.\  {\bf 1305}, E02 (2013)]
  [arXiv:1208.0267 [astro-ph.HE]].

\bibitem{Aguilar:2013qda}
  M.~Aguilar {\it et al.}  [AMS Collaboration],
  Phys.\ Rev.\ Lett.\  {\bf 110}, 141102 (2013).

\bibitem{Adriani:2008zr}
  O.~Adriani {\it et al.}  [PAMELA Collaboration],
  Nature {\bf 458}, 607 (2009)  [arXiv:0810.4995 [astro-ph]].

\bibitem{FermiLAT:2011ab} 
  M.~Ackermann {\it et al.}  [Fermi LAT Collaboration],
  Phys.\ Rev.\ Lett.\  {\bf 108}, 011103 (2012)
  [arXiv:1109.0521 [astro-ph.HE]].

\bibitem{Ackermann:2010ij}
  M.~Ackermann {\it et al.}  [Fermi LAT Collaboration],
  Phys.\ Rev.\ D {\bf 82}, 092004 (2010) [arXiv:1008.3999 [astro-ph.HE]].

\bibitem{Adriani:2010rc} 
  O.~Adriani {\it et al.}  [PAMELA Collaboration],
  Phys.\ Rev.\ Lett.\  {\bf 105}, 121101 (2010)
  [arXiv:1007.0821 [astro-ph.HE]].

\bibitem{Aprile:2012nq} 
  E.~Aprile {\it et al.}  [XENON100 Collaboration],
  Phys.\ Rev.\ Lett.\  {\bf 109}, 181301 (2012)
  [arXiv:1207.5988 [astro-ph.CO]].

\bibitem{Gustafsson:2007pc} 
  M.~Gustafsson, E.~Lundstrom, L.~Bergstrom and J.~Edsjo,
  Phys.\ Rev.\ Lett.\  {\bf 99}, 041301 (2007)
  [astro-ph/0703512 [ASTRO-PH]].

\bibitem{Ref:PeskinTakeuchi}
M.~E.~Peskin and T.~Takeuchi,
  Phys.\ Rev.\  D {\bf 46} (1992) 381.

\bibitem{Baak:2012kk} 
  M.~Baak, M.~Goebel, J.~Haller, A.~Hoecker, D.~Kennedy, R.~Kogler, K.~Moenig and M.~Schott {\it et al.},
  Eur.\ Phys.\ J.\ C {\bf 72}, 2205 (2012)
  [arXiv:1209.2716 [hep-ph]].

\bibitem{pdg} 
  K.~Nakamura {\it et al.}  [Particle Data Group Collaboration],
  J.\ Phys.\ G G {\bf 37}, 075021 (2010).

\bibitem{Pierce:2007ut} 
  A.~Pierce and J.~Thaler,
  JHEP {\bf 0708}, 026 (2007)
  [hep-ph/0703056 [HEP-PH]].

\bibitem{Acciarri:1999km} 
  M.~Acciarri {\it et al.}  [L3 Collaboration],
  Phys.\ Lett.\ B {\bf 472}, 420 (2000)
  [hep-ex/9910007];
R.~Barate {\it et al.}  [ALEPH Collaboration],
  Phys.\ Lett.\ B {\bf 499}, 67 (2001)
  [hep-ex/0011047];
J.~Abdallah {\it et al.}  [DELPHI Collaboration],
  Eur.\ Phys.\ J.\ C {\bf 31}, 421 (2003)
  [hep-ex/0311019];
G.~Abbiendi {\it et al.}  [OPAL Collaboration],
  Eur.\ Phys.\ J.\ C {\bf 35}, 1 (2004)
  [hep-ex/0401026].

\bibitem{higgs-invisible-atlas}
  ATLAS Collaboration,
  ATLAS-CONF-2013-011.

\bibitem{higgs-invisible-cms}
CMS Collaboration, CMS-PAS-HIG-13-018.

\bibitem{higgs-invisible-cmsVBF}
CMS Collaboration, CMS-PAS-HIG-13-013.

\bibitem{ATLAS:2013oma} 
  ATLAS Collaboration,
  ATLAS-CONF-2013-012.

\bibitem{Palmer:2013xza} 
  C.~Palmer [CMS Collaboration],
  arXiv:1305.3654 [hep-ex];
S.~Chatrchyan \textit{et al.} [CMS Collaboration], CMS-PAS-HIG-13-001.

\bibitem{monojet1} 
CMS Collaboration,
  CMS-PAS-EXO-12-048.

\bibitem{monojet2} 
ATLAS Collaboration,
  ATLAS-CONF-2012-147.

\bibitem{micro} 
 G.~Belanger, F.~Boudjema, A.~Pukhov and A.~Semenov,
  Comput.\ Phys.\ Commun.\  {\bf 180}, 747 (2009)
  [arXiv:0803.2360 [hep-ph]];
G.~Belanger, F.~Boudjema, A.~Pukhov and A.~Semenov,
  Nuovo Cim.\ C {\bf 033N2}, 111 (2010)
  [arXiv:1005.4133 [hep-ph]].

\bibitem{Larson:2010gs} 
D.~Larson, J.~Dunkley, G.~Hinshaw, E.~Komatsu, M.~R.~Nolta, C.~L.~Bennett, B.~Gold and M.~Halpern {\it et al.},
  Astrophys.\ J.\ Suppl.\  {\bf 192}, 16 (2011)
  [arXiv:1001.4635 [astro-ph.CO]].

\bibitem{Griest:1990kh} 
  K.~Griest and D.~Seckel,
  Phys.\ Rev.\ D {\bf 43}, 3191 (1991).

\bibitem{Binetruy:1983jf} 
  P.~Binetruy, G.~Girardi and P.~Salati,
  Nucl.\ Phys.\ B {\bf 237}, 285 (1984).


\bibitem{GeringerSameth:2011iw} 
  A.~Geringer-Sameth and S.~M.~Koushiappas,
  Phys.\ Rev.\ Lett.\  {\bf 107}, 241303 (2011)
  [arXiv:1108.2914 [astro-ph.CO]].

\bibitem{Tsai:2012cs} 
  Y.~-L.~S.~Tsai, Q.~Yuan and X.~Huang,
  JCAP {\bf 1303}, 018 (2013)
  [arXiv:1212.3990 [astro-ph.HE]].

\bibitem{Fermi:2011bm}
  Fermi-LAT Collaboration,
  Astrophys.\ J.\ Suppl.\  {\bf 199}, 31 (2012)
  [arXiv:1108.1435 [astro-ph.HE]].

\bibitem{Chang:2008aa}
  J.~Chang, J.~H.~Adams, H.~S.~Ahn, G.~L.~Bashindzhagyan, M.~Christl, O.~Ganel, T.~G.~Guzik and J.~Isbert {\it et al.},
  Nature {\bf 456}, 362 (2008).  

\bibitem{Aharonian:2008aa}
  F.~Aharonian {\it et al.}  [H.E.S.S. Collaboration],
  Phys.\ Rev.\ Lett.\  {\bf 101}, 261104 (2008) [arXiv:0811.3894 [astro-ph]];
  F.~Aharonian {\it et al.}  [H.E.S.S. Collaboration],
  Astron.\ Astrophys.\  {\bf 508}, 561 (2009) [arXiv:0905.0105 [astro-ph.HE]].


\bibitem{BorlaTridon:2011dk} 
  D.~Borla Tridon {\it et al.}  [MAGIC Collaboration],
  arXiv:1110.4008 [astro-ph.HE].

\bibitem{Adriani:2011xv}
  O.~Adriani {\it et al.}  [PAMELA Collaboration],
  Phys.\ Rev.\ Lett.\  {\bf 106}, 201101 (2011)  
  [arXiv:1103.2880 [astro-ph.HE]].

\bibitem{review}
  X.~-G.~He,
  Mod.\ Phys.\ Lett.\ A {\bf 24}, 2139 (2009)
  [arXiv:0908.2908 [hep-ph]];
  Y.~-Z.~Fan, B.~Zhang and J.~Chang,
  Int.\ J.\ Mod.\ Phys.\ D {\bf 19}, 2011 (2010)
  [arXiv:1008.4646 [astro-ph.HE]];
  P.~D.~Serpico,
  Astropart.\ Phys.\  {\bf 39-40}, 2 (2012)
  [arXiv:1108.4827 [astro-ph.HE]];
  M.~Cirelli,
  Pramana {\bf 79}, 1021 (2012)
  [arXiv:1202.1454 [hep-ph]].

\bibitem{Bertone:2008xr}
  G.~Bertone, M.~Cirelli, A.~Strumia and M.~Taoso,
  JCAP {\bf 0903}, 009 (2009)
  [arXiv:0811.3744 [astro-ph]];
  L.~Bergstrom, G.~Bertone, T.~Bringmann, J.~Edsjo and M.~Taoso,
  Phys.\ Rev.\ D {\bf 79}, 081303 (2009)
  [arXiv:0812.3895 [astro-ph]];
  M.~Cirelli, P.~Panci and P.~D.~Serpico,
  Nucl.\ Phys.\ B {\bf 840}, 284 (2010)
  [arXiv:0912.0663 [astro-ph.CO]];
  M.~Papucci and A.~Strumia,
  JCAP {\bf 1003}, 014 (2010)
  [arXiv:0912.0742 [hep-ph]].

\bibitem{Grasso:2009ma}
  D.~Grasso {\it et al.}  [FERMI-LAT Collaboration],
  Astropart.\ Phys.\  {\bf 32}, 140 (2009)
  [arXiv:0905.0636 [astro-ph.HE]];
  T.~Linden and S.~Profumo,
  Astrophys.\ J.\  {\bf 772}, 18 (2013)
  [arXiv:1304.1791 [astro-ph.HE]];
  P.~-F.~Yin, Z.~-H.~Yu, Q.~Yuan and X.~-J.~Bi,
  Phys.\ Rev.\ D {\bf 88}, 023001 (2013)
  [arXiv:1304.4128 [astro-ph.HE]].

\bibitem{Bergstrom:2013jra} 
  L.~Bergstrom, T.~Bringmann, I.~Cholis, D.~Hooper and C.~Weniger,
  Phys.\ Rev.\ Lett.\  {\bf 111}, 171101 (2013)
  [arXiv:1306.3983 [astro-ph.HE]].

\bibitem{Yuan:2013eja}
  Q.~Yuan, X.~-J.~Bi, G.~-M.~Chen, Y.~-Q.~Guo, S.~-J.~Lin and X.~Zhang,
  arXiv:1304.1482 [astro-ph.HE];
  Q.~Yuan and X.~-J.~Bi,
  Phys.\ Lett.\ B {\bf 727}, 1 (2013)
  [arXiv:1304.2687 [astro-ph.HE]].
  
\bibitem{Gleeson:1968zza}
  L.~J.~Gleeson and W.~I.~Axford,
  Astrophys.\ J.\  {\bf 154}, 1011 (1968).

\bibitem{Adriani:2011cu} 
  O.~Adriani {\it et al.}  [PAMELA Collaboration],
  Science {\bf 332}, 69 (2011)
  [arXiv:1103.4055 [astro-ph.HE]].

\bibitem{DeSimone:2013fia} 
  A.~De Simone, A.~Riotto and W.~Xue,
  JCAP {\bf 1305}, 003 (2013)
  [arXiv:1304.1336 [hep-ph]];
  H.~-B.~Jin, Y.~-L.~Wu and Y.~-F.~Zhou,
  JCAP {\bf 1311}, 026 (2013)
  [arXiv:1304.1997 [hep-ph]].

\bibitem{Adriani:2008zq} 
  O.~Adriani, G.~C.~Barbarino, G.~A.~Bazilevskaya, R.~Bellotti, M.~Boezio, 
  E.~A.~Bogomolov, L.~Bonechi and M.~Bongi {\it et al.},
  Phys.\ Rev.\ Lett.\  {\bf 102}, 051101 (2009)
  [arXiv:0810.4994 [astro-ph]].

\bibitem{Donato:2008jk} 
  F.~Donato, D.~Maurin, P.~Brun, T.~Delahaye and P.~Salati,
  Phys.\ Rev.\ Lett.\  {\bf 102}, 071301 (2009)
  [arXiv:0810.5292 [astro-ph]];
  I.~Cholis,
  JCAP {\bf 1109}, 007 (2011)
  [arXiv:1007.1160 [astro-ph.HE]].

\bibitem{Aguilar:2002ad} 
  M.~Aguilar {\it et al.}  [AMS Collaboration],
  Phys.\ Rept.\  {\bf 366}, 331 (2002)
  [Erratum-ibid.\  {\bf 380}, 97 (2003)].

\bibitem{Asaoka:2001fv} 
  Y.~Asaoka, Y.~Shikaze, K.~Abe, K.~Anraku, M.~Fujikawa, H.~Fuke, M.~Imori and S.~Haino {\it et al.},
  Phys.\ Rev.\ Lett.\  {\bf 88}, 051101 (2002)
  [astro-ph/0109007].

\bibitem{BESS02}
Haino \textit{et al.},
Proc. 29th Int. Cosmic Ray Conf. 3, 13 (2005).

\bibitem{de Austri:2006pe} 
  R.~R.~de Austri, R.~Trotta and L.~Roszkowski,
  JHEP {\bf 0605}, 002 (2006)
  [hep-ph/0602028].

\bibitem{Roszkowski:2012uf} 
  L.~Roszkowski, E.~M.~Sessolo and Y.~-L.~S.~Tsai,
  Phys.\ Rev.\ D {\bf 86}, 095005 (2012)
  [arXiv:1202.1503 [hep-ph]].

\bibitem{Alwall:2011uj} 
  J.~Alwall, M.~Herquet, F.~Maltoni, O.~Mattelaer and T.~Stelzer,
  JHEP {\bf 1106}, 128 (2011)
  [arXiv:1106.0522 [hep-ph]].

\bibitem{GCexcess} 
  D.~Hooper and L.~Goodenough,
  Phys.\ Lett.\ B {\bf 697}, 412 (2011)
  [arXiv:1010.2752 [hep-ph]];
  A.~Boyarsky, D.~Malyshev and O.~Ruchayskiy,
  Phys.\ Lett.\ B {\bf 705}, 165 (2011)
  [arXiv:1012.5839 [hep-ph]];
  D.~Hooper and T.~Linden,
  Phys.\ Rev.\ D {\bf 84}, 123005 (2011)
  [arXiv:1110.0006 [astro-ph.HE]];
  K.~N.~Abazajian and M.~Kaplinghat,
  Phys.\ Rev.\ D {\bf 86}, 083511 (2012)
  [arXiv:1207.6047 [astro-ph.HE]];
  C.~Gordon and O.~Macías,
  arXiv:1306.5725 [astro-ph.HE].

\bibitem{Stahov:2012ca} 
  J.~Stahov, H.~Clement and G.~J.~Wagner,
  Phys.\ Lett.\ B {\bf 726}, 685 (2013)
  [arXiv:1211.1148 [nucl-th]].

\bibitem{Feroz:2008xx} 
  F.~Feroz, M.~P.~Hobson and M.~Bridges,
  Mon.\ Not.\ Roy.\ Astron.\ Soc.\  {\bf 398}, 1601 (2009)
  [arXiv:0809.3437 [astro-ph]].

\bibitem{Rolke:2004mj} 
  W.~A.~Rolke, A.~M.~Lopez and J.~Conrad,
  Nucl.\ Instrum.\ Meth.\ A {\bf 551}, 493 (2005)
  [physics/0403059].

\bibitem{Kanemura:2010sh} 
  S.~Kanemura, S.~Matsumoto, T.~Nabeshima and N.~Okada,
  Phys.\ Rev.\ D {\bf 82}, 055026 (2010)
  [arXiv:1005.5651 [hep-ph]].

  \bibitem{Djouadi:2012zc} 
  A.~Djouadi, A.~Falkowski, Y.~Mambrini and J.~Quevillon,
  Eur.\ Phys.\ J.\ C {\bf 73}, 2455 (2013)
  [arXiv:1205.3169 [hep-ph]].
  
\bibitem{Klasen:2013btp}
  M.~Klasen, C.~E.~Yaguna and J.~D.~Ruiz-Alvarez,
  Phys.\ Rev.\ D {\bf 87} (2013) 075025
  [arXiv:1302.1657 [hep-ph]].
  
\bibitem{Pato:2010ih} 
  M.~Pato, D.~Hooper and M.~Simet,
  JCAP {\bf 1006}, 022 (2010)
  [arXiv:1002.3341 [astro-ph.HE]].
  
\bibitem{geofac}
A. Malinlin,
\url{http://www.atic.umd.edu/pub/ams/pub07/Malinin.LomonosovConf07.pdf}

\bibitem{Garcia-Cely:2013zga} 
  C.~Garcia-Cely and A.~Ibarra,
  JCAP {\bf 1309}, 025 (2013)
  [arXiv:1306.4681 [hep-ph]].

\bibitem{Ball:2007zza}
 G.~L.~Bayatian {\it et al.}  [CMS Collaboration],
 J.\ Phys.\ G {\bf 34}, 995 (2007).


\end{thebibliography}
\end{document}